\documentclass[12pt]{iopart}
\usepackage{bm,url}
\usepackage{iopams}
\usepackage{graphicx}

\usepackage[all]{xy} 

\usepackage{color}
\usepackage{soul}

\begin{document}

\title[Vortex-bound solitons in topological superfluid $^3$He]{Vortex-bound solitons in topological superfluid $^3$He}

\author{J. T. M\"akinen$^1$, K. Zhang$^{1,2,3}$, and V. B. Eltsov$^1$}

\address{$^1$ Department of Applied Physics, Aalto University, FI-00076 Aalto, Finland}
\address{$^2$ Department of Physics and Helsinki Institute of Physics, P.O. Box 64, FI-00014 University of Helsinki, Finland}
\address{$^3$ Department of Physics and Astronomy, University of Sussex, Falmer, Brighton BN1 9QH, United Kingdom}
\ead{vladimir.eltsov@aalto.fi}

\begin{abstract}
The different superfluid phases of $^3$He are described by $p$-wave order parameters that include anisotropy axes both in the orbital and spin spaces. The anisotropy axes characterize the broken symmetries in these macroscopically coherent quantum many-body systems. The systems' free energy has several degenerate minima for certain orientations of the anisotropy axes. As a result, spatial variation of the order parameter between two such regions, settled in different energy minima, forms a topological soliton. Such solitons can terminate in the bulk liquid, where the termination line forms a vortex with trapped circulation of mass and spin superfluid currents. Here we discuss possible soliton-vortex structures based on the symmetry and topology arguments and focus on the three structures observed in experiments: solitons bounded by spin-mass vortices in the B phase, solitons bounded by half-quantum vortices in the polar and polar-distorted A phases, and the composite defect fomed by a half-quantum vortex, soliton and the Kibble-Lazarides-Shafi wall in the polar-distorted B phase. The observations are based on nuclear magnetic resonance (NMR) techniques and are of three types: first, solitons can form a potential well for trapped spin waves, observed as an extra peak in the NMR spectrum at shifted frequency; second, they can increase the relaxation rate of the NMR spin precession; lastly, the soliton can present the boundary conditions for the anisotropy axes in bulk, modifying the bulk NMR signal. Owing to solitons' prominent NMR signatures and the ability to manipulate their structure with external magnetic field, solitons have become an important tool for probing and controlling the structure and dynamics of superfluid $^3$He, in particular half-quantum vortices with core-bound Majorana modes.
\end{abstract}

%\maketitle

\section{Introduction}

Solitary waves, or solitons, often refer to localized wave packets that maintain their shape as they propagate through a medium or interact with other solitons. In the context of this paper, the word 'soliton' refers to continuous variation of the superfluid order parameter between two regions settled in distinct yet degenerate energy minima. Within the soliton the free energy is not minimized -- instead, it is stabilized by the topology of the uderlying superfluid. Topological solitons are protected by the same mechanism as linear topological defects, quantized vortices, on which solitons can terminate in bulk liquid. Such composite objects are in the focus of this work.

We discuss vortex-bound solitons in the $p$-wave superfluid $^3$He \cite{vollhardt2013superfluid}, where complex spontaneous symmetry breaking (SSB) patterns allow for the existence of a plethora of different topological objects \cite{Volovik1book}. The topological protection of composite objects is described by relative homotopy groups \cite{MineevBook1998}, which reflect on the relation between residual symmetries in two ordered phases. These phases may appear in a sequence of SSB phase transitions, similar to some scenarios of the evolution of the early Universe. Alternatively, they may refer to the same ordered state viewed at two distinct length scales set by different orientational energies. Both scenarios are realized in superfluid $^3$He and discussed in this paper: Several superfluid phases are observed in bulk and more are engineered with nanostructured confinement. In these phases condensation, magnetic anisotropy, and spin-orbit interaction energies provide the hierarchy of relevant length scales ranging from tens of nanometers to a fraction of a millimeter. Diversity of superfluid $^3$He even allows for the existense of an object where a quantized vortex both terminates a planar topological soliton and simultaneously serves as a staring line of another planar object, the Kibble-Lazarides-Shafi wall -- we call such composite defect a nexus.

Vortex-bound solitons provide an experimental handle to  probing topological defects, as the characteristic size of a soliton is often a few orders of magnitude larger than that of the accompanying linear defects. The nuclear magnetic resonance (NMR) signature of solitons may be used to identify such defects, including half-quantum vortices (HQVs) in the polar and polar-distorted superfluid phases \cite{Autti2016}. Interesting on their own right, HQVs can also provide experimental insight into grand unified theories (GUT) when taken through successive phase transitions \cite{Makinen2019}, or host Majorana zero modes in the chiral superfluid or superconducting phases \cite{PhysRevLett.86.268}. Experimental systems with known and sufficiently complex order parameters for testing the predictions of GUTs are extremely rare and therefore valuable, while Majorana zero modes are highly sought after for their promise in topological quantum computing scenarios \cite{KITAEV20032}. The superfluid phases of $^3$He provide experimental access to such scenarios; to date the polar-distorted A (PdA) phase of $^3$He is the only experimental platform known to host HQVs in a chiral superfluid. Therefore, understanding vortex-bound solitons is important for many research directions.

This work aims to provide the necessary theoretical background for understanding vortex-bound solitons in the superfluid phases of $^3$He, as well as review the relevant experimental results. In section \ref{sec:Residual} we discuss the symmetries that govern the order parameters of the normal fluid, as well as superfluid phases encountered in bulk or confined samples. We further discuss how these symmetries are linked to topological defects, and how they are affected by the presence of orienting energy terms. In section \ref{CompositeTopologicalObjects} we introduce relative homotopy groups and their connection to topological and composite defects, and solitons. The rest of the paper focuses on experimental aspects: Section \ref{SD} lays out the theoretical background for understanding the NMR measurements and solitons' effect on the observed NMR spectrum, while the experimental observations of different types of vortex-bound solitons in various superfluid phases of $^3$He are discussed in section \ref{sec:experiments}. Finally, we provide concluding remarks in section \ref{sec:conclusions}.

\section{Residual symmetries in $^3$He} \label{sec:Residual}

\subsection{Superfluid phases in the bulk liquid}
\label{BulkPhases}

At temperatures well below the Fermi temperature $T_{\mathrm{F}} \sim 1$~K but above the superfluid transition temperature $T_{\mathrm{c}} \sim 1\,$mK, $^3$He behaves as Fermi liquid, as described by Landau \cite{LandauFermiLiquid}. Properties of the Fermi liquid can be modeled through a weakly interacting gas of excitations in the vicinity of the Fermi surface. According to the Bardeen-Cooper-Schrieffer (BCS) theory \cite{PhysRev.106.162}, the presence of any attractive interaction, in the case of $^3$He arising from Van der Waals interaction accompanied by spin-fluctuation exchange mechanism \cite{vollhardt2013superfluid}, there exists a temperature below which the fermions tend to form a coherent macroscopic condensate via Cooper pairs. The spin-exchange mechanism favors $p$-wave pairing with angular momentum $l=1$. Simultaneously, the antisymmetry of the fermionic wave function requires $s=1$, where $s$ is the spin quantum number for the Cooper pair.

\begin{figure}[!ht] 
\begin{center}
\includegraphics[width=0.4\linewidth]{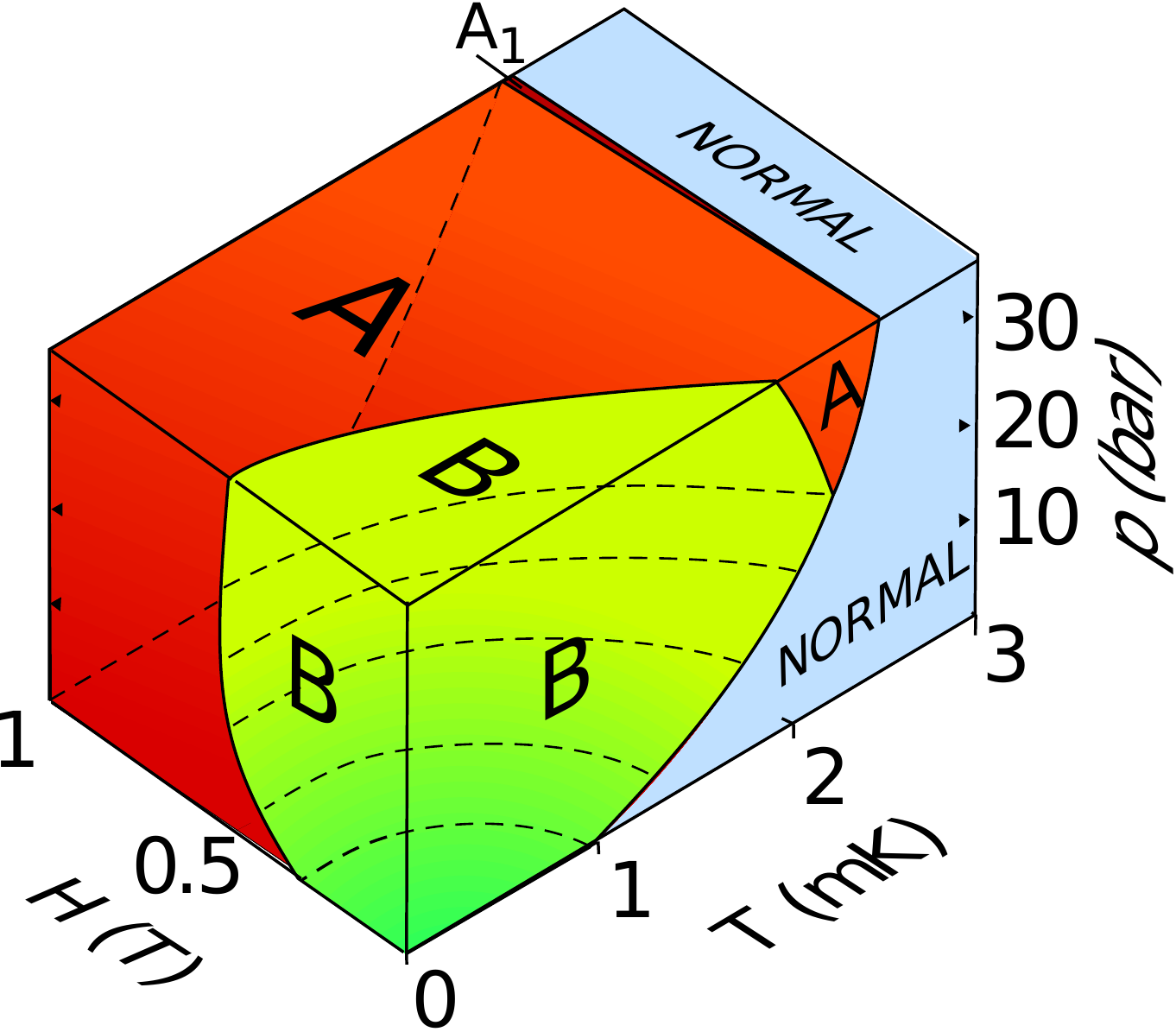}
\end{center}
\caption{\label{fig:bulk_phase_diagram}\textbf{Superfluid phase diagram of bulk $^3$He.} At 1\,bar pressure $^3$He liquefies at $\sim 3.2$~K. It remains liquid all the way to the absolute zero temperature for pressures $P \lesssim 30$~bar. In bulk fluid, three different superfluid phases are observed below the critical temperature -- the time-reversal symmetric B phase, the chiral equal spin-pairing A phase, and at high magnetic fields the A$_{1}$ phase which only allows Cooper pairs with both spins oriented along the magnetic field.}
\end{figure}

The formation of Cooper pairs becomes energetically favorable below the critical temperature $T_{\mathrm{c}}$, and the liquid undergoes a phase transition to the superfluid state. Above the superfluid transition bulk $^3$He is described by the symmetry group \cite{0953-8984-27-11-113203}
\begin{equation} \label{eq:bulk_symmetry}
G = SO(3)_{\mathbf{L}} \times SO(3)_{\mathbf{S}} \times U(1)_{\phi} \times T \times C  \times P,
\end{equation}
which includes continuous symmetries, three-dimensional rotations of coordinates $SO(3)_{\mathbf{L}}$, rotations of the spin space $SO(3)_{\mathbf{S}}$, and the global phase transformation group $U(1)_{\phi}$, as well as discrete symmetries, $T$ is the time-reversal symmetry, $C$ is the particle-hole conversion symmetry, and $P$ is the space parity symmetry. The transitions from normal fluid to superfluid phases as well as transitions between different superfluid phases are accompanied by spontaneous breaking of continuous and/or discrete symmetries in $G$. In bulk $^3$He three superfluid phases have been realized \cite{vollhardt2013superfluid}, Fig.~\ref{fig:bulk_phase_diagram}; the fully-gapped superfluid B phase characterized by broken relative spin-orbit symmetry, the chiral $p_{\mathrm{x}}+ip_{\mathrm{y}}$ superfluid A phase, and finally, the spin-polarized A$_1$ phase close to $T_{\rm c}$ in high magnetic fields. In the rest of the section~\ref{BulkPhases}, we discuss the symmetry-breaking patterns in the B and A phases, while the infomation on $A_{1}$ phase is included for completeness in \ref{A1Phase}.

%%%%%%%%%%%%%%%%%%%%%%%%%%%%%%%%%%%%%%%%%%%%%%%%%%%%%%%%%%%%%%%%%%%%%%%%%%%%%%%%%%%%%%%%%%%%%
%%%%%%%%%%%%%%%%%%%%%%%%%%%%%%%%%%%%%%%%%%%%%%%%%%%%%%%%%%%%%%%%%%%%%%%%%%%%%%%%%%%%%%%%%%%%%
%%%%%%%%%%%%%%%%%%%%%%%%%%%%%%%%%%%%%%%%%%%%%%%%%%%%%%%%%%%%%%%%%%%%%%%%%%%%%%%%%%%%%%%%%%%%%
%%%%%%%%%%%%%%%%%%%%%%%%%%%%%%%%%%%%%%%%%%%%%%%%%%%%%%%%%%%%%%%%%%%%%%%%%%%%%%%%%%%%%%%%%%%%%

\subsubsection{B phase}

In the absence of magnetic field, the superfluid B phase has the total angular momentum $\mathbf{j} = \mathbf{l} + \mathbf{s} = 0$. The requirement $\mathbf{j}=0$ allows three spin configurations for Cooper pairs, i.e. $s_{\mathrm{z}} = \{ -1,0,1 \}$ and, respectively, $l_{\mathrm{z}} = \{ 1,0,-1 \}$. The B phase is characterized by broken spin-orbit symmetry, i.e.\ the relative orientation of the spin- and orbital vectors becomes locked. It is reflected in the order parameter of the B phase~\cite{vollhardt2013superfluid}
\begin{equation}
A_{\mu j}^{\mathrm{B}} = \Delta_{\mathrm{B}} e^{i \phi} \mathbf{R}_{\mu j} (\mathbf{\hat{n}}, \theta),
\label{Bop}
\end{equation}
where $\Delta_{\mathrm{B}}$ is the pressure- and temperature-dependent superfluid gap characterizing the energy required to break a Cooper pair, $\phi$ is the superfluid phase, and matrix $\mathbf{R}(\mathbf{\hat{n}}, \theta)$ describes the rotation of spins with index $\mu$ relative to the orbital coordinates with index $j$ around the vector $\mathbf{\hat{n}}$ by angle $\theta$. In equilibrium the spin vector $\hat{\mathbf{s}}$ can be obtained from the orbital vector $\hat{\mathbf{l}}$ by rotating $\hat{\mathbf{l}}$ around axis $\hat{\mathbf{n}}$ by the ``Leggett angle'' $\theta_{\rm L} = \arcsin(-1/4) \approx 104^{\circ}$, minimizing the spin-orbit interaction \cite{RevModPhys.47.331}.

Taking into account the broken relative symmetry between the spin and orbital parts of the order parameter and the broken phase symmetry, the remaining symmetry in the B phase is \cite{vollhardt2013superfluid,0953-8984-27-11-113203}
\begin{equation}
 H_{\mathrm{B}} = SO(3)_{\mathbf{J}} \times T \times C \times PU_\pi,
\end{equation}
where $SO(3)_{\mathbf{J}}$ denotes the joint three-dimensional rotation in the orbital and spin spaces, and $PU_\pi$ denotes the joint discrete symmetry of the parity $P$ and $\pi$ phase rotation $U_\pi$.

\subsubsection{A phase}

The A phase is an equal-spin-pairing state, where the Cooper pairs consist of spins with the same sign and thus $s_{\mathrm{z}} = \{ -1,1 \}$. The order parameter of the A phase can be written as
\begin{equation}
 A_{\mu j}^{\mathrm{A}} = \Delta_{\mathrm{A}} e^{i \phi} \hat{\mathbf{d}}_\mu (\hat{\mathbf{m}}_{j} + i\hat{\mathbf{n}}_{j}),
\end{equation}
where the vectors $\hat{\mathbf{m}}$ and $\hat{\mathbf{n}}$ (different from the $\hat{\mathbf{n}}$ vector in the B phase) form an orthogonal triad with the Cooper pair orbital angular momentum axis $\hat{\mathbf{l}} = \hat{\mathbf{m}} \times \hat{\mathbf{n}}$, $\hat{\mathbf{d}}$ is the spin anisotropy vector along which the total spin of a Cooper pair vanishes, and $\Delta_{\mathrm{A}}$ is the maximum superfluid gap in the A phase. In the A phase, the gauge symmetry $U(1)_\phi$ is broken and any change of the order parameter phase $\phi \rightarrow \phi + \Delta \phi$ may be compensated by rotating the orbital component of the order parameter about $\hat{\mathbf{l}}$ by an angle $- \Delta \phi$, leaving the order parameter invariant under that particular combination of transformations. This combination corresponds to a remaining relative $U(1)_{\phi+\mathbf{L}}$ symmetry. Additionally, rotation of the spin space about vector $\hat{\mathbf{d}}$ leaves the order parameter unchanged, corresponding to a remaining $SO(2)_{\mathbf{S}}$ symmetry. The order parameter is also symmetric under the simultaneous rotation about perpendicular axis and $\pi$ phase change, i.e. under transformation $(\hat{\mathbf{d}},\phi) \rightarrow (-\hat{\mathbf{d}},\phi+\pi)$, corresponding to a discrete symmetry described by the cyclic group $\mathbb{Z}_{2}$ (denoted $\mathbb{Z}_{2 (\phi+\mathbf{S})}$) of order~2. In the A phase, the time-reversal symmetry is partially broken due to the non-zero imaginary part in the orbital space, reducing it to the combined discrete symmetry $\mathbb{Z}_{2 (T+\mathbf{L})}$ corresponding to simultaneous time reversal and $\pi$-rotation of the orbital space about $\hat{\mathbf{m}}$. The remaining symmetries are described by the group \cite{vollhardt2013superfluid}
\begin{equation}
 H_{\mathrm{A}} = U(1)_{\phi+\mathbf{L}} \times \tilde D_{\infty {\mathbf S}} \times \mathbb{Z}_{2 (T+\mathbf{L})} \times C \times PU_\pi,
\end{equation}
where we have combined the symmetries $\tilde D_{\infty \mathbf{S}} \hat{=} SO(2)_{\mathbf{S}} \rtimes \mathbb{Z}_{2 (\phi+\mathbf{S})}$.

\subsection{$^3$He under confinement by nafen}
\label{NafenConfinement}

The presence of nanostructured confinement, i.e. thin slabs \cite{3heslab,Levitin841} or various aerogels \cite{PhysRevLett.112.115303, Wiman2014, PhysRevLett.115.165304,Askhadullin2012,Ikeda2014}, modifies the superfluid phase diagram. Anisotropic confinement may also alter the symmetry group of the normal phase of $^3$He by explicitly breaking the three-dimensional rotational symmetry in the coordinate space. Here we consider the effect of commercially available nematically ordered material called nafen \cite{PhysRevLett.115.165304}, which breaks the three-dimensional continuous rotational symmetry $SO(3)_{\mathbf{L}}$ in Eq.~(\ref{eq:bulk_symmetry}). The total symmetry group of the normal phase is reduced to \cite{0953-8984-27-11-113203}
\begin{equation}
 G' = D_{\infty \mathbf{L}} \times SO(3)_{\mathbf{S}} \times U(1)_{\phi} \times T \times C  \times P \,,
\label{G'group}
\end{equation}
where $D_{\infty \mathbf{L}}$ contains rotations about the nafen anisotropy axis $\hat{\mathbf{z}}$ and $\pi$ rotations about perpendicular axes. This symmetry may also be written as a product of rotations of the space around the anisotropy axis and reflection with respect to the perpendicular plane, i.e. $D_{\infty \mathbf{L}} \hat{=} SO(2)_{\mathbf{L}} \rtimes \mathbb{Z}_{2\mathbf{L}}$.

\begin{figure}
\begin{center}
\includegraphics[width=0.8\linewidth]{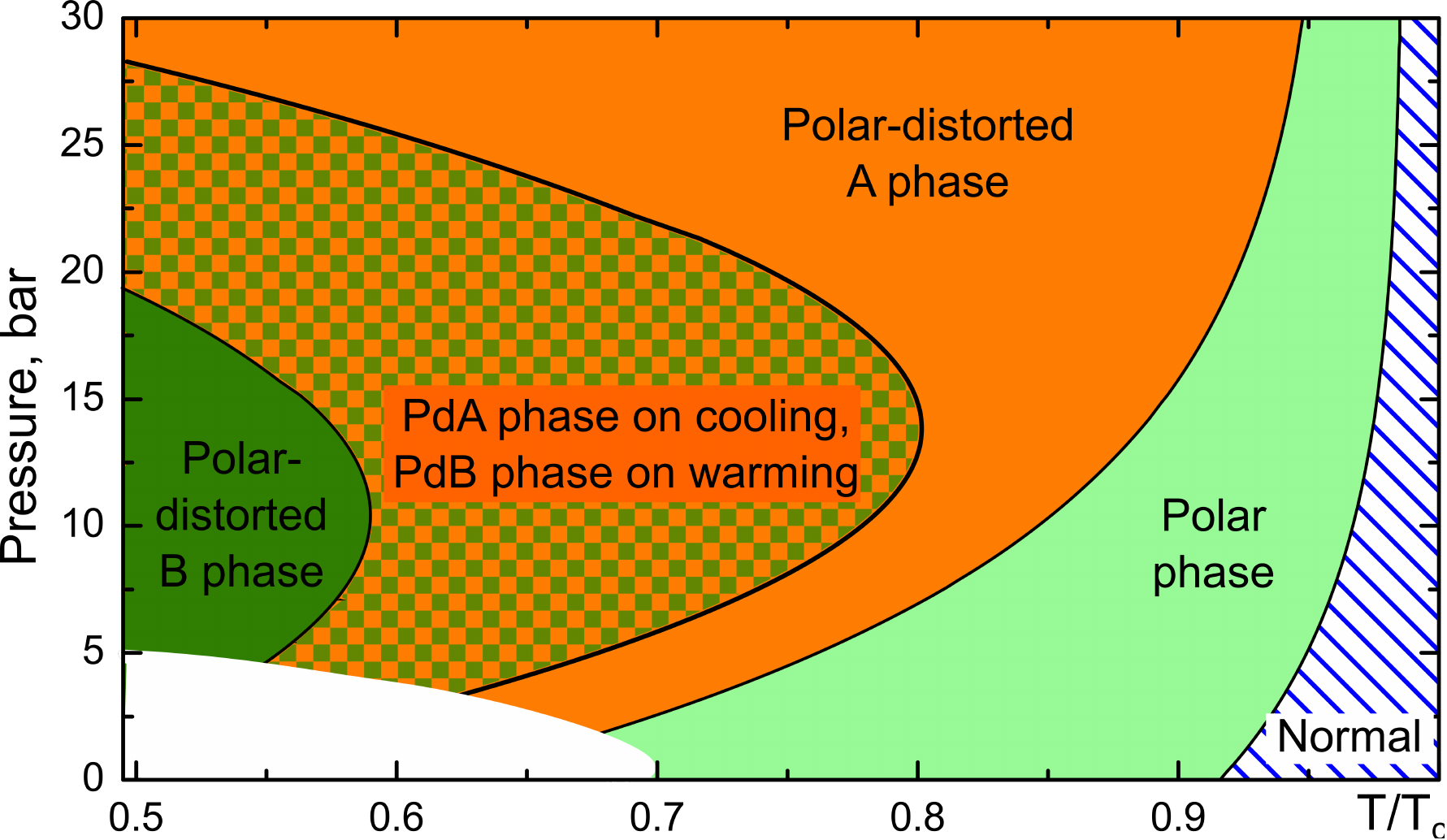}
\end{center}
\caption{\label{fig:nafen_90_phase_diagram} \textbf{Schematic phase diagram of superfluid $^3$He under confinement by nafen.} Nanostructured confinement by uniaxial cylinders modifies the symmetries of $^3$He in the normal fluid and below the superfluid transition temperature. The superfluid transition with the highest critical temperature under these conditions occurs to the polar phase \cite{PhysRevLett.115.165304}. The phase diagram shown in this figure sketches the measured phase diagram in a sample of nematically ordered aerogel called nafen-90, which consists of nearly uniaxial strands of Al$_2$O$_3$ with density $90$~mg/cm$^3$. In addition to the polar phase, polar-distorted A (PdA) and polar-distorted B (PdB) phases are encountered at lower temperatures. The PdA phase can be supercooled, as illustrated by the patterned area. The phase diagram in the low-pressure and low-temperature region (white) is not well known and there is a possibility that the direct transition from the polar to the PdB phase is realized in this area.}
\end{figure}

\begin{figure}
\begin{center}
\includegraphics[width=0.8\linewidth]{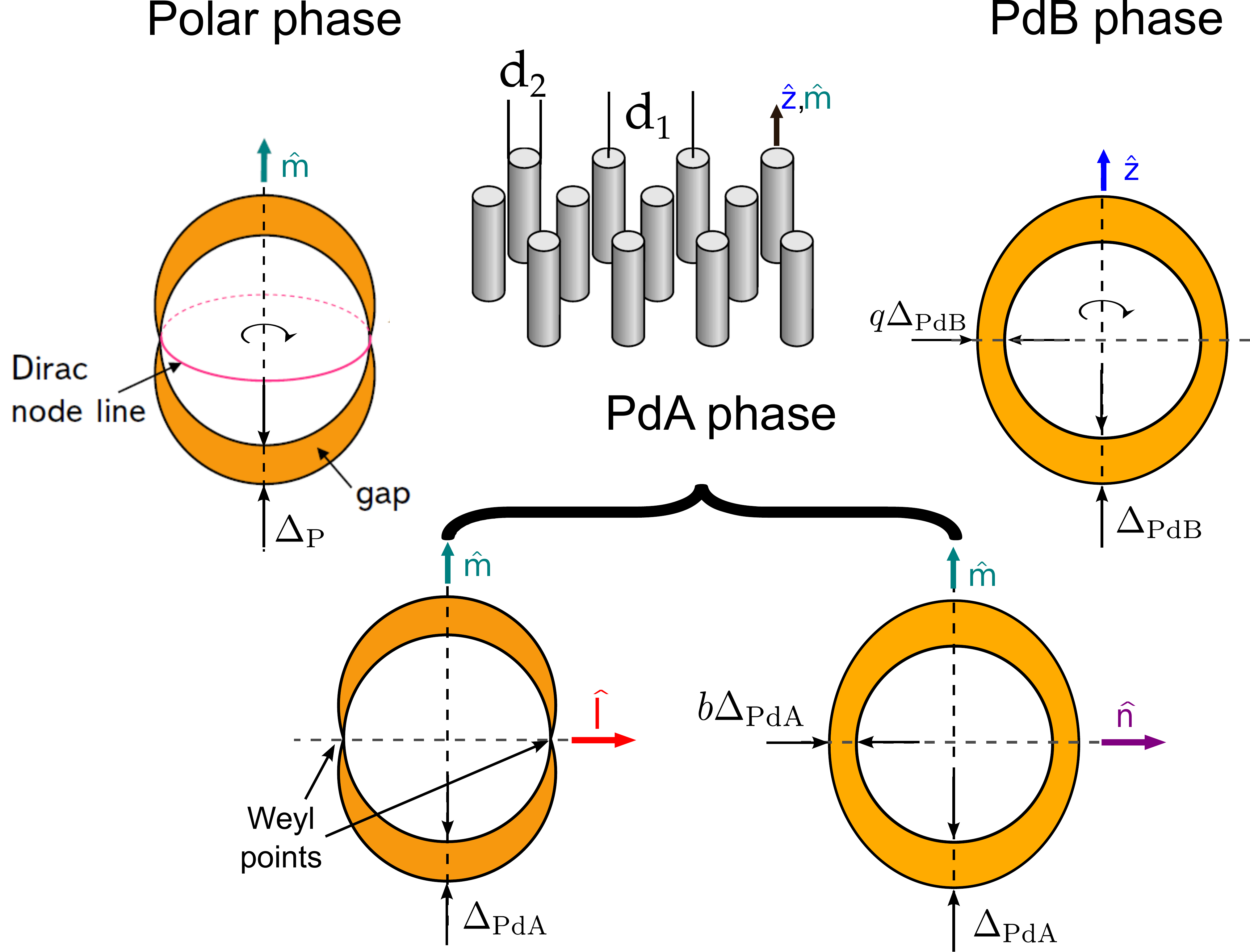}
\end{center}
\caption{\label{fig:gaps}\textbf{Schematic illustration of superfluid gaps in superfluid phases presented in Fig.~\ref{fig:nafen_90_phase_diagram}.} The polar phase and PdB phase gaps are symmetric under rotation by the nafen axis, and the PdA phase gap is shown in two projections as it lacks the rotational symmetry. For each phase, the maximum gap is oriented along the anisotropy axis of nafen strands, which have characteristic diameter $d_{1} \approx 9$~nm and separation $d_2 \sim 50$~nm. Gap asymmetry is not drawn to scale.}
\end{figure}

The resulting phase diagram in nafen with $90\,$mg/cm$^{3}$ density \cite{PhysRevLett.115.165304} is shown in Fig.~\ref{fig:nafen_90_phase_diagram}. There are notable differences to the bulk phase diagram as novel superfluid phases -- the polar, polar-distorted A (PdA), and polar-distorted B (PdB) phases -- are observed. In all cases the superfluid gap becomes anisotropic due to the effect of the confinement. Schematic illustrations of the superfluid gap in these phases are shown in Fig.~\ref{fig:gaps}. It is interesting that despite strong scattering by the confining strands, the critical temperature is suppressed only by a few percent compared to the bulk liquid. The robustness of the polar phase which appears immediately below transion is provided by extension of the Anderson theorem. This theorem has been derived in the case of $p$-wave pairing for the specular scattering from parallel strands \cite{fomin_theorem}. This robustness of the polar phase is also demonstrated by the characteristic $T^3$ low-temperature dependence of the gap resulting from the presence of the nodal line \cite{PolarT3}. In realistic samples scattering is never completely specular, nor are the strands perfectly aligned, and a small suppression of $T_{\mathrm{c}}$ is thus observed.

%%%%%%%%%%%%%%%%%%%%%%%%%%%%%%%%%%%%%%%%%%%%%%%%%%%%%%%%%%%%%%%%%%%%%%%%%%%%%%%%%%%%%%%%%%%%%
%%%%%%%%%%%%%%%%%%%%%%%%%%%%%%%%%%%%%%%%%%%%%%%%%%%%%%%%%%%%%%%%%%%%%%%%%%%%%%%%%%%%%%%%%%%%%
%%%%%%%%%%%%%%%%%%%%%%%%%%%%%%%%%%%%%%%%%%%%%%%%%%%%%%%%%%%%%%%%%%%%%%%%%%%%%%%%%%%%%%%%%%%%%
%%%%%%%%%%%%%%%%%%%%%%%%%%%%%%%%%%%%%%%%%%%%%%%%%%%%%%%%%%%%%%%%%%%%%%%%%%%%%%%%%%%%%%%%%%%%%

\subsubsection{Symmetry breaking patterns and fibrations of vacuum manifolds}
\label{SSBpatternAndFibration}

Besides stabilizing novel superfluid phases, confinement by nafen features even more complicated symmetry breaking patterns not possible in bulk samples. Such symmetry-breaking phase transitions can be used to study cosmological and dark matter models with spontaneous symmetry breaking (SSB) \cite{Lazarides1982,Lazarides2019,Lazarides2021,Lazarides2021b,Niita2021a,Niita2021b,Ryosuke2018,AxionSolitons} and, in particular, topological defects that appear in such theories. In nafen samples, the following symmetry breaking pattern can be realized:
\begin{equation}
  \xymatrix@R+1.5pc@C+1.5pc{
    &G' \ar[r] &{H_{\mathrm{P}} \subset G'} \ar[r] \ar[d] &{H_{\mathrm{PdA}} \subset H_{\mathrm{P}} \subset G'} \ar@{-->}[ld]\\
    & &{H_{\mathrm{PdB}} \subset H_{\mathrm{P}} \subset G'} \ar@{-->}[ru] &
  }
\label{SSBpattern}  
\end{equation}
Here solid arrows represent the directions of symmetry reduction through SSB, and the dashed line represents the first order transition. The sequence of transitions from normal $\to$ polar $\to$ PdA $\to$ PdB phases is seen on the phase diagram in Fig.~\ref{fig:nafen_90_phase_diagram}, while a direct transition form the polar to the PdB phase remains a possibility at low pressures and temperatures.

%%%%%%%%%%%%%%%%%%%%%%%%%%%%%%%%%%%%%%%%%%%%%%%%%%%%%%%%%%%%%%%%%%%%%%%%%%%%%%%%%%%%%%%%%%%%%
%%%%%%%%%%%%%%%%%%%%%%%%%%%%%%%%%%%%%%%%%%%%%%%%%%%%%%%%%%%%%%%%%%%%%%%%%%%%%%%%%%%%%%%%%%%%%
%%%%%%%%%%%%%%%%%%%%%%%%%%%%%%%%%%%%%%%%%%%%%%%%%%%%%%%%%%%%%%%%%%%%%%%%%%%%%%%%%%%%%%%%%%%%%
%%%%%%%%%%%%%%%%%%%%%%%%%%%%%%%%%%%%%%%%%%%%%%%%%%%%%%%%%%%%%%%%%%%%%%%%%%%%%%%%%%%%%%%%%%%%%

The SSB pattern, Eq.~(\ref{SSBpattern}), suggests that, starting from the normal phase, both the PdA and PdB phases are accessible via two consecutive second-order steps. Such situation cannot be realized in bulk superfluid $^3$He. One can calculate the vacuum manifolds $R$ of the confined phases from their respective residual symmetry groups $H$ by taking the quotient space of $G'$ over $H$. Denoting the vacuum manifolds generated through SSB from the normal phase as
\begin{equation}
R^{\rm P} = G'/H_{\rm P}, \; R_{1}^{\rm PdA} =G'/H_{\rm PdA}, \;
R_{1}^{\rm PdB} =G'/H_{\rm PdB}
\,,
 \label{RpR1}
\end{equation}
and the vacuum manifolds accessible via the polar phase as
\begin{equation}
R_{2}^{\rm PdA} =H_{\rm P}/H_{\rm PdA},\;
R_{2}^{\rm PdB} =H_{\rm P}/H_{\rm PdB}\,,
  \label{R2}
\end{equation}
we find $R_{2} \subset R_{1}$. In the rest of the subsection we omit PdA or PdB indices at $R$ and $H$, since equations apply in both phases. Detailed discussion of residual symmetries in each of the confined phases is presented in the following subsections.

Applying the third isomorphism theorem $R_{1}/R_{2}=(G'/H)/(H_{\rm P}/H)$  \cite{MartinBook2009}, we get
\begin{equation}
R_{1}/R_{2} \cong R^{\rm P} = G'/H_{\rm P}\,.
\label{RIsomophsm}
\end{equation}
Eq.~(\ref{RIsomophsm}) suggests that the normal phase vacuum manifold $R_{1}$ consists of infinite disjoint subspaces (cosets) $R^{\rho}_{2}$, which relate to $R_{2}$ though action of $\rho \in R_{1}$ i.e., $R^{\rho}_{2}=\rho R_{2}$. We also notice that every $R^{r}_{2}$ is isomorphically mapped to an element of the vacuum manifold of the polar phase. Mathematically speaking, this signifies that the vacuum manifold $R_{1}$ covers $R^{\rm P}$ \cite{Nakahara2003}. The resulting fibration $p$ between the vacuum manifolds $R_{1}$ and $R^{\rm P}$ is
\begin{equation}
\xymatrix@1@R=10pt@C=13pt{
R_{2} \,\, \ar@{^{(}->}[r] \,\, & \,\, R_{1} \,\, \ar[r]^-{p} \,\, & \,\, R^{\rm P}\,,\\
}\,
\label{RFibration}
\end{equation}
leading to the relationship between the homotopy groups of different vacuum manifolds
\begin{equation}
\pi_{n}(R_1, R_2) \cong \pi_{n}(R^{\rm P})\,.
\label{RelativeGroup}
\end{equation}
Consequently, for $n=1$ this relationship suggests that the polar phase linear topological defects are conserved as linear topological objects in the subsequent second order phase transition. These defects, as we will see later, are HQVs and singly quantized vortices. To assess whether the resulting linear objects are composite topological defects and serve as a termination line of a planar object, one must calculate the exact sequences of $\pi_{1}(R_1, R_2)$. Similarly, one can extract information about composite defects where monopole terminates a string \cite{volovik2000} by setting $n=2$ in Eq.~(\ref{RelativeGroup}). The resulting topological objects are discussed in section~\ref{CompositeTopologicalObjects}.

The presence of various effects, such as liquid-surface interactions, Zeeman splitting, and spin-orbit interaction affect the underlying symmetries of the superfluid phases. Since these orientational energies have characteristic length scales (see \ref{LengthScales}), both residual symmetry $H$ and vacuum manifold $R$ of a superfluid depend on the length scale at which superfluid is considered. The maximum $H$ is found at the smallest length scales, limited by coherence length from below and, depending on conditions, by the magnetic $\xi_{\rm H}$ or the spin-orbit $\xi_{\rm D}$ lengths or by the confining geometry from above. At larger scales the symmetry is reduced. For polar and PdA phases below we provide one example of such reduction relevant to the soliton bounded by HQVs found in those phases. For the PdB phase we discuss reduced symmetries in more details in section \ref{RReductionPdB} as they are relevant for analysis of composite objects in section~\ref{CompositeTopologicalObjects}.

\subsubsection{Polar phase}

The order parameter of the polar phase can be written as
\begin{equation} \label{eq:polarop}
 A_{\mu j}^{\mathrm{P}} = \Delta_{\mathrm{P}} e^{i \phi} \hat{\mathbf{d}}_{\mu} \hat{\mathbf{m}}_{j},
\end{equation}
where $\Delta_{\mathrm{P}}$ is the maximum superfluid gap in the polar phase. The order parameter is invariant with respect to rotations of the spin space about $\hat{\mathbf{d}}$ and rotations of the orbital space about $\hat{\mathbf{m}}$, as well as about combined rotations about perpendicular axes and $\pi$ phase changes, i.e. about transformations $(\hat{\mathbf{d}},\phi) \rightarrow (-\hat{\mathbf{d}},\phi + \pi)$ and $(\hat{\mathbf{m}},\phi) \rightarrow (-\hat{\mathbf{m}},\phi + \pi)$. Similarly to other superfluid phases, the gauge symmetry $U(1)_{\phi}$ is broken in the transition. The maximum group of remaining symmetries in the polar phase is
\begin{equation}
H_{\mathrm{P}} = \tilde D_{\infty \mathbf{L}} \times \tilde D_{\infty \mathbf{S}}  \times T \times C \times PU_\pi \,,
\label{HP}
\end{equation}
where $\tilde D_{\infty \mathbf{L}}$ and $\tilde D_{\infty \mathbf{S}}$ are the symmetries involving $\pi$ rotations about axes transverse to $\hat{\mathbf{m}}$ or $\hat{\mathbf{d}}$ in the orbital and spin spaces, respectively, combined with a phase rotation by $e^{\pi i}$, i.e. $\tilde D_{\infty \mathbf{L}} \hat{=} SO(2)_{\mathbf{L}} \rtimes \mathbb{Z}_{2 (\phi+\mathbf{L})}$ and $\tilde D_{\infty \mathbf{S}} \hat{=} SO(2)_{\mathbf{S}} \rtimes \mathbb{Z}_{2 (\phi+\mathbf{S})}$. The vacuum manifold $R^{\mathrm{P}}$ of the polar phase is \cite{vollhardt2013superfluid}
\begin{equation}
  R_{\mathrm{P}} = G'/H_{\mathrm{P}} = (S^2_{\mathbf{S}} \times U(1)_\phi)/\mathbb{Z}_{2(\phi + \mathbf{S})}\,,
  \label{RPolar}
\end{equation}  
where $S^2_{\mathbf{S}}$ denotes inversion.

The spin-orbit coupling (SOC) energy in the polar phase is proportional to $(\hat{\mathbf{d}} \cdot \hat{\mathbf{m}})^2$, which favors $\hat{\mathbf{d}}$ orientation in the plane perpendicular to the nafen anisotropy axis. In addition, there is a magnetic energy term proportional to $(\mathbf{B} \cdot \hat{\mathbf{d}})^2$, which orients $\hat{\mathbf{d}} \perp \mathbf{B}$ at sufficiently large magnetic fields above the dipole field $H_{\rm D} \sim 3$\,mT. For a tilted field $\mathbf{B} \nparallel \hat{\mathbf{m}}$, the combination of the spin-orbit and magnetic energies breaks the $SO(2)_{\mathbf{S}}$ symmetry of $H_{\rm P}$ in Eq.~(\ref{HP}), resulting in the reduced residual symmetries at scales larger than $\xi_{\rm H}$ and $\xi_{\rm D}$
\begin{equation}
\tilde H_{\rm P} = SO(2)_{\mathbf{L}} \times \mathbb{Z}_{2 (\phi+\mathbf{S})} \times T \times C \times PU_\pi\,.
\label{tildeHP}
\end{equation}
Here we additionally took into account strong orienting effect of nafen strands on the orbital space which fix $\hat{\mathbf{m}}$ along the strands (common direction of  $\hat{\mathbf{m}}$ can be assumed since $(\hat{\mathbf{d}}, \hat{\mathbf{m}})$ and $(-\hat{\mathbf{d}}, -\hat{\mathbf{m}})$ is the same state), while keeping the symmetry $SO(2)_{\mathbf{L}}$ of orbital rotation about the strand direction unaffected by any other orientational energy.

The existence of the symmetry group $\mathbb{Z}_{2 (\phi+\mathbf{S})}$ in Eq.~(\ref{tildeHP}) signifies that there are two degenerate minimum energy states which can be obtained via transformation $(\hat{\mathbf{d}},\phi) \rightarrow (-\hat{\mathbf{d}},\phi + \pi)$. Such degenerate regions are connected by topological $\hat{\mathbf{d}}$ solitons, which can be terminated in bulk by HQVs with $\pi$ phase winding, discussed in detail in section~\ref{sec:polarsolitons}. 

\subsubsection{Polar-distorted A phase}

In the $90\,$mg/cm$^{3}$ nafen sample, the polar-distorted A (PdA) phase is reached on cooling via a second-order phase transition from the polar phase \cite{PhysRevLett.115.165304}. The order parameter of the PdA phase is
\begin{equation}
A_{\mu j}^{\mathrm{PdA}} = \Delta_{\mathrm{PdA}} e^{i \phi} \hat{\mathbf{d}}_\mu (\hat{\mathbf{m}}_{j} + ib\hat{\mathbf{n}}_{j}),
\label{opPdA}
\end{equation}
where $0 < b < 1$ is a dimensionless parameter characterizing the gap distortion and $\Delta_{\mathrm{PdA}}(b)$ is the maximum gap in the PdA phase. For $b=0$ the order parameter of the polar phase is obtained, while $b=1$ recovers the order parameter of the bulk A phase. The order parameter is symmetric under rotations of the spin space about $\hat{\mathbf{d}}$, about the combined transformation $\mathbb{Z}_{2 (\phi+\mathbf{S})}$, and about $\pi$ rotation about $\hat{\mathbf{l}}$ and $\pi$ phase change, i.e. about $(\hat{\mathbf{m}}, \hat{\mathbf{n}},\phi) \rightarrow (-\hat{\mathbf{m}}, -\hat{\mathbf{n}}, \phi + \pi)$. Moreover, the $\tilde D_{\infty \mathbf{S}}$ symmetry is preserved in the polar-PdA transition due to similar spin structure of the order parameter. In the orbital space, $\pi$ rotation about the axis $\hat{\mathbf{l}}$, in combination with a phase rotation $e^{\pi i}$, form a symmetry, i.e. the orbital symmetry is reduced to a discrete symmetry $U_{\pi (\phi+\mathbf{L})}$. Similar to the bulk A phase, the time-reversal symmetry is partially broken in the PdA phase and the PdA phase is invariant under the combined discrete symmetry $\mathbb{Z}_{2 (T+\mathbf{L})}$. The maximum residual symmetry group in the PdA phase is
\begin{equation}
 H_{\mathrm{PdA}} =  U_{\pi (\phi+\mathbf{L})} \times \tilde D_{\infty \mathbf{S}} \times \mathbb{Z}_{2 (T + \mathbf{L})} \times C \times PU_\pi \,.
\end{equation}
Assuming phase transition from the normal phase, the vacuum manifold of the PdA phase becomes
\begin{equation}
  R^{\mathrm{PdA}}_{1} = G'/H_{\mathrm{PdA}} = SO(2)_{\mathbf{L}} \times S^{2}_{\mathbf{S}} \times U(1)_\phi\,.
  \label{RNToPdA}
\end{equation}
In addition, the group $H_{\mathrm{PdA}}$ is also the subgroup of $H_{\mathrm{P}}$, reflecting the fact that the PdA-polar phase transition is of the second order.  For the polar-PdA transition, the vacuum manifold becomes
\begin{equation}
  R^{\rm PdA}_{2} = H_{\mathrm{P}}/H_{\mathrm{PdA}} = SO(2)_{\mathbf{L}}\,.
  \label{RPolarToPdA}
\end{equation}

The spin-orbit coupling $\propto (\hat{\mathbf d}\cdot \hat{\mathbf{l}})^2$ in the PdA phase affects distribution of both spin $\hat{\mathbf d}$ and orbital $\hat{\mathbf{l}}$ anisotropy axes since, unlike the polar phase, the orbital degrees of freedom are not fully fixed by the confinement. The effect of the magnetic anisotropy is similar to that in the polar phase. The reduced symmetry at scales larger than $\xi_{\rm H}$ and $\xi_{\rm D}$ is
\begin{equation}
 \tilde H_{\rm PdA} = \mathbb{Z}_{2 (\phi+\mathbf{S})} \times \mathbb{Z}_{2 (T + \mathbf{L})} \times C \times PU_\pi \,.       
\end{equation}
The similar consideration holds for the spin part of the bulk A phase, where the residual symmetries in the presence of magnetic and spin-orbit energies become
\begin{equation}
 \tilde H_{\rm A} = U(1)_{\phi+\mathbf{L}} \times \mathbb{Z}_{2 (\phi+\mathbf{S})} \times \mathbb{Z}_{2 (T + \mathbf{L})}  \times C \times PU_\pi \,.
\end{equation}
In both cases of bulk A and confined PdA phases the residual symmetry groups  contain the two degenerate ground states connected by the $\mathbb{Z}_{2 (\phi+\mathbf{S})}$ symmetry and thus support HQVs, as in the polar phase.

It is worth noting that the symmetry considerations above concern an ideal confinement without random inhomogeneity. In practical samples, fluctuations of the density of confining strands leads to formation of the Larkin-Imry-Ma (LIM) orbital glass state in the PdA phase. Originally, LIM state was discovered in the A phase confined in the isotropic aerogels \cite{Dmitriev2010,Li2013}, where random anisotropy breaks long-range order of the orbital vector $\hat{\mathbf{l}}$ and orientation of $\hat{\mathbf{l}}$ fluctuates over all possible directions in 3D space with a characteristic length scale significantly smaller that the dipolar length $\xi_{\rm D}$. In the PdA phase under strong anisotropic columnar confinement, $\hat{\mathbf{l}}$ is fixed into the plane perpendicular to the confining strands. Experimental evidence \cite{spinglass,Dmitriev2020} suggests, that also in this case the long-range orientational order of $\hat{\mathbf{l}}$ within this plane is lost, and the 2D LIM state is formed. Formation of LIM state does not eliminate the $\mathbb{Z}_{2 (\phi+\mathbf{S})}$ symmetry responsible for the existence of solitons considered in this work, but it does modify the NMR response.

Another way to introduce a discrete $\mathbb{Z}_{2}$ symmetry is to confine superfluid $^3$He into a narrow slab. Independently of the original superfluid phase, the boundary orients the orbital momentum of the Cooper pairs along the surface normal. Two states connected with $\mathbb{Z}_{2}$ symmetry correspond to the orbital momentum oriented towards or away from the wall. The corresponding solitons or domain walls connecting two states were observed both in the B \cite{Levitin2019} and A \cite{Walmsley2004,Ikegami2015,Kasai2018} phases. At the conditions near the phase transition between A and B phases such walls can proliferate and create a spatially modulated order parameter \cite{Levitin2019}. In the A phase, reorientation of the orbital momentum can take place independently or in combination with $\hat{\mathbf d}$, owing to the spin-orbit coupling. In the latter case the soliton is relatively thick, of the order of the slab thickness, and has been directly visualized by magnetic resonance imaging \cite{Kasai2018}.

\subsubsection{Polar-distorted B phase}
\label{RPdB}

Depending on the confinement, the phase transition to the polar-distorted B phase (PdB) may occur directly from the normal phase \cite{DistB}, via a first-order transition from the PdA phase \cite{PhysRevLett.115.165304} (as realized in the nafen-90 sample), or, in principle, via a second-order phase transition from the polar phase. The order parameter of the PdB phase in zero magnetic field can be written as
 \begin{equation}
A_{\mu j}^{\mathrm{PdB}} = \Delta_{\mathrm{PdB}} e^{i\phi} (\hat{\mathbf{d}}_\mu \hat{\mathbf{z}}_{j} + q_{1} \hat{\mathbf{e}}^1_\mu \hat{\mathbf{x}}_j 
+ q_{2} \hat{\mathbf{e}}^2_\mu \hat{\mathbf{y}}_j ) \,,
\label{eq:distBop}
\end{equation}
where $| q_1 | = | q_2 | \equiv q \in (0,1)$ describes the relative gap size in the plane perpendicular to the strands. Vectors $\hat{\mathbf{e}}^1$ and $\hat{\mathbf{e}}^2$ are unit vectors in the spin space and form an orthogonal triad with vector $\hat{\mathbf{d}}$. The maximum gap $\Delta_{\mathrm{PdB}}(q)$ is achieved along the nafen anisotropy axis $\hat{\mathbf{z}}$. For $q=0$, the order parameter of the polar phase is obtained while $q=1$ recovers the order parameter of the bulk B phase. The order parameter is symmetric under the joint rotation of the spin and orbital spaces about the nafen anisotropy axis $\hat{\mathbf{z}}$. The remaining symmetries in the PdB phase are
\begin{equation}
 H_{\mathrm{PdB}} = SO(2)_{\mathbf{J}} \times T \times C \times PU_{\pi} \,,
\end{equation}
where the subscript notation $\mathbf{J}$ refers to the symmetry of the simultaneous rotation of orbital and spin spaces. 

The vacuum manifold for the normal-PdB phase is obtained from
\begin{equation}
  R^{\mathrm{PdB}}_{1} = G'/H_{\mathrm{PdB}} = SO(3)_{\mathbf{J}} \times U(1)_\phi\,.
  \label{RNToPdB}
\end{equation}        
The group $H_{\mathrm{PdB}}$ is again a subgroup of $H_{\mathrm{P}}$, reflecting the fact that also the polar-PdB phase transition is of second order. In this case, the resulting vacuum manifold becomes
\begin{equation}
  R^{\mathrm{PdB}}_{2} = H_{\mathrm{P}}/H_{\mathrm{PdB}} = SO(2)_{\mathbf{J}} \times \mathbb{Z}_{2 (\phi + \mathbf{S})}\,.
  \label{RPolarToPdB}
\end{equation}
We note that $R^{\rm PdB}_{2}$ is a disconnected space characterized by states $(\mathbf{d}, \mathbf{e}^{1}, \mathbf{e}^{2}, \phi)$ and $(-\mathbf{d}, -\mathbf{e}^{1}, \mathbf{e}^{2}, \pi+\phi)$. These right-handed and left-handed spaces can not be made equal by any action from $H_{\mathrm{P}}$. Moreover, calculating $\pi_{0} (R^{\rm PdB}_{2})$ yields $\mathbb{Z}_{2 (\mathbf{S}+\phi)}$ describing a domain wall structure (whereas solitons are described by the $\pi_1$ homotopy group). As we will see further, the domain wall may combine with various types of solitons to form composite topological defects.

\subsubsection{Residual symmetries in the presence of orientation energies in the PdB phase}
\label{RReductionPdB}

The orientational energies with different coupling strengths and hierarchy of characteristic length scales reduce the normal phase symmetry group, resulting in reduced degrees of freedom for the order parameter vacuum manifolds after SSB phase transitions. The reduced vacuum manifolds introduce a different set of topological objects than those generated through fibration, discussed in Sec.~\ref{SSBpatternAndFibration}. For example, in the PdB phase the orientation energies allow the topological objects described by the relative homotopy groups $\pi_{n}(R_{1}^{\rm PdB},R_{2}^{\rm PdB})$ to expand into mesoscopic length scales, forming a nexus \cite{Zhang2020b, Zhang2020, Volovik2009Book}. In this section we will discuss the reduction of the vacuum manifolds at mesoscopic length scales.

\begin{figure*}%[tb!]
\centerline{\includegraphics[width=0.5\linewidth]{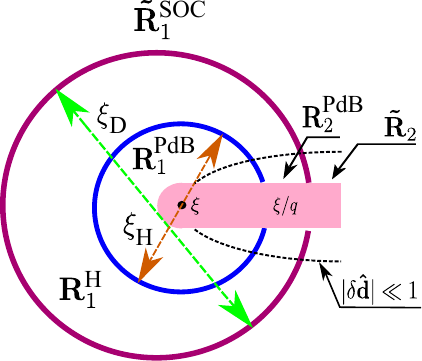}}
\caption{{\bf Illustration of relevant length scales in the PdB phase.} The smallest length scale in the PdB phase is provided by the coherence length $\xi$, which sets the size of the vortex core. The KLS wall, described by $\pi_{1}(R_{1},R_{2})$, is larger than the the coherence length by a factor $q^{-1}$. In the region $r<\xi_{\rm H}$ the vacuum manifolds are described by the groups $R_{1}^{\rm PdB}$ and $R_{2}^{\rm PdB}$, while the presence of magnetic and SOC energies extends the hierarchy of length scales to mesoscopic lengths. In the region $\xi_{\rm H} < r < \xi_{\rm D}$, $R_{1}^{\rm PdB}$ reduces to $R_{1}^{\rm H}$ and the spin vector $\hat{\mathbf{d}}$ becomes fixed perpendicular to the static magnetic field $\mathbf{H}^{(0)}$, while the $R_{2}^{\rm PdB}$ remains unchanged. When the SOC energy is taken into account, $R_{1}^{\rm H}$ further reduces to $\tilde{R}_{1}^{\rm SOC}$ and $R_{2}^{\rm PdB}$ reduces to $\tilde{R}_{2}$.}
\label{Degenerate_Spaces}
\end{figure*}

In $^3$He the relevant length scales are the magnetic length $\xi_{\rm H}$ and the dipole length $\xi_{\rm D}$ \cite{vollhardt2013superfluid, Autti2016}, which characterize the spatial ranges in which the gradient energy dominates over the orientational energy. When the length scale of the spatial variation is larger than these characteristic length scales, the vacuum manifolds are reduced to minimize the orientation energies. See \ref{LengthScales} for details on calculation of values of $\xi_{\rm H}$ and~$\xi_{\rm D}$.

If a static magnetic field $\mathbf{H}^{(0)}$ with fixed direction is turned on \cite{Makinen2019}, the degenerate space of the PdB order parameter reduces to 
\begin{equation}
R^{\rm H}_{1} = S^{1}_{\bf S} \times U(1)_{\phi}
\end{equation}
in the region where length scale of spatial variation is larger than $\xi_{\rm H}$. Here $S^{1}_{\bf S}$ is the reflection symmetry. Since the magnetic energy locks the $\hat{\mathbf{d}}$ vector into the plane perpendicular to $\mathbf{H}^{(0)}$, $R_{2}^{\rm PdB}$ retains its form inside the region characterized by the length scale $\xi_{\rm H}$. It follows that $R_{2}^{\rm PdB} = SO(2)_{{\bf J}} \times \mathbb{Z}_{2(\phi + {\bf S})}$ in the region where $\hat{\mathbf{d}}$ is approximately constant. In Fig.~\ref{Degenerate_Spaces}, we illustrate the $R_{1}^{\rm H}$ and $\xi_{\rm H}$ in the presence of the KLS wall.

When the SOC is taken into account, the vacuum manifolds are further reduced from $R^{\rm H}_{1}$ and $R_{2}^{\rm PdB}$. In general, the requirement of minimizing the SOC energy in the region with $r > \xi_{\rm D}$ fixes the relative directions between spin and orbital vectors, resulting  in the broken relative spin-orbit symmetry \cite{vollhardt2013superfluid}. Thus $R_{1}^{\rm H}$ reduces to
\begin{equation}
\tilde{R}_{1}^{\rm SOC} = R_{\bf S}^{\rm SOC}\times U(1)_{\phi}
\end{equation}
in the region with $r > \xi_{\rm D}$, where $R_{\bf S}^{\rm SOC}$ is the reduced vacuum manifold of the spin degree of freedom. Generally speaking, $R_{\bf S}^{\rm SOC}$ has a complicated form, which may be simplified by using the following parametrization
\begin{equation}
\hat \mathbf{{d}} =\hat{\bf x} \cos \theta-\hat{\bf z} \sin \theta, \,\,
\hat \mathbf{{e}}^{1} =-\hat{\bf x} \sin \theta-\hat{\bf z} \cos \theta, \,\, 
\hat \mathbf{{e}}^{2} = \hat{\bf y}, \,\, \mathbf{H}^{(0)} = H\hat{\bf y} \,,
\label{PARA1}  
\end{equation}
where $\theta$ is the angle between $\hat{\mathbf{d}}$ and local orbital-coordinate frame \cite{Makinen2019}. In this case, we find 
\begin{equation}
R_{\bf S}^{\rm SOC} = \{\theta_{0}, \pi-\theta_{0}, -\theta_{0}, \pi+\theta_{0} \}\,,
\end{equation}
where $\theta_{0}= \arcsin \left( q/(1-|q|) \right)$. In the region where $\hat{\mathbf{d}}$ is approximately constant, the SOC energy fixes the relative rotation of $SO(2)_{{\bf J}}$, reducing $R_{2}^{\rm PdB}$ to 
\begin{equation}
\tilde{R}_{2} = \mathbb{Z}_{2(\phi + {\bf S})}
\end{equation}
in the region $r > \xi_{\rm D}$. The regions described by groups $R_{1}^{\rm H}$, $\tilde{R}_{1}^{\rm SOC}$, and $\tilde{R}_{2}$ are illustrated in Fig.~\ref{Degenerate_Spaces}.

\section{Composite topological objects}
\label{CompositeTopologicalObjects}

The relative homotopy groups may be utilized to investigate topological objects in the presence of multiple characteristic length scales \cite{ MineevBook1998, NashBook1988, Mineev1978}. In addition to the PdB phase discussed in the previous section, examples of such systems include e.g. solitons terminated by HQVs in spinor Bose condensates with quadratic Zeeman energy \cite{Seji2019,Liu2020}. In this section, we will discuss various composite objects connected by solitons in the superfluid phases of $^3$He.

\subsection{Relative homotopy groups -- from polar to polar-distorted phases}
\label{ExactSequence}

In experimentally reachable magnetic field $\mathbf{H}^{(0)}$ the magnetic healing length is much larger than the coherence length,  $\xi_{\rm H} \gg \xi$. The symmetry group $G'$ and the symmetry breaking pattern described by Eq.~(\ref{SSBpattern}) are valid in the region $\xi \ll r \leq \xi_{\rm H}$. In Sec.~\ref{SSBpatternAndFibration}, we have seen that the polar-phase topological objects survive a second symmetry-breaking transition into either PdA or PdB phases. In this section, we will study how these objects change in the transition to the PdA and PdB by analyzing the relative homotopy groups $\pi_{n}(R_{1},R_{2})$. The structures of the relative homotopy groups $\pi_{n}(R_{1},R_{2})$ can be extracted from their short exact sequences (SES). The SESs are calculated by splitting the long exact sequences of $\pi_{n}(R_{1},R_{2})$. In this section we use the results calculated in \ref{SESsPdA} and \ref{SESsPdB}.

The SES of $\pi_{1}(R^{\rm PdA}_{1},R^{\rm PdA}_{2})$ is
\begin{equation}
\xymatrix@R=8pt{
0 \ar[r] & \mathbb{Z}_{\phi} \ar[r] & \pi_{1}( R^{\rm PdA}_{1},R^{\rm PdA}_{2}) \ar[r]^-{\partial^{*}} & 0 \ar[r] & 0\\
}
\label{SESPi1PdA}
\end{equation}
and SES of $\pi_{2}(R^{\rm PdA}_{1},R^{\rm PdA}_{2})$ is
\begin{equation}
\xymatrix@R=8pt{
0 \ar[r] & \mathbb{Z}_{\mathbf{S}} \ar[r] & \pi_{2}( R^{\rm PdA}_{1},R^{\rm PdA}_{2}) \ar[r]^-{\partial^{*}} & 0 \ar[r] & 0\\
},
\label{SESP2PdA}
\end{equation}
where the vacuum manifolds $R^{\rm PdA}_{1}$ and $R^{\rm PdA}_{2}$ are given by Eqs.~(\ref{RNToPdA}) and (\ref{RPolarToPdA}). From the SESs of $\pi_{1(2)}(R^{\rm PdA}_{1},R^{\rm PdA}_{2})$ in Eqs.~(\ref{SESPi1PdA}) and (\ref{SESP2PdA}), we find that the boundary homomorphisms $\partial^{*}$ are trivial both for $\pi_{1}(R_{1}^{\rm PdA},R_{2}^{\rm PdA})$ and $\pi_{2}(R_{1}^{\rm PdA},R_{2}^{\rm PdA})$. In other words, linear and point-like defects (HQVs and $\mathbf{d}$-monopoles, respectively), which PdA phase inherits from polar phase, are simple (i.e. not composite) topological objects. Due to similar spin structure, the HQVs and $\mathbf{d}$-monopoles are similar in both phases \cite{Zhang2020b}.

In contrast, the symmetry breaking pattern resulting in the PdB phase yields the SESs for $\pi_{1}(R_{1}^{\rm PdB},R_{2}^{\rm PdB})$ and $\pi_{2}(R_{1}^{\rm PdB},R_{2}^{\rm PdB})$, which have non-trivial boundary homomorphisms. Specifically, they are
\begin{equation}
\xymatrix@R=8pt{
0 \ar[r] & \mathbb{Z}_{\phi} \ar[r] & \pi_{1}( R^{\rm PdB}_{1},R^{\rm PdB}_{2}) \ar[r]^-{\partial^{*}} & \mathbb{Z}_{2 (\phi + \mathbf{S})} \ar[r] & 0\\
}
\label{SESPi1PdB}
\end{equation}
and
\begin{equation}
\xymatrix@R=8pt{
0 \ar[r] & 0 \ar[r] & \pi_{2}( R^{\rm PdB}_{1},R^{\rm PdB}_{2}) \ar[r]^-{\partial^{*}} & 2\mathbb{Z}_{\mathbf{J}} \ar[r] & 0\\
},
\label{SESPi2PdB}
\end{equation}
where $R^{\rm PdB}_{1}$, $R^{\rm PdB}_{2}$ are given by Eqs.~(\ref{RNToPdB}) and (\ref{RPolarToPdB}). The homotopy group $\mathbb{Z}_{2 (\phi + \mathbf{S})} = \pi_{0}(R^{\rm PdB}_{2})$ gives rise to the Kibble-Lazarides-Shafi (KLS) domain walls \cite{Kibble1982a,Kibble1982b} owing to the disconnected vacuum manifold $R^{\rm PdB}_{2}$ discussed in Sec.~\ref{RPdB}, while the homotopy group $2\mathbb{Z}_{\mathbf{J}} \subset \pi_{1}(R^{\rm PdB}_{2})$ describes spin vortices with an even winding number \cite{Zhang2020b}.

Non-trivial boundary homomorphisms $\partial^{*}$ in Eqs.~(\ref{SESPi1PdB}) and (\ref{SESPi2PdB}) describe how low-dimensional objects, e.g. KLS walls and spin vortices, connect to objects with higher dimensionality. Here we have considered HQVs and $\mathbf{d}$-monopoles, inherited from the polar phase. We refer to the resulting composite objects consisting of $D$-dimensional and $(D+1)$-dimensional objects, as string-walls and string-monopoles, respectively, based on the rank of their relative homotopy groups \cite{Zhang2020b}.

\subsection{Vortex-bound solitons in the PdB phase and the nexus object}
\label{SolitonNexusPdB}

We now know that the PdB phase supports several composite topological objects with different dimensions, contained in length scales smaller than $\xi_{\rm H}$. However, typical NMR experiments probe superfluid properties at length scales larger than $\xi_{\rm H}$, which poses a question: what happens to composite objects at length scales approaching $\xi_{\rm H}$? This question is both theoretically interesting and relevant for understanding experimental observations, as composite objects with characteristic length scales $\xi$ or $\xi/|q|$ cannot be observed directly with NMR methods.

Composite defects, such as walls bounded by strings studied in Ref.~\cite{Makinen2019}, can be identified from the NMR signature of the related spin solitons. HQVs can therefore be considered as one dimensional (1D) nexuses connecting defects with different characteristic sizes \cite{Zhang2020,Volovik2009Book}. In this section, we will discuss the relative homotopy groups for the reduced vacuum manifolds analyzed in Sec.~\ref{RReductionPdB}. We will see how HQVs and KLS walls with characteristic length scales $\xi$ and $\xi/|q|$, respectively, connect to spin solitons with much larger characteristic length scale, set by the dipole length $\xi_{\rm D}$. In this problem the length scales are well separated, $\xi_{\rm D} \gg \xi/|q| \gg \xi$ \cite{Zhang2020}. For technical details, we refer the reader to \ref{SESpi1RHRSOC}.

\begin{figure}%[tb!]
\centerline{\includegraphics[width=0.5\linewidth]{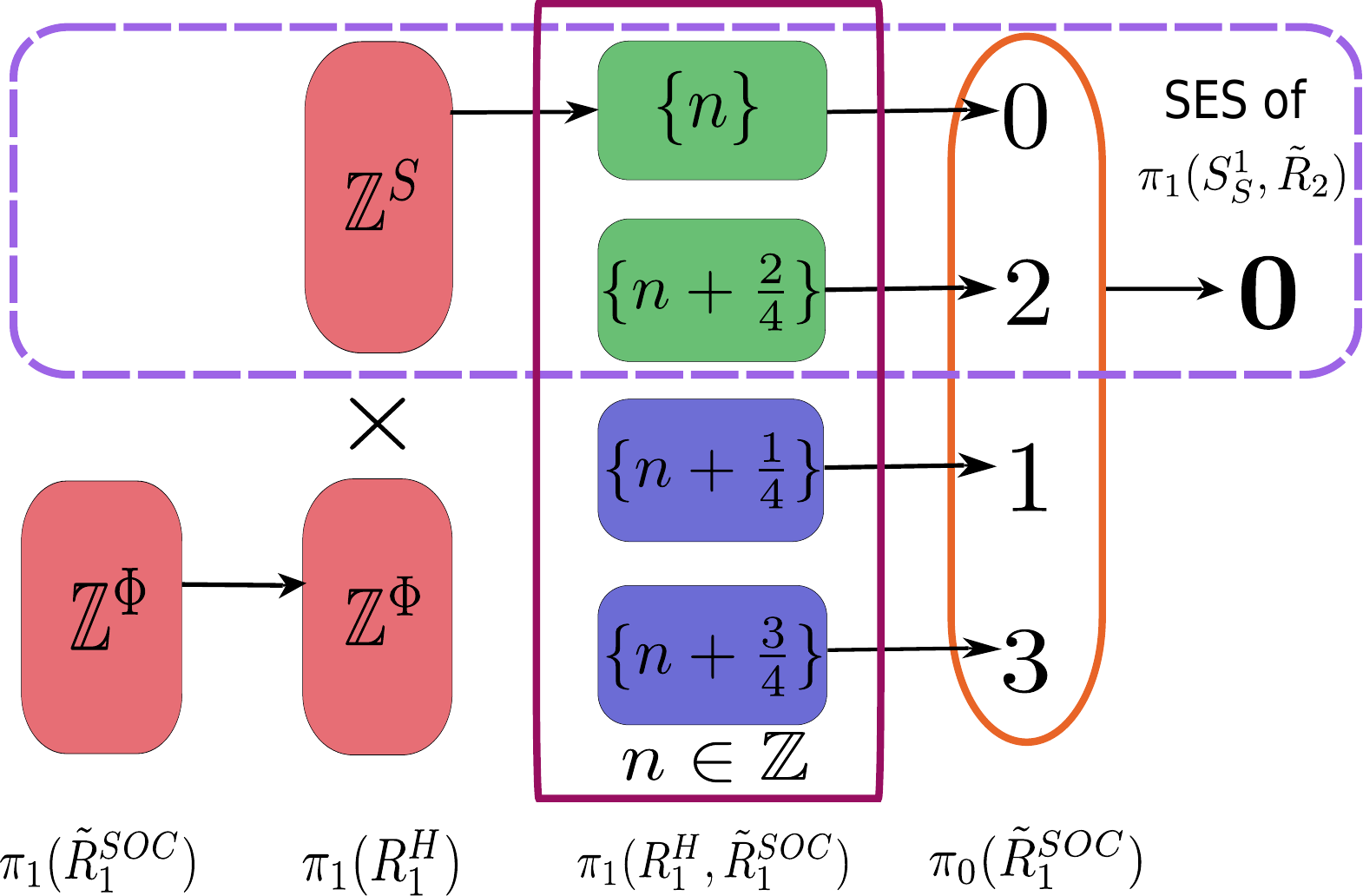}}
\caption{{\bf LES, SES, and spin solitons in the PdB phase.} The mapping diagram drawn here demonstrates that the linear objects given by $\pi_{1}(R_{1}^{\rm H},\tilde{R}_{1}^{\rm SOC})$ are spin solitons. This is because the mapping between $\pi_{1}(\tilde{R}_{1}^{\rm SOC})$ and $\pi_{1}(R_{1}^{\rm H})$ is a projection, the image of homomorphism $i^{*} : \pi_{1}(\tilde{R}_{1}^{\rm SOC}) \rightarrow \pi_{1}(R_{1}^{\rm H}) = \mathbb{Z}_{\phi}$, i.e. the topological invariant of all phase vortices. As a result, all the trivial linear objects given by $\pi_{1}(R_{1}^{\rm H},\tilde{R}_{1}^{\rm SOC})$ are phase vortices as ${\rm Im} [i^{*}] \cong \ker [j^{*}]$. Since $\ker [k^{*}] \cong {\rm Im} [\partial^{*}] = \mathbb{Z}_{4}$, there are three types of spin solitons (and one type of spin vortex). In addition, the subgroup $G=\{[n],[n+2/4]\}$ is an extension of $\mathbb{Z}_{\bf S}$ by $\pi_{0}(\tilde{R}_{2}^{\rm SOC}) = \mathbb{Z}_{2}$ and therefore isomorphic to $M$. The corresponding SES is marked by the dashed line. The HQV acts as a 1D nexus between the spin soliton given by the coset $[2/4]$ and the KLS wall \cite{Zhang2020}.}
\label{MappingDiagramAndSES}
\end{figure}

\subsubsection{Walls bounded by strings}

In the region $\xi_{\rm H} < r < \xi_{\rm D}$, the SES of $\pi_{}(R^{\rm H}_{1},R_{2}^{\rm PdB})$ is
\begin{equation}
\xymatrix@R=8pt{
0 \ar[r] & \mathbb{Z}_{\phi} \ar[r] &
\pi_{1}( R_{1}^{\rm H},R_{2}^{\rm PdB}) \ar[r]^-{\partial^{*}} & \mathbb{Z}_{2(\phi + \mathbf{S})} \ar[r] & 0\\
}\,.
\label{SES1RH1R2}
\end{equation}
Eq.~(\ref{SES1RH1R2}) determines how linear defects with a characteristic length scale $\xi_{\rm H} < r < \xi_{\rm D}$ connect with a possible domain wall. Eq.~(\ref{SES1RH1R2}) suggests 
\begin{equation}
  \pi_{1}( R_{1}^{\rm H},R_{2}^{\rm PdB}) \cong \tilde{\mathbb{Z}},
  \label{pi1RH1R2}
\end{equation}
which is isomorphic to $\pi_{1}(R_{1}^{\rm PdB},R_{2}^{\rm PdB})$ within $\xi_{\rm H}$. In other words the KLS wall, determined by two length scales $\xi$ and $\xi/q$, extends into the region $\xi_{\rm H} \leq r \leq \xi_{\rm D}$. However, Eq.~(\ref{SES1RH1R2}) only contains the phase factor $\phi$. In particular, all information about the spin degrees of freedom is lost as they are trivial elements of $\pi_{1}(R_{1}^{\rm H}, R_{2}^{\rm PdB})$. To restore the spin part of the KLS wall, we recall that the domain wall connects regions with $(\mathbf{d}, \mathbf{e}^{1}, q, \phi)$ and $(-\mathbf{d}, -\mathbf{e}^{1}, -q, \pi+\phi)$ \cite{Volovik1990}. However, Eq.~(\ref{pi1RH1R2}) only catches the phase degree of freedom, i.e. phase vortices. Similarly, the spin degree of freedom can be described by the group
\begin{equation}
M \equiv \{ n_{\bf S}/2 | n_{\bf S} \in \mathbb{Z} \}\,,
\end{equation}
such that $M/\pi_{1}(S^{1}_{\bf S}) \cong \mathbb{Z}_{2} = \{ [0], [1/2]\}$, where $\pi_{1}(S^{1}_{\bf S}) \subset \pi_{1}(R_{1}^{\rm H})$ represents free spin vortices. The cosets $[0]$ and $[1/2]$ correspond to the absence or presence of the KLS string wall in the region $\xi_{\rm H} < r \leq \xi_{\rm D}$, respectively. Coset $[0] \cong 2\mathbb{Z}$ contains all integer spin vortices, while the Coset $[1/2] \cong \{n + 1/2 | n \in \mathbb{Z}\}$ contains all half-integer spin vortices, including the HQV. These properties are equivalent to SES
\begin{equation}
\xymatrix@R=8pt{
0 \ar[r] & \mathbb{Z}_{\mathbf{S}} \ar[r] &
M \ar[r]^-{\partial^{*}} & \mathbb{Z}_{2(\phi + \mathbf{S})} \ar[r] & 0
}.
\label{SESM}  
\end{equation}

\subsubsection{Spin solitons}
\label{FourSpinSolitons}

\begin{figure*}[!t]
\centerline{\includegraphics[width=0.7\linewidth]{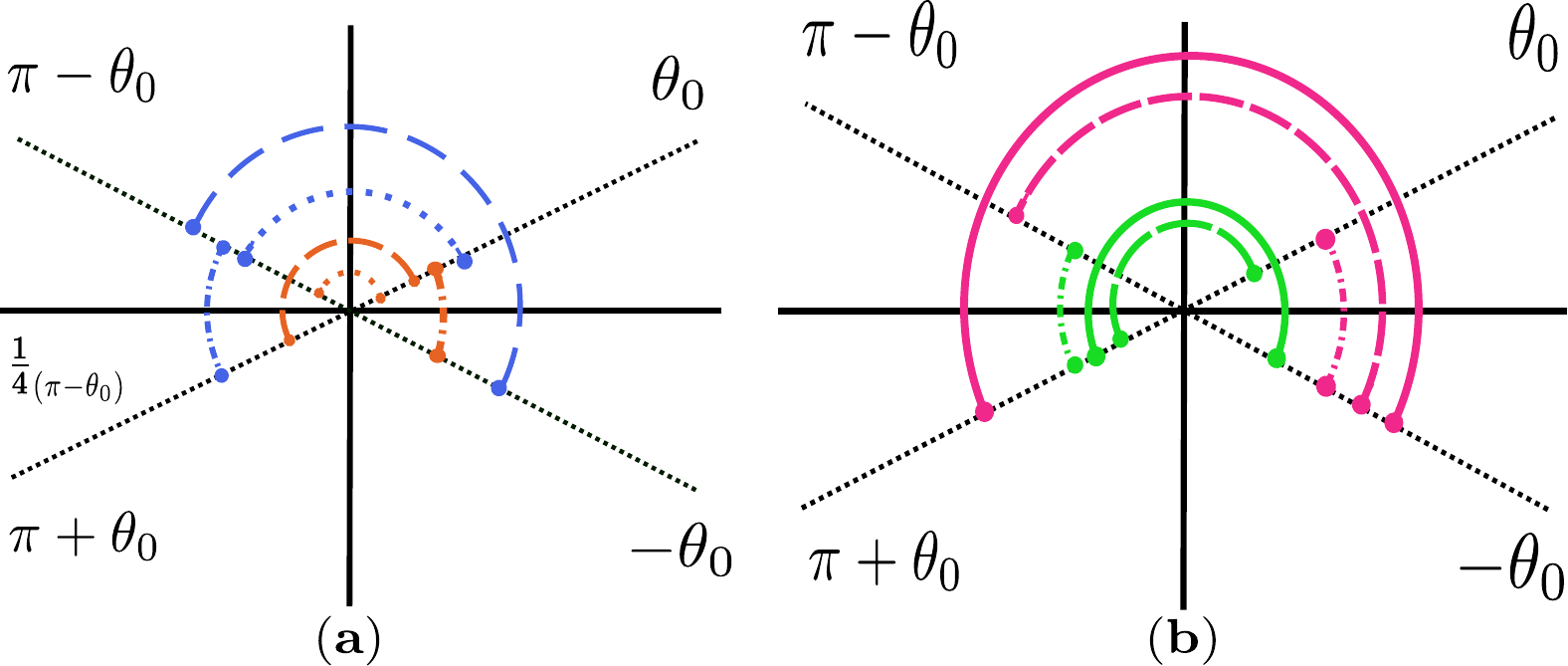}}
\caption{{\bf Soliton structures in the PdB phase.} The black fine dotted lines represent the four elements of $R_{\bf S}^{\rm SOC}$ i.e., $\pm \theta_{0}$ and $\pi \pm \theta_{0}$. The colored dashed, dotted, dash-dotted, and solid lines correspond to $\pi$-soliton ($|\Delta{\theta}| = \pi$), soliton ($|\Delta{\theta}| = \pi -2\theta_{0}$), KLS-soliton ($|\Delta{\theta}| = 2\theta_{0}$) and big-soliton ($|\Delta{\theta}| = \pi+2\theta_{0}$), respectively. Panel (a) sketches the spin solitons with topological invariants $1/4$, $2/4$ and $3/4$ originating at $\theta_{0}$ (orange) and $\pi-\theta_{0}$ (blue), respectively, and panel (b) sketches similar spin solitons originating at $-\theta_{0}$ (pink) and $\pi+\theta_{0}$ (green).}
\label{ClassesOfSolitons}
\end{figure*}

For $r > \xi_{\rm D}$, $R_{1}^{\rm H}$ reduces to $\tilde{R}_{1}^{\rm SOC} = R_{\bf S}^{\rm SOC} \times U(1)_\phi$. The resulting linear objects are classified by the homotopy group $\pi_{1}(R_{1}^{\rm H}, \tilde{R}_{1}^{\rm SOC})$ which has the SES
\begin{equation}
\xymatrix@R=8pt{
0 \ar[r] & \mathbb{Z}_{\bf S} \ar[r] &
\pi_{1}( R_{1}^{\rm H},\tilde{R}_{1}^{\rm SOC}) \ar[r]^-{{\partial}^{*}} & \mathbb{Z}_{4} \ar[r] & 0\\
}.
\label{SES2}
\end{equation}
It follows that 
\begin{equation}
\pi_{1}( R_{1}^{\rm H},\tilde{R}_{1}^{\rm SOC}) = \{n_{\bf S}/4 | n_{\bf S} \in \mathbb{Z}\} \cong \mathbb{Z}\,,
\end{equation}
and
\begin{equation}
\pi_{1}( R_{1}^{\rm H},\tilde{R}_{1}^{\rm SOC})/\mathbb{Z}_{\bf S} \cong \mathbb{Z}_{4}.
\end{equation}
Because Eq.~(\ref{SES2}) is determined by $ \mathbb{Z}_{\bf S} = \pi_{1}(S^{1}_{\bf S})$ and $\mathbb{Z}_{4} = \pi_{0}(R_{\bf S}^{\rm SOC})$ (see details in \ref{SESpi1RHRSOC}), we have
\begin{equation}
\pi_{1}(R_{1}^{\rm H}, \tilde{R}_{1}^{\rm SOC}) = \pi_{1}(S^{1}_{\bf S}, R_{\bf S}^{\rm SOC})\,.
\label{isomorphism1}
\end{equation}

Eq.(\ref{isomorphism1}) signifies that the linear objects classified by $\pi_{1}(R_{1}^{\rm H}, \tilde{R}_{1}^{\rm SOC})$ only involve the spin degree of freedom, i.e. they are spin solitons and spin vortices \cite{Zhang2020b, Mineev1978}. The four cosets of $\pi_{1}(S^{1}_{\bf S}, R_{\bf S}^{\rm SOC})$ are 
\begin{equation}
\hspace{-2cm}
[0] = \left\{ n_{\bf S} \right\} \,\,, \left[ \frac{1}{4} \right] = \left\{n_{\bf S}+\frac{1}{4} \right\}, \,\, \left[\frac{2}{4} \right]  =  \left\{ n_{\bf S}+\frac{2}{4} \right\} \,\,, \,\,{\rm and} \left[ \frac{3}{4} \right] = \left\{n_{\bf S}+\frac{3}{4} \right\}\,.
\label{cosets}
\end{equation} 
These cosets give the topological invariants of the four types of linear objects corresponding to spin vortices and three types of spin solitons. Fig.~\ref{ClassesOfSolitons} shows the representatives of spin solitons of $\pi_{1}(S^{1}_{\bf S}, R_{\bf S}^{\rm SOC})$. Following the terminology in Ref.~\cite{Makinen2019}, they are the big-soliton ($|\Delta \theta| = \pi +2\theta_{0}$), the soliton ($|\Delta \theta_{0}|=\pi-2\theta_{0}$), the KLS-soliton ($|\Delta \theta_{0}|=2\theta_{0}$), and the $\pi$-soliton ($|\Delta \theta|=\pi$). Spin vortices, i.e. the coset $[0]$, are not discussed further as they are outside the scope of this manuscript.

\begin{figure*}%[tb!]
\centerline{\includegraphics[height=0.35\linewidth, width=0.9\linewidth]{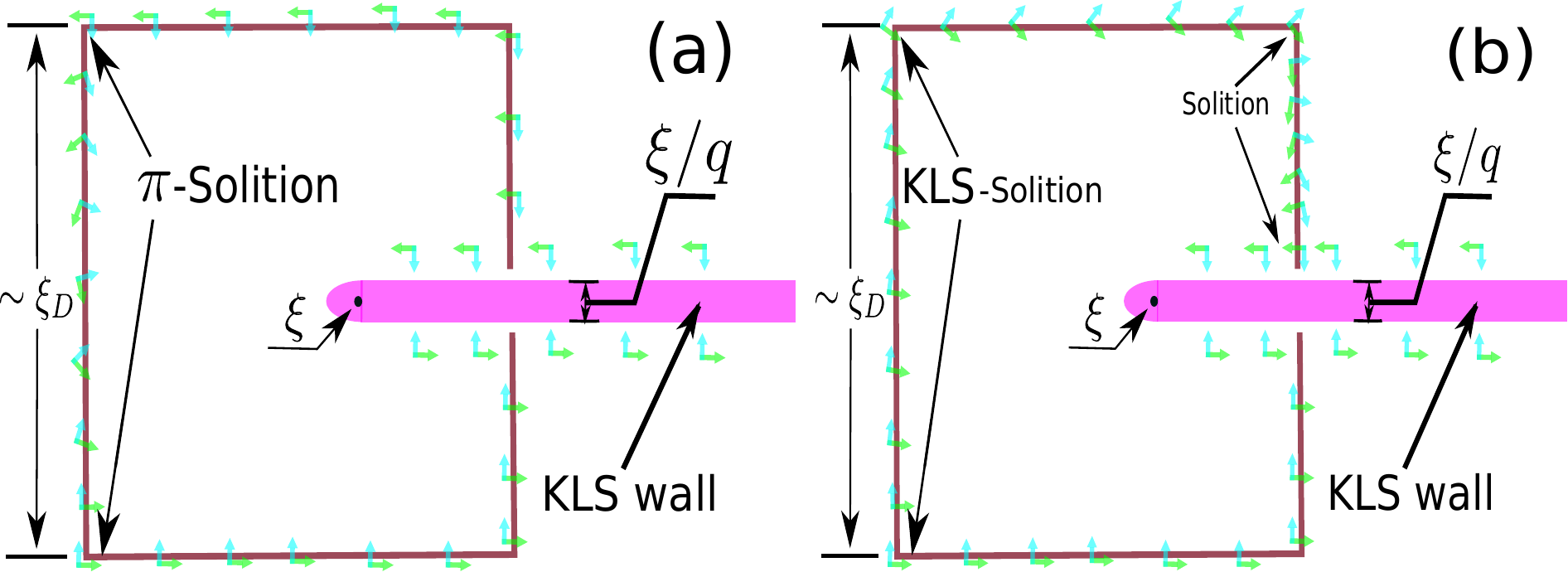}}
\caption{{\bf Nexus object configurations.} The 1D nexus object consists of spin soliton(s), a HQV and a KLS wall. The green and cyan arrows represent the $\hat{\mathbf{d}}$ and $\hat{\mathbf{e}}$ vectors respectively. (a) Configuration of the inseparable spin soliton. In this configuration, the topological invariant of spin soliton is $1/2$, which corresponds to $\pi$-soliton. (b) In this configuration there are two spin solitons with topological invariant $1/4$ when the group $\pi_{1}(S^{1}_{\bf S}, \tilde{R}_{2})$ is implemented in alternative way. Following the requirement of continuity of order parameter, these two spin solitons are the KLS-soliton ($\Delta{\theta}=2\theta_{0}$) and the soliton ($\Delta{\theta}=\pi-2\theta_{0}$).
\label{IllustrationsOfInseparableAndSeparableSolitons}}
\end{figure*}

A significant property of $\pi_{1}(S^{1}_{\bf S},R_{\bf S}^{\rm SOC})$ is that it has a subgroup $G_{\rm sub} \equiv \{[0],[2/4]\}$ such that $G_{\rm sub}/\mathbb{Z}_{\bf S} \cong \mathbb{Z}_{2}$. It follows that the SES of $G_{\rm sub}$ is given by Eq.~(\ref{SES2}) as
\begin{equation}
\xymatrix@R=8pt{
0 \ar[r] & \mathbb{Z}_{\bf S} \ar[r] &
G_{\rm sub} \ar[r]^{\partial^{*}} & \mathbb{Z}_{2} \ar[r] & 0 \,.
}\,
\label{SES3}
\end{equation}
The mapping diagram of Eq.~(\ref{SES3}) is shown as the dashed panel in Fig.~\ref{MappingDiagramAndSES}. Comparing Eqs.~(\ref{SESM}) and (\ref{SES3}) leads to
\begin{equation}
G_{\rm sub} = \pi_{1}(S^{1}_{\bf S}, \tilde{R}_{2}) \cong \hat{\mathbb{Z}} = M,
\label{isomorphism2}
\end{equation}
where $\hat{\mathbb{Z}} \equiv \{n_{\bf S}/2 | n_{\bf S} \in \mathbb{Z}\}$.

Eq.~(\ref{isomorphism2}) suggests that one can continuously transform spin solitons, classified by the $[2/4]$ coset of $\pi_{1}(S^{1}_{\bf S}, \tilde{R}_{2})$, to half spin vortices of $M$. In other words, the KLS wall smoothly connects to the $[2/4]$ spin soliton via a HQV. Similarly to a 2D nexus connecting the string monopole and the vortex skyrmion \cite{Zhang2020b,Volovik2009Book}, the HQV is a 1D nexus connecting the KLS wall and the $[2/4]$ spin soliton \cite{Zhang2020}. The composite object formed by the $[2/4]$ spin soliton and the KLS wall is called the 1D nexus object.

\begin{figure*}[!t]
\centerline{\includegraphics[width=1.0\linewidth]{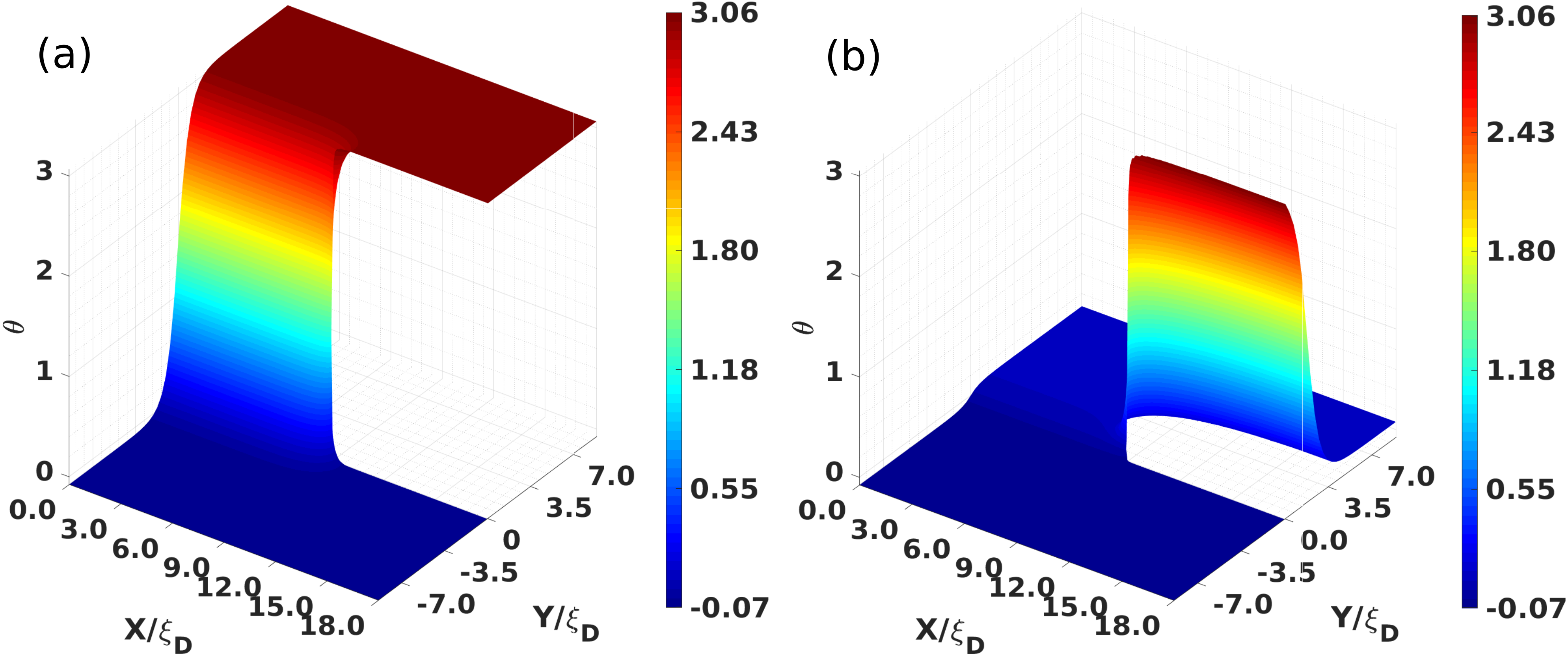}}
\caption{{\bf Minimum energy configurations for different spin solitons.} Numerically calculated equilibrium configurations of inseparable and separable spin solitons in one-half unit cell with $|q|=0.18$ and $D=18\xi_{\rm D}$. (a)~Inseparable configuration with the $\pi$-soliton, corresponding to the sketch in Fig.~\ref{IllustrationsOfInseparableAndSeparableSolitons}a. (b)~Separable configuration with combination of he KLS-soliton and the soliton, corresponding to the sketch in Fig.~\ref{IllustrationsOfInseparableAndSeparableSolitons}b.
\label{OneHalf}   
} 
\end{figure*}

\subsubsection{The 1D nexus object}
\label{TwoDifferentSolitonConfiguretions}

We note that since $\pi_{1}(S^{1}_{\bf S},R^{\rm SOC}_{\bf S})/\mathbb{Z}_{\bf S} \cong \mathbb{Z}_{4}$, we have $[2/4] = [1/4] + [1/4]$. It follows that the relative homotopy group $\pi_{1}(S^{1}_{\bf S},\tilde{R}_{2})$ can be represented both as $\{[0],[1/4]+[1/4]\}$ and as $\{[0],[2/4]\}$. In other words, there are two types of spin soliton configurations connecting to a KLS wall via a HQV for a given element of $\pi_{1}(S^{1}_{\bf S},\tilde{R}_{2})$. When the topological invariant is $2/4$, the spin soliton is spatially inseparable and forms a $\pi$-soliton sketched in Fig.~\ref{IllustrationsOfInseparableAndSeparableSolitons}(a).

On the other hand, when the topological invariant is $1/4+1/4$, the spin soliton is a combination of two spatially separable spin solitons, each described by the topological invariant $1/4$. These two spatially separated spin solitons are the KLS soliton ($|\Delta{\theta}|= 2\theta_{0}$) and the soliton ($|\Delta{\theta}| = \pi - 2\theta_{0}$). As illustrated in Fig.~\ref{IllustrationsOfInseparableAndSeparableSolitons}(b), the 1D nexus object contains two spin solitons. As we will see in Sec.~\ref{SD}, the free energies and spin textures of the two cases are quite different, resulting in distinguishable dynamic spin response properties and NMR frequency shifts. The corresponding minimum energy configurations based on 2D calculations \cite{Zhang2020} for both configurations are shown in Fig.~\ref{OneHalf}.

\subsection{Vortex-bound solitons in the polar phase}
\label{sec:polarsolitons}

In the polar phase, the solitons are bound by HQVs. The existence of HQVs was predicted decades ago in $^3$He-A \cite{VolovikMineev1976} and observed recently in the polar \cite{Autti2016} and polar-distorted \cite{Makinen2019} phases of $^3$He. Previously, HQVs have been observed in grain boundaries of $d$-wave cuprate superconductors \cite{PhysRevLett.76.1336}, in superconductor rings \cite{Jang186}, and in Bose condensates \cite{Lagoudakis974,PhysRevLett.115.015301}. In $p$-wave superfluids such as $^3$He, HQVs provide access to vortex-core-bound fermion states, which have been predicted to harbor non-Abelian anyons in 2D $p_{x} + ip_y$ superconductors and superfluids \cite{PhysRevLett.86.268, KITAEV20032}.

In the polar phase the presence of magnetic field larger than the dipole field, $H > 3\,$mT, fixes the spin anisotropy vector $\hat{\mathbf{d}} = \hat{\mathbf{i}} \cos \theta (\mathbf{r}) + \hat{\mathbf{j}} \sin \theta (\mathbf{r})$ in Eq.~(\ref{eq:polarop}) via the spin-orbit interaction $F_{\mathrm{so}} \propto (\hat{\mathbf{d}} \cdot \hat{\mathbf{m}})^2$ to the plane perpendicular to ${\bf H}$. Vectors $\hat{\mathbf{i}} $ and $ \hat{\mathbf{j}}$ are mutually orthogonal unit vectors in the plane normal to $\mathbf{H}$. The orbital anisotropy vector $\hat{\mathbf{m}}$ is pinned parallel to nafen strands, $\hat{\mathbf{m}} \parallel \hat{\mathbf{z}}$. The combined effect of the confinement and magnetic field affects the distribution of the $\hat{\mathbf{d}}$ vector such that $\theta$ is governed by the Sine-Gordon equation
\begin{equation}\label{SineGordonEq}
 \nabla^2 \theta =  \frac{1}{2\xi_{\mathrm{D}}^2 } \sin^2 \mu\,\sin 2 \theta\,.
\end{equation}
Here $\xi_{\mathrm{D}}\sim 10\,\mu$m is the dipole length and $\mu$ is the angle of the magnetic field with respect to $\hat{\mathbf{z}}$.

%%%%%%%%%%%%%%%%%%%%%%%%%%%%%%%%%%%%%%%%%
%%%%%%%%%%%%%%%%%%%%%%%%%%%%%%%%%%%%%%%%%
\begin{figure*}[tb!]
\centerline{\includegraphics[width=1\linewidth]{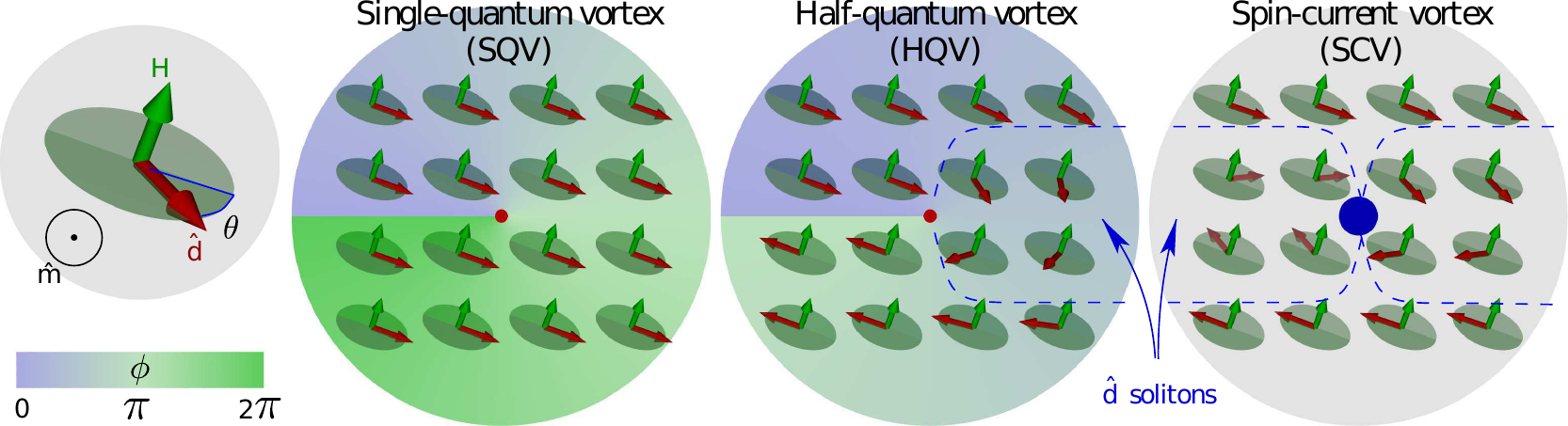}}
\caption{\label{PolarPh_vortices} \textbf{Vortex types in the polar phase.} The phase of the order parameter $\phi$ is shown as the background color. The spin anisotropy vector $\hat{\mathbf{d}}$ is locked to the plane (green disks) perpendicular to the magnetic field $\mathbf H$ by magnetic energy. In this plane, the vector $\hat{\mathbf{d}}$ rotates by $\pi$ around the HQV core and by $2 \pi$ around the spin vortex core. In a tilted magnetic field ($\mathbf{H} \nparallel \hat{\mathbf{m}}$) the reorientation of the $\hat{\mathbf{d}}$ vector is concentrated in one or two solitons (illustrated with dashed lines) with $\pi$ winding, terminating at the HQV or at the spin vortex core, respectively. The nafen strands, oriented along $\hat{\mathbf{m}}$, and the vortex lines are orthogonal to the figure. The phase vortex and the HQV have hard cores (red discs) of the size of coherence length $\xi \sim 40\,$nm, while the spin vortex has a soft core (blue disc) of much larger dipolar size $\xi_{\mathrm{D}} \sim 10$~$\mu$m.}
\end{figure*}
%%%%%%%%%%%%%%%%%%%%%%%%%%%%%%%%%%%%%%%%%
%%%%%%%%%%%%%%%%%%%%%%%%%%%%%%%%%%%%%%%%%

The polar phase order parameter supports three different vortex types illustrated in Fig. \ref{PolarPh_vortices} -- the singly quantized phase vortices, the spin-current vortices (SCV), and the HQVs. Reorientation of $\hat{\mathbf{d}}$ outside of SCV and HQV cores is governed by solitonic solutions of Eq.~(\ref{SineGordonEq}). In tilted ${\bf H} \nparallel \hat{\bf z}$ field the SCV terminates two $\pi$-solitons, while the HQV terminates just one. Additionally, solitons may be terminated at the sample boundary. The soliton width is fixed by the characteristic length scale $\sim \xi_{\mathrm{D}} / \sin \mu$. For ${\bf H} \parallel \hat{\bf z}$ or at zero magnetic field, all states with $\hat{\bf d} \perp \hat{\bf m}$ are degenerate, and solitons are not created.

\subsection{Composite spin-mass vortex with soliton tail in $^3$He-B}

\begin{figure}
\includegraphics[width=\linewidth]{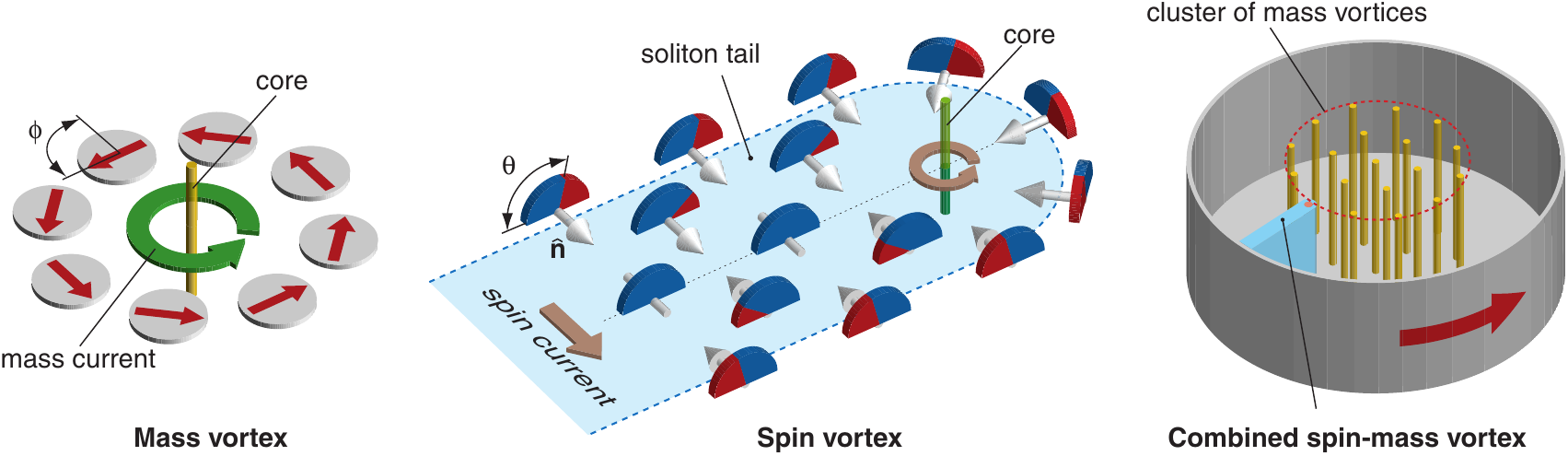}
\caption{{\bf Structure of the spin-mass vortex (SMV) in $^{\bf 3}$He-B.} \textit{(Left)} Mass vortex traps winding of phase $\phi$ around a hard core. \textit{(Center)} Disclination in the orbital rotation field takes form of a soliton tail terminated by a spin vortex with a hard core. \textit{(Right)} When cores of mass and spin vortices merge, a stable configuration of SMV in rotating container is possible.}
\label{SMV_struct}
\end{figure}

The existence of the spin-mass vortex -- a composite defect in superfluid $^3$He-B \cite{Kondo1992, Eltsov2000} -- may be seen from the B-phase order parameter,  Eq.~(\ref{Bop}). The B phase order parameter supports different types topological defects sketched in Fig.~\ref{SMV_struct}. In rotation the most usual defect is a mass-current vortex, a conventional quantized vortex line. Around it the phase $\phi$ winds by $2\pi$, which results in a supercurrent proportional to $\nabla \phi$ around its singular core. Within the core the amplitude of the order parameter $|A_{\alpha j}|$ is depleted from its equilibrium value $\Delta_{\rm B}$. A second defect is a disclination in the $R_{\alpha j}$ field. It also has a singular core which is encircled by a spin current. On moving once around the core $\hat{\mathbf{n}}$ reverses its direction twice: First by smooth rotation while the angle $\theta$ remains at the equilibrium value $\theta_{\rm L} \approx 104^\circ$, which minimizes the spin-orbit interaction energy. Later by increasing $\theta$ to $180^\circ$, where both directions of $\hat{\mathbf{n}}$ are equivalent, and then decreasing back to $\theta_{\rm L}$. The second leg in the direction reversal does not minimize the spin-orbit interaction and hence it becomes confined in space within a planar structure, a soliton sheet, which terminates on the linear singular core or on the wall of the container. This structure becomes possible through the existence of two different energy (and length) scales: The superfluid condensation energy defines the scale of the coherence length $\xi \sim 10\,$--$100\,$nm, which is roughly the radius of the singular core. The much weaker spin-orbit interaction defines the scale of the dipolar healing length $\xi_{\rm D} \sim 10\,\mu$m, which is approximately the width of the soliton sheet. The detailed topological analysis of the B-phase soliton structures can be found in Ref.~\cite{MineevVolovik1978}, where experimentally realized configurations correspond to $(-+1)$ or $(+-1)$ classes in terms of that work.

The spin vortex by itself is an unstable structure: The surface tension of the soliton leads to its annihilation and disappearance. Another composite defect -- the combined spin-mass vortex (SMV) -- can be stabilized in the rotating container. It has both phase $\phi$ winding and a $R_{\alpha j}$ disclination trapped on the same core. If the number of mass-current vortex lines in the container is less than that in the equilibrium state in rotation, then the existing lines are confined to a cluster in the center of the container by the Magnus force from the uncompensated normal-superfluid counterflow. Due to the trapped mass current the SMV also experiences the Magnus force which pulls it towards the vortex cluster. This force is opposed by the surface tension of the soliton and as a result its equilibrium position is slightly outside the cluster of mass vortices. The soliton tail of the SMV has a characteristic NMR absorption response which allows its identification.

Another possible stable configuration is the soliton sheet bounded by two SMVs and embedded within the vortex cluster. The size of such pair is set by competition of the surface tension of the soliton sheet, which tries to pull vortices together and repulsion of two mass vortices of the same circulation. The equilibrium size is estimated to be about $6\xi_{\rm D}$ and it is too small to identify contribution of a single pair in the NMR spectrum.

\section{Spin dynamics and NMR}
\label{SD}
NMR techniques are used as a non-invasive way to probe the properties of different superfluid phases of $^3$He. NMR methods are particularly useful for extracting information about the SOC. This is because the long range coherence in superfluid states enhances the SOC energy \cite{Anderson1973,AndersonVarma1973}. In the superfluid state the observable NMR resonance frequency is often shifted from the Larmor value since the precessing spin experiences an additional torque from SOC. The topological solitons form potential wells for bound spin-wave states, which result in satellite peaks with characteristic frequency shifts. Since the characteristic time scale of SOC is much longer than the microscopic time scales of superfluid, i.e., ${\hbar}\Delta^{-1}$, the microscopic processes remain in equilibrium under weak magnetic perturbation. In other words, the NMR response is the hydrodynamic response of the spin densities $\delta{S_{a}}$ and spin vectors \cite{vollhardt2013superfluid, chaikin1995}.

Under weak magnetic perturbation ${\delta}H_{a} \equiv {\delta}\mathbf{H}$ along $y$-axis, the linear response of spin density is
\begin{equation}
{\delta}S_{+}({\mathbf{r}},t)={\int}{{\rm d}{\sigma}'}{\int}{{\rm d}t'}{\frac{{\delta}{S_{+}}}{{\delta}{H_{a}}}}(\mathbf{r},t,\mathbf{r}',t'){\delta}{H_{a}}({\mathbf{r}}',t')+O({\delta}{H_{a}}^{2}),
\label{LinearResponse}
\end{equation}
where ${\delta}S_{+}=[{\delta}S_{1}+i{\delta}S_{3}]/\sqrt{2}$ is transverse spin density and $a=1,2,3$ are spatial coordinate indices.
To calculate the response function and its poles, we use method based on hydrodynamic equations \cite{chaikin1995}. In the hydrodynamic limit, the system of dynamic equations of spin densities $S_{\alpha}$ and spin vectors are a system of Liouville equations 
\begin{equation}
{\frac{\partial{S_{\alpha}}}{\partial{t}}} =\{F_{\rm hy},S_{\alpha}\}, \,\,
 {\frac{\partial{V_{\alpha}^{a}}}{\partial{t}}}  =\{F_{\rm hy},V_{\alpha}^{a}\},%\,\,
\label{LiouvilleEquations}
\end{equation}
where $V_{\alpha}^{a=1} = \hat{\bf d}$,$ V_{\alpha}^{a=2} = \hat{\bf e}^1 $ and $V_{\alpha}^{a=3} = \hat{\bf e}^2$, such that $\hat{\bf e}^2 = \hat{\bf d} \times \hat{\bf e}^1$ is directed along spin polarization, denote the three spin vectors of order parameter.
These equations are known as the Leggett equations  \cite{vollhardt2013superfluid}. The hydrodynamic free energy of a superfluid dominated by the SOC energy is
\begin{equation}
F_{\rm hy}=\int\nolimits_{\Sigma} (f_{\rm H} + f_{\rm soc} + f_{\rm grad}){\rm d}\Sigma.
\label{FreeEnergyOfHydrodynamics}
\end{equation}
Eq.~(\ref{LiouvilleEquations}) can be rewritten as
\begin{equation} \hspace{-1cm}
  \eqalign{
  {\frac{\partial{S_{\alpha}}}{\partial{t}}} &= {\int_{\Sigma}}{{{\rm d}^{3}}r'}\frac{{\delta}F_{\rm hy}}{{\delta}{S_{\beta}}}(r')\{S_{\beta}(r'),{S_{\alpha}}(r)\} 
  + {\int_{\Sigma}}{{{\rm d}^{3}}r'}\frac{{\delta}F_{\rm hy}}{{\delta}{V_{\beta}^{a}}}(r')\{V_{\beta}^{a}(r'),{S_{\alpha}}(r)\}
  }
\label{LiouvilleEquations1}
\end{equation}
and
\begin{equation}
{\frac{\partial{V_{\alpha}^{a}}}{\partial{t}}}={\int_{\Sigma}}{{{\rm d}^{3}}r'}\frac{{\delta}F_{\rm hy}}{{\delta}{S_{\beta}}}(r')\{S_{\beta}(r'),{V_{\alpha}^{a}}(r)\},
\label{LiouvilleEquations2}
\end{equation}
where $\alpha,\beta=1,2,3$ are indexes of spatial components of hydrodynamic variables. The Poisson brackets between $S_{\alpha}$ and $V_{\alpha}^{a}$ are \cite{Dzyaloshinskii1980} 
\begin{equation}
\eqalign{
\{{S_{\alpha}}(r_{1}),{{S_{\beta}}(r_{2})}\} &= {\epsilon}_{{\alpha}{\beta}{\gamma}}{S_{\gamma}}{\delta}(r_{1}-r_{2}) \cr
\{{S_{\alpha}}(r_{1}),{{V_{\beta}^{a}}(r_{2})}\} &= {\epsilon}_{{\alpha}{\beta}{\gamma}}{V_{\gamma}^{a}}{\delta}(r_{1}-r_{2}),
}
\label{PoissonBrackets1}
\end{equation}
where $r_{1}$ and $r_{2}$ are the spatial coordinates and ${\epsilon}_{{\alpha}{\beta}{\gamma}}$ is the Levi-Civita symbol.

\subsection{Satellite shifts induced by spin solitons in the PdB phase}
\label{PdBshift}

After plugging Eq.~(\ref{PoissonBrackets1}) into Eqs.~(\ref{LiouvilleEquations1}) and (\ref{LiouvilleEquations2}), the coupled first order dynamic equations for spin densities $S_{\alpha}$ and $V_{\alpha}^{a}$ for the PdB phase are given as
\begin{equation}
\frac{{\partial}{S_{\alpha}}}{{\partial}{t}} ={\gamma}{H_{\beta}}{\epsilon_{{\alpha}{\beta}{\gamma}}}{S_{\gamma}} -{\frac{6}{5}}{g_{\rm D}}{V_{j}^{d}}{V_{\gamma}^{b}}{{\epsilon}_{{\alpha}{\beta}{\gamma}}}{Q_{{\beta}j}^{bd}}+({{\partial}_{i}}{{\partial}_{j}}{V_{\beta}^{b}}){V_{\gamma}^{a}}{\epsilon_{{\alpha}{\beta}{\gamma}}}{K_{ij}^{ba}}, 
\label{1stOderEquationsA}
\end{equation}
\begin{equation}
\frac{{\partial}{V_{\alpha}^{a}}}{{\partial}t} = {\gamma}{H_{\beta}}{\epsilon_{{\alpha}{\beta}{\gamma}}}{V_{\gamma}^{a}} -{\delta}{\gamma^{2}}{\chi_{\perp}^{-1}}{S_{\eta}V_{\eta}^{3}}{V_{\beta}^{3}}{\epsilon_{{\alpha}{\beta}{\gamma}}}{V_{\gamma}^{a}}-{\gamma^{2}}{\chi_{\perp}^{-1}}{S_{\beta}}{\epsilon_{{\alpha}{\beta}{\gamma}}}{V_{\gamma}^{a}},
\label{1stOderEquationsB}
\end{equation}
where $\delta  = (\chi_{\bot} -\chi_{\|})/\chi_{\|}$ in which $\chi_{\bot}$ and $\chi_{\|}$ are the PdB phase transverse and longitudinal magnetic susceptibility, respectively. Here
\begin{equation}
  \eqalign{
K_{ij}^{ba} =K_{1}{\delta}_{ij}{X_{m}^{b}}{X_{m}^{a}}+K_{2}{X_{j}^{a}}{X_{i}^{b}}+K_{3}{X_{j}^{b}}{X_{i}^{a}},\cr
Q_{{\beta}{j}}^{bd} ={X_{\beta}^{b}}{X_{j}^{d}}+{X_{\beta}^{d}}{X_{j}^{b}}
}
\end{equation}
with ${X_{i}^{1}}={{\Delta}_{{\perp}1}}{\hat{x}_{i}}, {X_{i}^{2}}={{\Delta}_{{\perp}2}}{\hat{y}_{i}}, {X_{i}^{3}}={{\Delta}_{{\parallel}}}{\hat{z}_{i}}$.
Starting from the first order equations of spin densities and degenerate parameters in Eq.~(\ref{1stOderEquationsA}) and  Eq.~(\ref{1stOderEquationsB}), we can further derive the second order spin dynamic response equations of $\delta{S_{\alpha}}$ under weak magnetic drive $\delta{H_{\alpha}}$ by plugging 
\begin{equation}
  \eqalign{
    H_{\alpha}=H_{\alpha}^{(0)}+{\delta}{H_{\alpha}(t)} \cr
    S_{\alpha}=S_{\alpha}^{(0)}+{\delta}{S_{\alpha}(\mathbf{r},t)} \cr
    V_{\alpha}^{a}=V_{\alpha}^{a(0)}+{\delta}{V_{\alpha}^{a}(\mathbf{r},t)}
    }
\end{equation}
into Eqs.~(\ref{1stOderEquationsA}) and (\ref{1stOderEquationsB}). Here the $S_{\alpha}^{(0)}$ and $V_{\alpha}^{a(0)}$ are the equilibrium spin densities and equilibrium degenerate parameters respectively. While the ${\delta}{S_{\alpha}(\mathbf{r},t)}$ and ${\delta}{V_{\alpha}^{a}(\mathbf{r},t)}$ are the dynamic parts of the perturbed spin densities and degenerate parameters. The $H_{\alpha}^{(0)}$ is the static magnetic field and ${\delta}{H_{\alpha}(t)} = |\delta\mathbf{H}|\hat{x}e^{-i{\omega}t}$ is the RF drive. The derived spin dynamic response equations are
\begin{equation}
  \eqalign{
i{\omega}{{\delta}{S_{\alpha}}({\omega})} =& {\gamma}{\epsilon_{{\alpha}{\beta}{\gamma}}}{H_{\beta}^{(0)}}{{\delta}S_{\gamma}(\omega)} +{\gamma}{\epsilon_{{\alpha}{\beta}{\gamma}}}S_{\gamma}^{(0)}{{\delta}{H_{\beta}}}(\omega) \nonumber \cr
&+{\frac{\Xi_{{\alpha}{\lambda}}}{i\omega}}{\delta}{S_{\lambda}}(\omega) +{\frac{C_{{\alpha}{\eta}}}{i{\omega}}}{\delta}{H_{{\eta}}}(\omega)}
\label{2ndOderEquations}
\end{equation}
and
\begin{equation}
  \eqalign{
{\Xi_{{\alpha}{\lambda}}} ={\frac{\gamma^{2}}{\chi_{\perp}}}{K_{ij}^{ba}}{\Lambda_{ij{\alpha}{\lambda}}^{ba}}+{\frac{6{g_{\rm D}}{\gamma^{2}}}{5{\chi_{\perp}}}}{R_{j{\lambda}{\alpha}{\beta}}^{db}}{Q_{{\beta}j}^{bd}}+{\frac{6{g_{\rm D}}{\gamma}^{2}}{5\chi_{\perp}}}{V_{\zeta}^{d{(0)}}}{V_{\gamma}^{b(0)}}{\epsilon_{j{\lambda}{\zeta}}}{\epsilon_{{\alpha}{\beta}{\gamma}}}Q_{{\beta}{j}}^{bd} \cr
C_{{\alpha}{\eta}} = {\gamma}G_{{i}{j}{\alpha}{\eta}}^{ba}{K_{{i}{j}}^{ba}} -{\frac{6{g_{\rm D}}{\gamma}}{5}}{R_{j{\eta}{\alpha}{\beta}}^{db}}{Q_{{\beta}j}^{bd}}-{\frac{6{g_{\rm D}}{\gamma}}{5}}{V_{\zeta}^{d(0)}}{V_{\gamma}^{b(0)}}{\epsilon_{j{\eta}{\zeta}}}{\epsilon_{{\alpha}{\beta}{\gamma}}}Q_{{\beta}{j}}^{bd},} 
\label{XiAndC}
\end{equation}
where
\begin{eqnarray}
  \eqalign{
R_{j{\eta}{\alpha}{\beta}}^{db} &= {V_{j}^{d(0)}}{V_{{\beta}}^{b(0)}}{\delta_{{\eta}{\alpha}}}-{V_{j}^{d(0)}}{V_{{\alpha}}^{b(0)}}{\delta_{{\eta}{\beta}}}, \cr
G_{ij{\alpha}{\gamma}}^{ba} &= ({\partial_{i}}{\partial_{j}}{V_{{\alpha}}^{b(0)}}){V_{{\gamma}}^{a(0)}}-({\partial_{i}}{\partial_{j}}{V_{{\beta}}^{b(0)}}){\delta_{{\beta}{\gamma}}}{V_{{\alpha}}^{a(0)}}, \cr
{\Lambda_{ij{\alpha}{\lambda}}^{ba}} &= ({\partial_{i}}{\partial_{j}}{V_{{\beta}}^{b(0)}}){{\delta}_{{\beta}{\lambda}}}{V_{{\alpha}}^{a(0)}} + ({V_{{\gamma}}^{b(0)}}{V_{{\gamma}}^{a(0)}}{\delta_{{\alpha}{\lambda}}}-{\delta_{{\gamma}{\lambda}}}{V_{{\alpha}}^{b(0)}}{V_{{\gamma}}^{a(0)}}){\partial_{i}}{\partial_{j}} \cr
&+ [({\partial_{i}}{V_{{\gamma}}^{b(0)}}){V_{{\gamma}}^{a(0)}}{\delta_{{\alpha}{\lambda}}}-({\partial_{i}}{V_{{\alpha}}^{b(0)}}){V_{{\gamma}}^{a(0)}}{\delta_{{\gamma}{\lambda}}}]{\partial_{j}} \cr
&+ [({\partial_{j}}{V_{{\gamma}}^{b(0)}}){V_{{\gamma}}^{a(0)}}{\delta_{{\alpha}{\lambda}}}-({\partial_{j}}{V_{{\alpha}}^{b(0)}}){V_{{\gamma}}^{a(0)}}{\delta_{{\gamma}{\lambda}}}]{\partial}_{i} \cr
&- {\delta_{{\gamma}{\lambda}}({\partial_{i}}{\partial_{j}}{V_{{\alpha}}^{b(0)}}){V_{{\gamma}}^{a(0)}}}.}
\label{RGandLambda}
\end{eqnarray}
The first two terms of Eq.~(\ref{2ndOderEquations}) correspond to the Larmor precession with frequency $\omega_{\rm L}=\gamma H^{(0)}$, while the last two terms, known as torque terms, describe the NMR response related to the superfluid order parameter texture.

\begin{figure*}%[tb!]
\centerline{\includegraphics[width=1.0\linewidth]{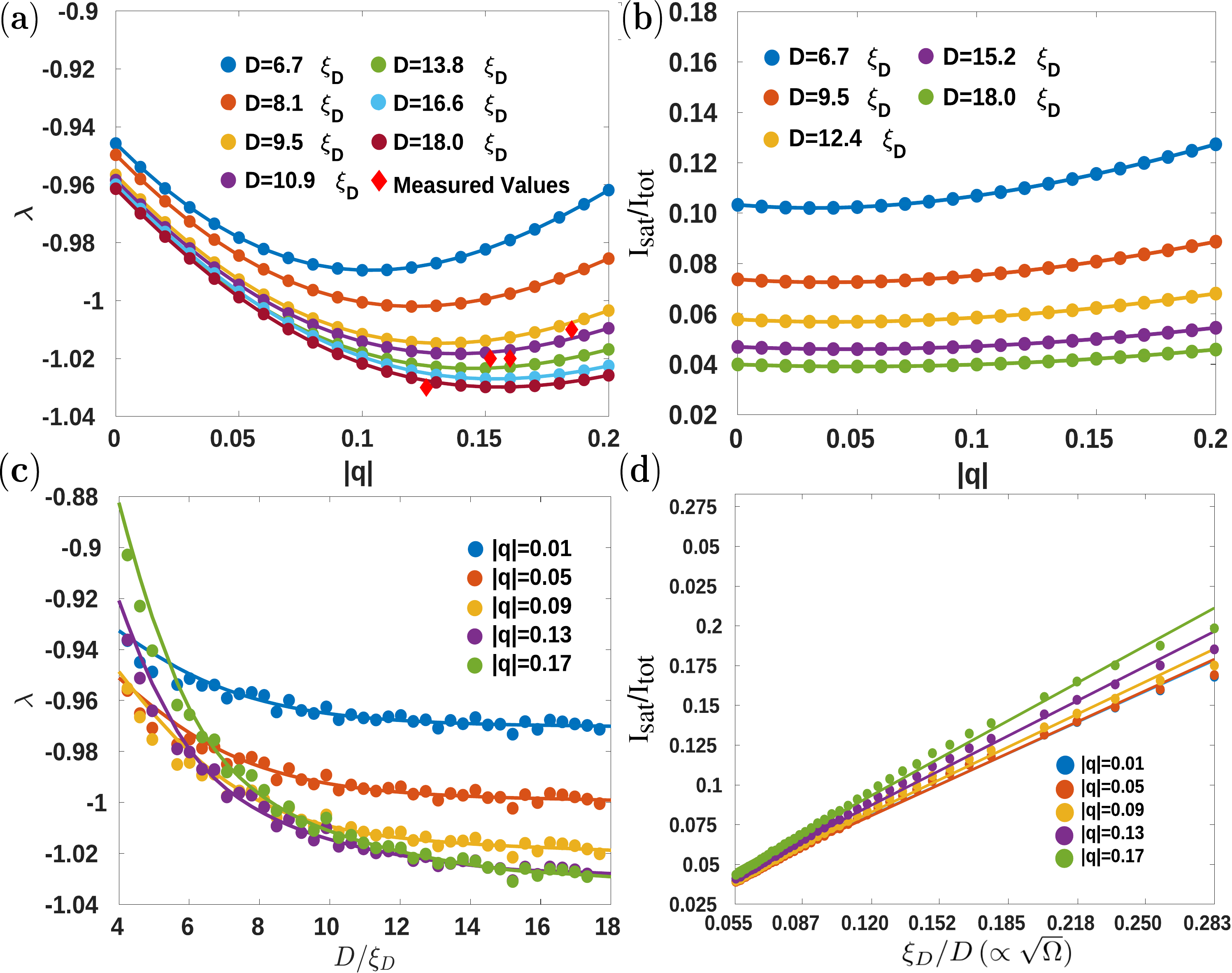}}
\caption{{\bf Intensity and frequency shift -- inseparable $\bf \pi$-soliton.} The relative frequency shift $\lambda$ corresponds to the lowest eigenvalue of Eq.~(\ref{NMREigenEquationDimensonless}) with the equilibrium texture of $\pi$-solitons in London limit. Filled circles represent the numerical results, while the lines are guides for the eye. (a) For experimentally relevant values of $|q| \leq 0.2$ the relative frequency shift $\lambda \approx -1$ for all tested inter-vortex separations $D$, in agreement with the experimental data from Ref.~\cite{Makinen2019} (red diamonds). (b) The relative satellite intensity remains fairly constant with $q$, but decreases with increasing inter-vortex distance. (c) The relative frequency shift increases when the inter-vortex distance is small and saturates to a value $\lambda \approx 1$ for larger separation, regardless of $q$. (d) The relative satellite intensity scales with intervortex separation approximately as $I_{\rm sat}/I_{\rm tot} \propto \sqrt{\Omega}$, in agreement with the experimental observations. \label{NMRLambdaAndRatioIntensityInseparableSoliton}}   
\end{figure*}

\begin{figure*}%[tb!]
\centerline{\includegraphics[width=1.0\linewidth]{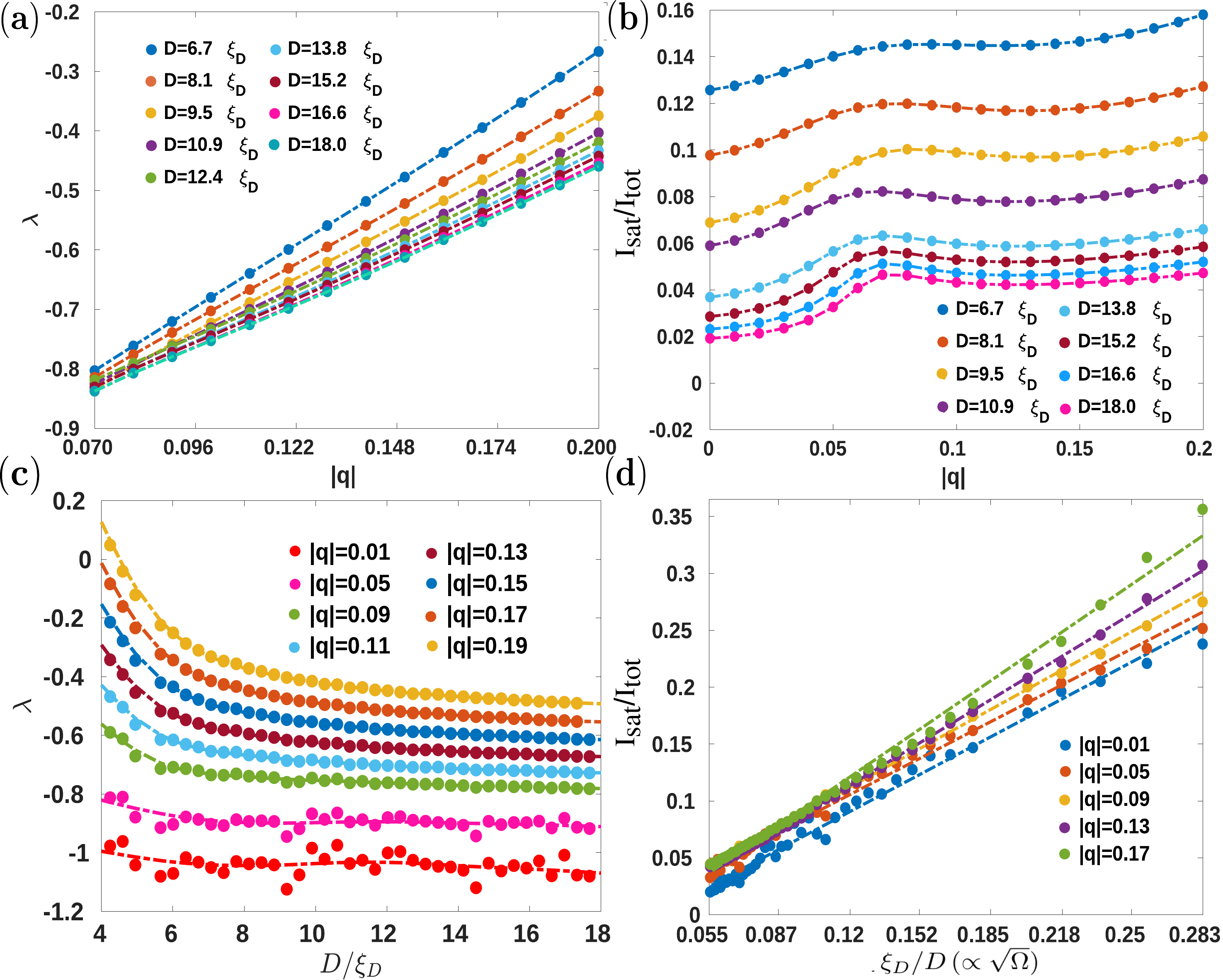}}
\caption{{\bf Intensity and frequency shift -- separable solitons.} The relative frequency shift $\lambda$ corresponds to the lowest eigenvalue of Eq.~(\ref{NMREigenEquationDimensonless}) with the equilibrium texture of $1/4+1/4$-solitons in London limit. Filled circles represent the numerical results, while the lines are guides for the eye. (a) The relative frequency shift $\lambda$ increases with increasing $|q|$, since only solitons ($|\Delta{\theta}|=\pi-2\theta_{0}$) contribute to the lowest transverse spin dynamic response mode. (b) The relative satellite intensity remains fairly constant with $q$, but decreases with increasing inter-vortex distance. (c) The relative frequency shift increases when the inter-vortex distance is small and saturates to a constant value determined by $|q|$ for larger separation. (d) The relative satellite intensity scales with intervortex separation approximately as $I_{\rm sat}/I_{\rm tot} \propto \sqrt{\Omega}$.
\label{NMRLambdaAndRatioIntensitySeparableSoliton}}
\end{figure*}

The torque terms in Eqs.~(\ref{XiAndC}) and (\ref{RGandLambda}) are fully determined by the equilibrium order parameter texture, i.e. the NMR frequency shift arises from the equilibrium texture of spin solitons. Taking the static magnetic field $\mathbf{H}^{(0)}=|\mathbf{H}^{(0)}|{\hat{\mathbf{y}}}$ and the parametrization Eq.~(\ref{PARA1}) into account, the poles of the dynamic response equations for the transverse spin density ${\delta}S_{+}=\frac{1}{\sqrt{2}}[{\delta}S_{1}(\omega)+i{\delta}S_{3}(\omega)]$ under weak magnetic drive $\delta{\mathbf{H}(t)}$ become
\begin{equation}
({\omega}^{2}-{\omega}_{\rm L}^{2}){\delta}S_{+}(\omega) =({\Xi}_{11}+{\Xi}_{33})+i({\Xi}_{13}-{\Xi}_{31}){\delta}S_{+}(\omega)\,.
\label{NMREigenEquation}
\end{equation}
In experiments, the observed transverse frequency shift $\lambda$ is found from an eigen-equation using the parametrization in Eq.~(\ref{PARA1}). For the full calculation, we refer the reader to Refs.~\cite{Zhang2020, ZhangThesis}. The end result of the calculations is
\begin{eqnarray}
  \lambda {\delta}S_{+}(\omega) &= \xi^{2}_{\rm D}[(6q_{2}^{2}+q_{1}^{2} + 1)\partial_{y}\partial_{y} + (3q_{1}^{2}+2q_{2}^{2} + 1)\partial_{x}\partial_{x}]\delta{S_{+}}(\omega)\cr
&-(2\xi^{2}_{\rm D}iV - U)\delta{S_{+}}(\omega)
\label{NMREigenEquationDimensonless}
\end{eqnarray}
with
\begin{eqnarray}
V &=  (1+3{q_{1}^{2}}\cos2\theta){\partial_{x}}\theta{\partial_{x}} + (1+q_{1}^{2}){\partial_{y}}\theta{\partial_{y}} \cr
  &+\frac{1}{2 \xi_{\rm D}^{2}} [(1 + q_{1})^{2}\sin2{\theta} - (1 + q_{1})q_{2}\cos{\theta}], \cr
U &= (1+q_{1})[-(1+q_{1})\cos2\theta-5q_{2}\sin\theta]+1+q_{1}^{2}+4q_{2}^{2},
\end{eqnarray}
and
\begin{equation}
\lambda=\frac{(\omega^{2}-\omega_{\rm L}^{2})}{\Omega_{\rm PdB}^{2}},\,
\Omega^{2}_{\rm PdB}=\left(\frac{6\gamma^{2}{\Delta_{\rm P}^{2}}g_{\rm D}}{5\chi_{\bot}}\right)\,,
\end{equation}
where $\Omega_{\rm PdB}$ is the Leggett frequency in the PdB phase. The Eigen-equation (\ref{NMREigenEquationDimensonless}) was solved through Galerkin Eigen-value method with a finite element mesh \cite{Zhang2020, ZhangThesis,ProjectPage,BrennerBook2008}. The Eigenvalues $\lambda$ resulting from stationary configuration of spin solitons are plotted in Figs.~\ref{NMRLambdaAndRatioIntensityInseparableSoliton} and \ref{NMRLambdaAndRatioIntensitySeparableSoliton}. The relative intensity of the soliton satellite, determined as the ratio of the area of the satellite $I_{\rm sat}$ to the total area $I_{\rm tot}$ under the absorption spectrum can be calculated using found eigenfunctions $\delta S_+$ as
\begin{equation}
\frac{I_{\rm sat}}{I_{\rm tot}} = n\,
\frac{\left|\int \delta S_+ \,{\rm d}V\right|^2}{\int |\delta S_+|^2 \,{\rm d}V}\,,
\end{equation}
where $n = D^{-2}/2$ is the density of the solitons.
The numerical results for $\lambda$ agree well with experimental observations, discussed in more detail in Sec.~\ref{PdBSolitonNMR}, while $I_{\rm sat}/I_{\rm tot}$ scales proportionally to $\sqrt{\Omega}$, also similar to the experiment.

\section{Experimental observations} \label{sec:experiments}

In many experiments, vortex-bound solitons result in observable signatures in the NMR spectrum, playing an integral role in distinguishing various topological and composite defects in superfluid phases of $^3$He. This section aims to provide a review on the experimental observations of such defects to date, underlining the role of solitons that led to these observations.

\subsection{Identification of the spin-mass vortex in $^3$He-B}

Spin-mass vortices are formed in $^3$He-B as a rare event in two different processes: When the front of the transition between A and B phases of superfluid $^3$He sweeps through the rotating sample \cite{Kondo1992} or by the Kibble-Zurek mechanism \cite{Eltsov2000}. Identification of the spin-mass vortex is based on the NMR spectrum.

\begin{figure}[b]
\centerline{\includegraphics[width=0.6\linewidth]{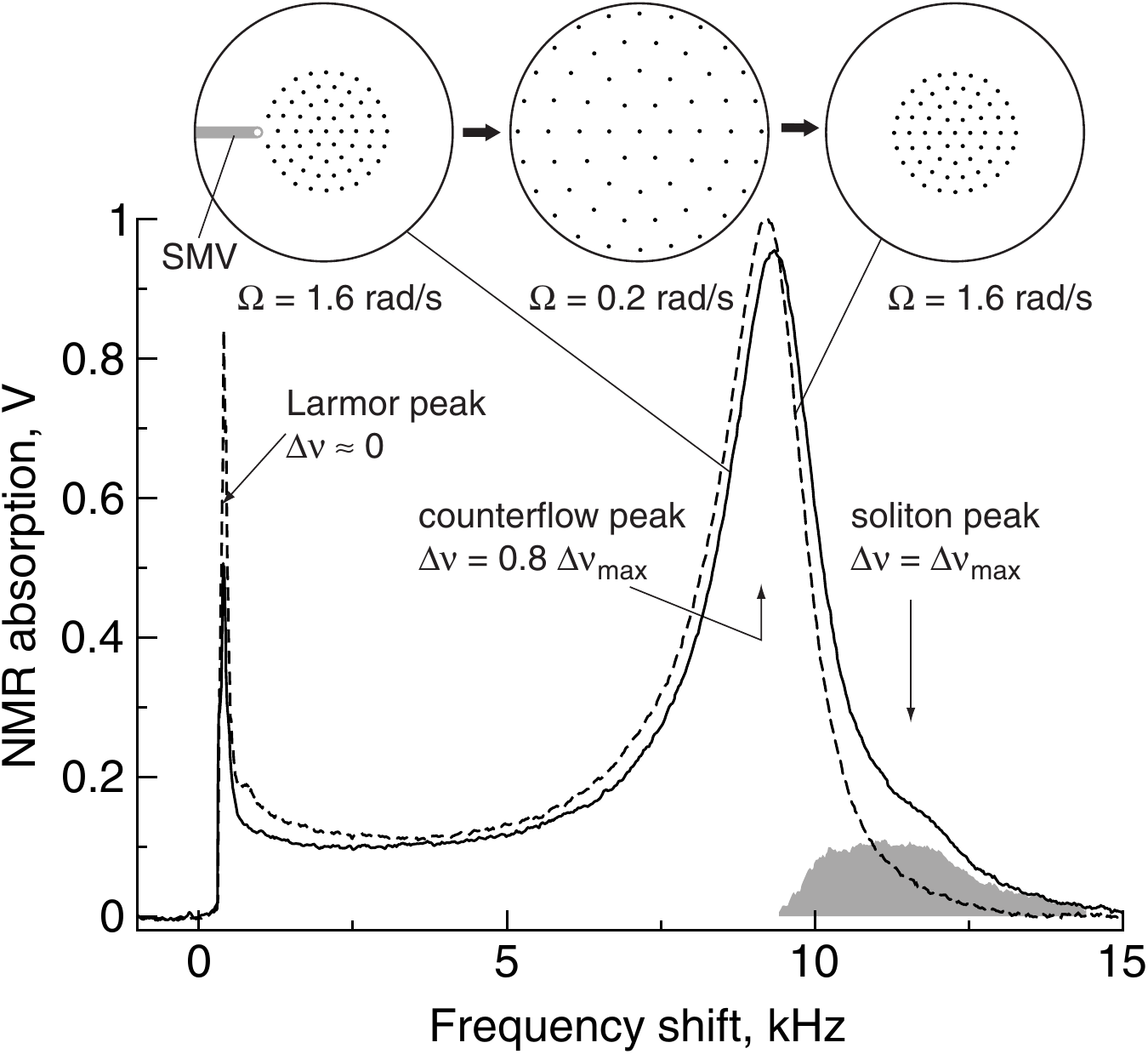}}
\caption{{\bf Observation of the spin-mass vortex in $^{\bf 3}$He-B.} The spin-mass vortex can be identified by comparing two NMR spectra \cite{Eltsov2000}. The first spectrum contains the signal from a spin-mass vortex created by neutron irradiation, and the second spectrum was measured after selectively removing the spin-mass vortex as illustrated at the top of the figure. The contribution of the spin-mass vortex (gray area) is centered around $\Delta \nu_{\rm max}$, distinguishable from the larger feature centered at $0.8 \Delta \nu_{\rm max}$, resulting from the mismatch of the normal fluid and superfluid velocities, i.e. from counterflow.}
\label{SMV_spec}
\end{figure}

Owing to spin-orbit interaction the NMR absorption in $^3$He-B is shifted from the Larmor frequency by an amount $\Delta\nu$ which depends on the local orientation of the anisotropy axis $\hat{\mathbf{n}}$ with respect to the applied magnetic field $\mathbf H$. In the experiment in Fig.~\ref{SMV_spec}, the field is oriented along the rotation axis $\mathbf{\Omega}$, which is also the symmetry axis of the sample cylinder. At the soliton sheet $\hat{\mathbf{n}} \perp \mathbf{H}$ (Fig.~\ref{SMV_struct}), which produces the maximum possible frequency shift $\Delta\nu_{\rm max}$. Everywhere else in the rotating cylinder the angle between $\hat{\mathbf{n}}$ and $\mathbf{H}$ (and thus the frequency shift) is smaller: The cluster of vortex lines in the center of the container gives rise to absorption with frequency shifts close to zero. The annular region with vortex-free counterflow between normal and superfluid  component around the cluster is responsible for the large absorption maximum at $0.8\,\Delta \nu_{\rm max}$. The height of this peak decreases with the decreas of the counterflow velocity, that is, when new vortex lines are added to the cluster or if rotation velocity decreased with given number of vortices. The spectrum drawn with the solid line was measured after the neutron irradiation and shows the absorption from the soliton sheet, centered around $\Delta\nu_{\rm max}$. A top view of the rotating cylinder with the vortex cluster is depicted on the top row of the figure. This illustrates how the SMV can be selectively removed by reducing $\Omega$ to where the cluster has almost expanded to the wall (at 0.2 rad/s) and the SMV as the outermost vortex has been pushed to the cylinder wall. After increasing $\Omega$ back to the original 1.6 rad/s the spectrum plotted with the dashed line was recorded. The difference from the original spectrum is the absence of the soliton signal (shown by the gray area).

%% Start new additions

\subsection{HQVs in the polar phase}

In the polar phase, the minimum energy configuration of the order parameter forms the main peak in the NMR spectrum at the frequency \cite{Mineev2016}
\begin{equation}
\Delta \omega_{\mathrm{P}}= \omega_{\mathrm{P}}  - \omega_{\mathrm L}  \approx
\frac{\Omega_{\mathrm{P}}^2}{2\omega_{\mathrm{L}}}  \cos^2\mu  .
\label{SatelliteGeneral1}
\end{equation}
Here $\Omega_{\mathrm{P}}$ is the Leggett frequency in the polar phase, which characterizes the spin-orbit torque. 

\begin{figure}[t]
\centerline{\includegraphics[width=\linewidth]{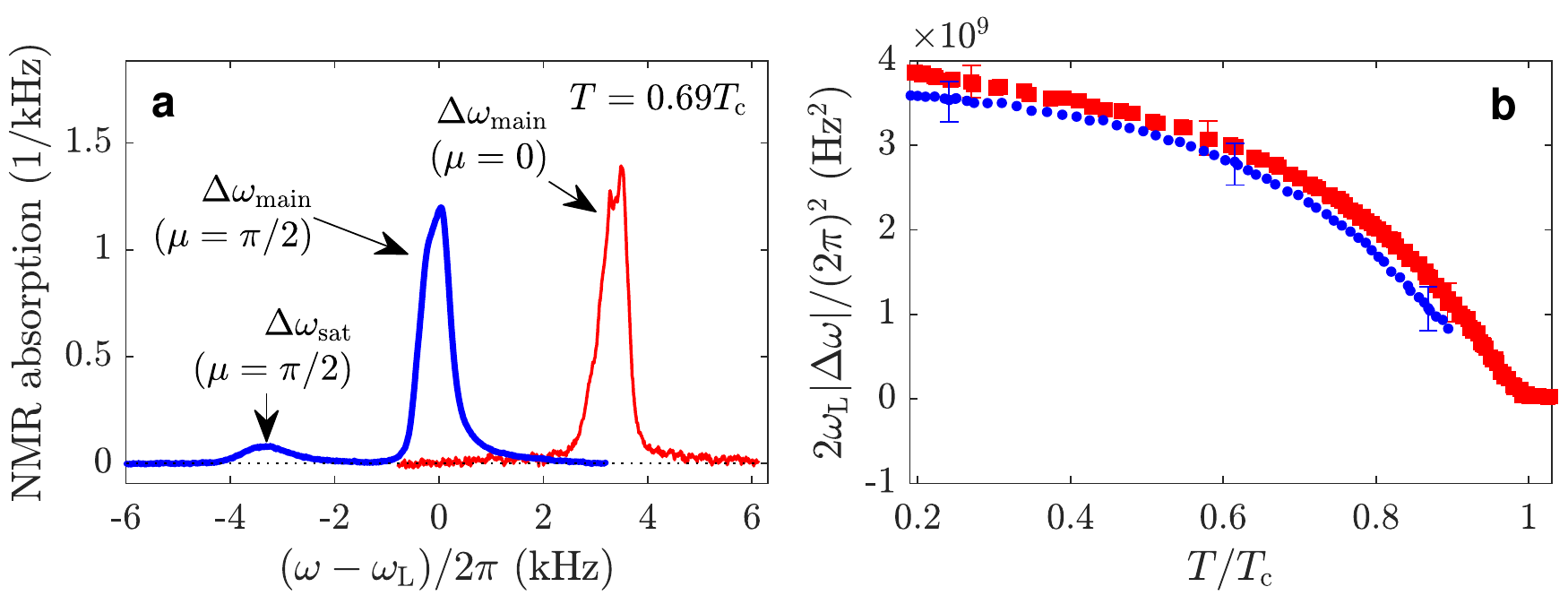}}
\caption{\label{SatelliteT} \textbf{NMR spectra in the polar phase in the presence of HQVs.} (\textbf{a}) Normalized spectra measured in transverse field $\mu=\pi/2$ (blue thick line) shows the HQV satellite at the negative frequency shift $\Delta\omega_{\mathrm{Psat}}$ and the main line at zero frequency shift \cite{Autti2016}. In the axial field $\mu=0$ (red thin line) only the main line at positive shift $\Delta\omega_{\mathrm P}$ is seen. This spectrum is not sensitive to presence of vortices as the spin-orbit interaction results in no trapping potential for spin waves. (\textbf{b}) Temperature dependencies of the satellite position in the transverse field $|\Delta\omega_{\mathrm{Psat}}(\mu=\pi/2)|$ (blue circles) and the main line position in the axial field $\Delta\omega_{\mathrm P}(\mu=0)$ (red squares) closely match as expected for HQVs. The error bars show full width at half maximum of the main line as an estimate of possible systematic error.}
\end{figure}

\begin{figure}[tb!]
\centering
\includegraphics[width=0.9 \linewidth]{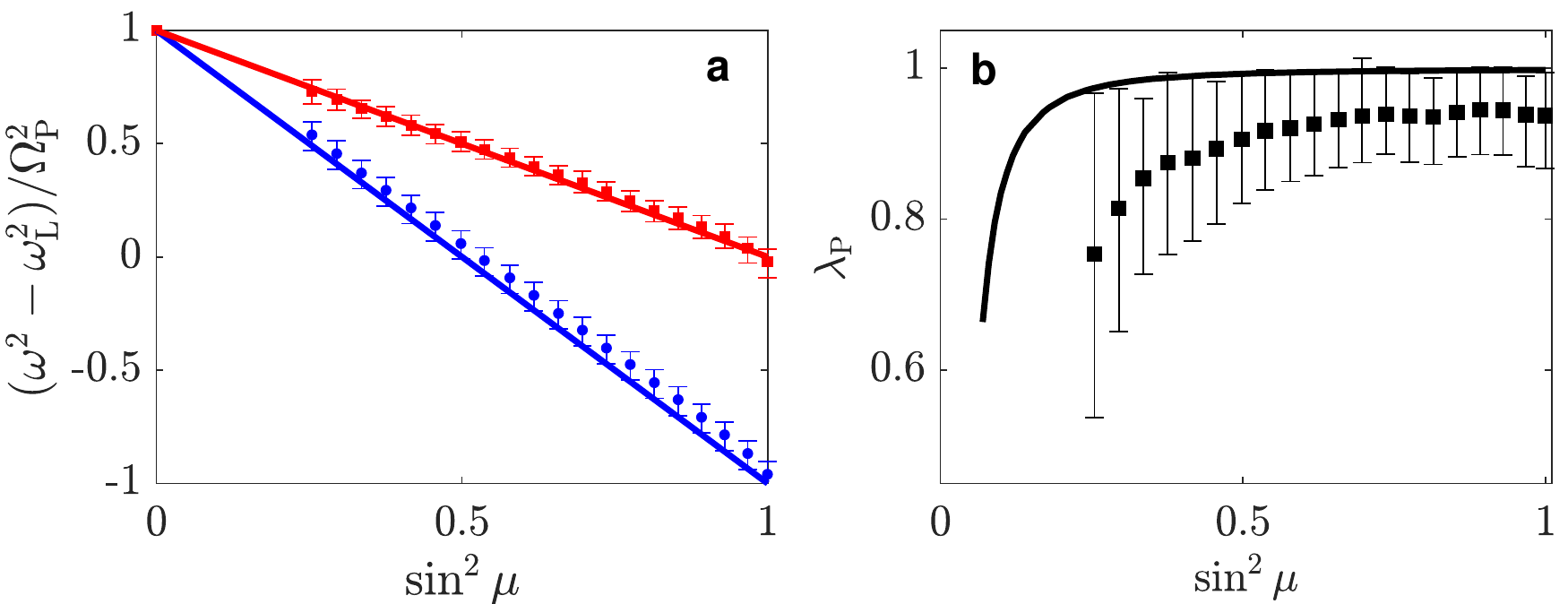}
\caption{\label{SatelliteTilt} \textbf{Frequency shift of the HQV satellite.} (\textbf{a}) Measured values of the dimensionless frequency shift $\lambda_{\mathrm{P}}$ as a function of the field tilt angle $\mu$ (symbols) are compared with numerical calculations for the uniform polar phase (solid line) using theoretical value of $\xi_{\mathrm{D}}=17 \mu$m \cite{Autti2016}. Leggett frequency $\Omega_{\mathrm{P}}$ is determined from a separate measurement at $\mu=0$. Deviation from the infinitely-long $\hat{\mathbf{d}}$ soliton value $\lambda_{\mathrm{P}} = -1$ increases towards small $\mu$. The disagreement between the experiment and calculations likely originates from disorder in the nafen strand orientation \cite{Asadchikov2015}, which leads to fluctuations of the spin-orbit interaction energy within the solitons. (\textbf{b}) Values of $\lambda_{\mathrm{P}}$ are found from positions of the HQV satellite $\Delta\omega_{\mathrm{Psat}}$ (blue circles) and of the main line $\Delta\omega_{\mathrm P}$ (red squares). The red and blue solid lines show results of Eqs.~(\ref{SatelliteGeneral1}) and (\ref{SatelliteGeneral2}), respectively, for $\lambda_{\mathrm{P}} = -1$. The bars show full width at half maximum of the spectral lines in both panels.}
\end{figure}

Winding of the $\hat{\mathbf{d}}$ vector, e.g. in the form of a soliton, provides an additional potential energy term for spin waves as the spin-orbit energy is not at minimum. Excitation of standing spin waves within these potential wells leads to a satellite peak in the NMR spectrum at frequency
\begin{equation}
\Delta \omega_{\mathrm{Psat}}= \omega_{\mathrm{Psat}}  - \omega_{\mathrm L}  \approx 
\frac{\Omega_{\mathrm{P}}^2}{2\omega_{\mathrm{L}}} \left( \cos^2 \mu + \lambda_{\mathrm{P}} \sin^2\mu \right)  \,,
\label{SatelliteGeneral2}
\end{equation}
where the parameter $\lambda_{\mathrm{P}}(\mu)$ is specific to the type of the topological object. For example, an infinite planar $\hat{\mathbf{d}}$ soliton gives $\lambda_{\mathrm{P}} = -1$ for the zero mode on the soliton \cite{vollhardt2013superfluid,Mineev2016,SlavaPolar}. The frequency shift for the satellite in this case becomes equal, but opposite than for the main peak, i.e. $\Delta \omega_{\mathrm{Psat}} (\mu=\pi/2) = -\Delta \omega_{\mathrm{P}}(\mu=0)$. In reality, the finite soliton length (similar to Fig.~\ref{NMRLambdaAndRatioIntensitySeparableSoliton}c) and disorder in the nafen confinement may lead to reduced shift, and $|\lambda_{\mathrm{P}}| < 1$ is expected \cite{Asadchikov2015}. In the experiments \cite{Autti2016}, a controlled amount of polarized HQVs was created by slowly cooling the sample in rotation with constant angular velocity $\Omega$ from above $T_{\mathrm c}$ in zero or axial (along $\hat{\bm{\Omega}} \parallel \hat{\bf m}$) magnetic field. The presence of HQVs is apparent from the NMR spectrum via the related spin soliton peak, Fig.~\ref{SatelliteT}a. The experimental results, summarized in Figs.~\ref{SatelliteT} and \ref{SatelliteTilt} indeed yield $\lambda_{\mathrm{P}}(\mu=\pi/2) = -0.93\pm 0.07 $, in good agreement with theoretical expectations. We note that the predicted spin polarization of the HQV core \cite{PhysRevLett.103.057003} does not affect the signal, as the winding of the $\hat{\mathbf{d}}$ vector (and thus the spin polarization) is always opposite for a pair of HQVs connected by the spin soliton.

%%%%%%%%%%%%%%%%%%%%%%%%%%%%%%%%%%%%%%%%%
%%%%%%%%%%%%%%%%%%%%%%%%%%%%%%%%%%%%%%%%%
%%%%%%%%%%%%%%%%%%%%%%%%%%%%%%%%%%%%%%%%%
%%%%%%%%%%%%%%%%%%%%%%%%%%%%%%%%%%%%%%%%%

In the polar phase HQVs are energetically preferable to SQVs in axial or zero magnetic field \cite{PhysRevB.98.094524,Mineev2014}. Application of tilted magnetic field changes the situation via the spin-orbit interaction related to winding of the $\hat{\bf d}$ vector within the HQV-bound spin solitons, making SQVs preferable to HQVs. In addition, SCVs can be created during the cooldown if strong time-dependent magnetic field is applied to generate a random distribution of vector $\hat{\mathbf d}$ \cite{Dmitriev2010,spinglass}.

\begin{figure}[tb!]
\centerline{\includegraphics[width=0.9\linewidth]{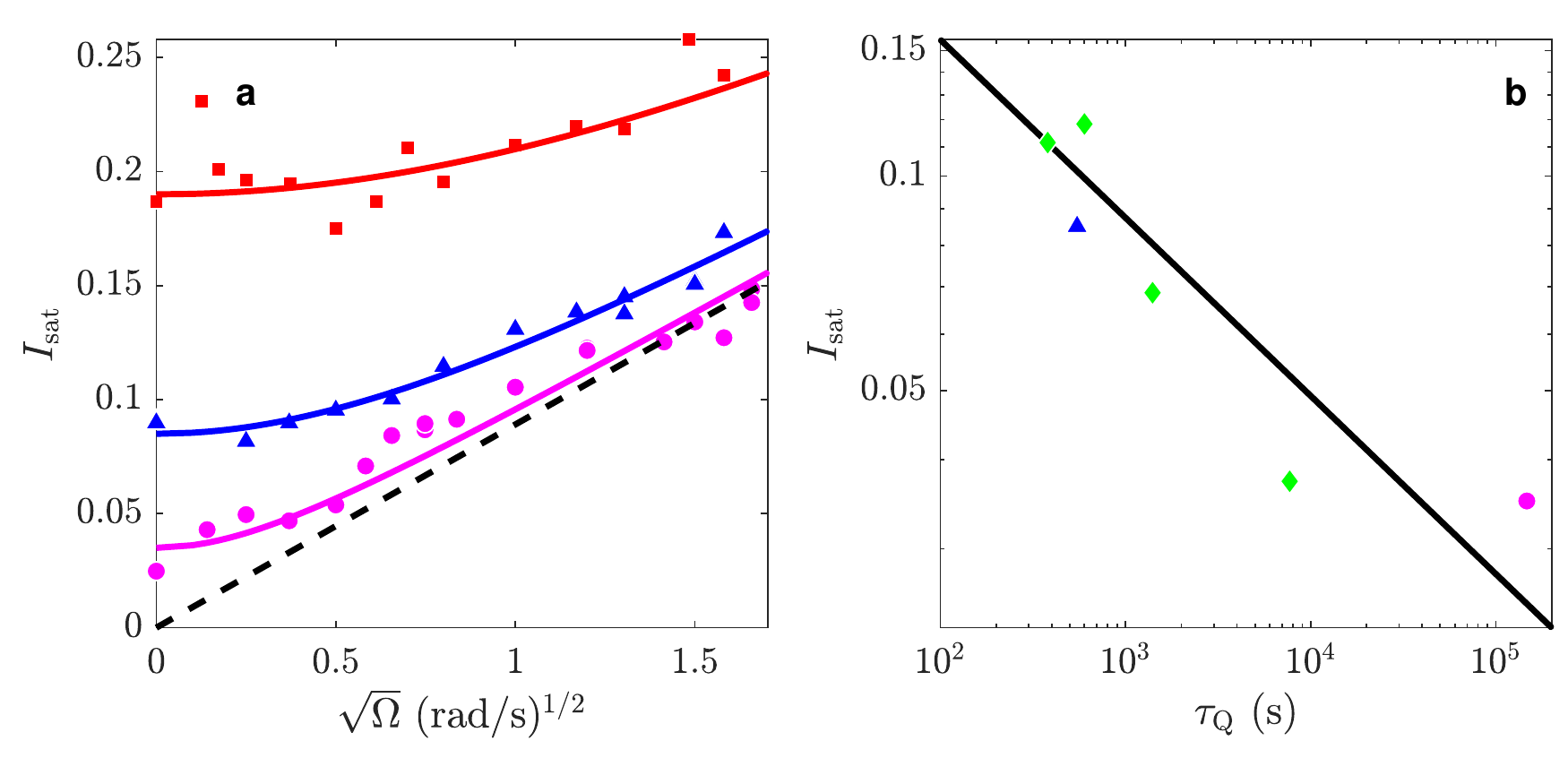}}
\caption{\label{intensityOmega_KZ} \textbf{Intensity of the HQV satellite.} \textbf{a} The satellite intensity $I_{\rm sat}$ measured in slow ($\tau_{\rm Q} \approx 1.5 \cdot 10^5\,$s, magenta circles) and fast ($\tau_{\rm Q} \approx 5.5 \cdot 10^2\,$s, blue triangles) zero-field cooldowns as a function of $\Omega$. The solid lines are theoretical fits assuming that HQV creation by the KZM is independent from rotation. The dash line shows fitted equilibrium $I_{\rm sat}$, corresponding to vanishing HQV density from KZM. Applying rf drive at the resonance during the cooldown in the axial field creates SCVs (red squares), seen as extra rotation-independent satellite intensity. \textbf{b} The satellite intensity $I_{\rm sat}$ measured in the absence of rotation and bias fields (green diamonds) follows the KZM power law $I_{\rm sat} \propto \tau_{\rm Q}^{-1/4}$ (solid line). The fitted satellite intensities at $\Omega = 0$ from panel \textbf{a} are marked with the corresponding symbols and colors.}
\end{figure}

HQVs can be distinguished from SCVs by their connection to rotation. In particular, the dependence of the relative satellite peak intensity $I_{\mathrm{sat}}$ on the angular velocity $\Omega$, shown in Fig.~\ref{intensityOmega_KZ}a, is expected to follow $\propto \sqrt{\Omega}$. This dependence follows from the following considerations; for solitons with their width set by the dipole length $\xi_{\mathrm{D}}$, the expected signal intensity is  $I_{\mathrm{sat}} = (n_{\mathrm v}/2)\, g_{\mathrm s} L \xi_{\mathrm{D}}$ \cite{HuMaki1987}. Here $L = b n_{\mathrm v}^{-1/2}$ is the average soliton length set by the average inter-vortex distance, while $g_{\mathrm s}\sim 1$ and $b\sim1$ are numerical factors. For a very low vortex density and long solitons $L \to \infty$ one has $g_{\mathrm s} \to 2$. Since the vortex density $n_{\rm v} \propto \Omega$, one expects $I_{\mathrm{sat}} \propto \Omega^{1/2}$, as indeed confirmed in Fig.~\ref{intensityOmega_KZ}a.

%%%%%%%%%%%%%%%%%%%%%%%%%%%%%%%%%%%%%%%%%
%%%%%%%%%%%%%%%%%%%%%%%%%%%%%%%%%%%%%%%%%
%%%%%%%%%%%%%%%%%%%%%%%%%%%%%%%%%%%%%%%%%
%%%%%%%%%%%%%%%%%%%%%%%%%%%%%%%%%%%%%%%%%

We note that the observation of solitons bounded by HQVs relies on a crucial experimental detail. Namely, the tension related to the energy cost of the solitons is overcome by the pinning of HQVs by the nafen strands. In fact, the pinning force is stronger than any relevant energy scale in the system. Each HQV core with a characteristic size given by the coherence length $\xi \sim 40\,$nm, is penetrated by a few nafen strands of $\sim 10\,$nm diameter, leading to reduced energy cost of the HQV core thus leading to effective pinning of HQVs in place as they are created. For example, after stopping the rotation the satellite in the NMR spectrum remained unchanged for weeks, while the Magnus force, pulling vortices towards  the sample boundary, exceeds the soliton tension by a large factor $10^3$. It is worth pointing out that the experiments shown in Fig.~\ref{intensityOmega_KZ} were performed with a stationary cryostat, after rotating it with constant angular velocity during the superfluid transition. Due to the strong pinning of HQVs by the nafen strands \cite{BFS}, the relation $I_{\mathrm{sat}} \propto \Omega^{1/2}$ holds for pinned vortices, as long as one takes $\Omega$ at the time of the superfluid transition.

%%%%%%%%%%%%%%%%%%%%%%%%%%%%%%%%%%%%%%%%%
%%%%%%%%%%%%%%%%%%%%%%%%%%%%%%%%%%%%%%%%%
%%%%%%%%%%%%%%%%%%%%%%%%%%%%%%%%%%%%%%%%%
%%%%%%%%%%%%%%%%%%%%%%%%%%%%%%%%%%%%%%%%%

\begin{figure}[tb!]
\centerline{\includegraphics[width=0.9\linewidth]{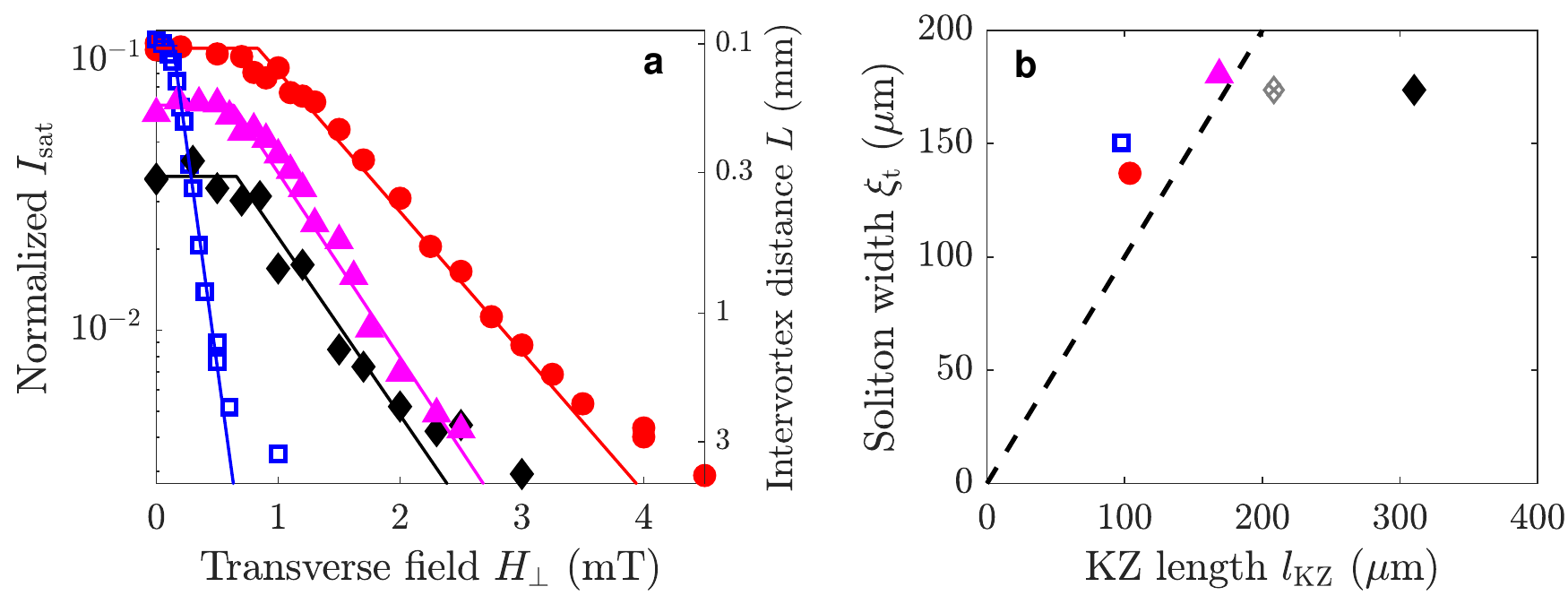}}
\caption{\label{SuppressedKZ} \textbf{Suppression of the HQV density created by the KZM under a symmetry-breaking bias.} \textbf{a} Filled red circles, magenta triangles, and black diamonds correspond to quench rates of $\tau_{\rm Q} \approx 3.8 \cdot 10^2\,$s, $\tau_{\rm Q} \approx 1.4 \cdot 10^3\,$s, and $\tau_{\rm Q} \approx 7.7 \cdot 10^3\,$s, respectively, while applying a constant $H = 11 \,$mT magnetic field~\cite{KZSupp}. The field is rotated to achieve different bias fields $H_\perp = H \sin \mu$. Open blue squares ($\tau_{\rm Q} \approx 6.0 \cdot 10^2\,$s) correspond to measurements with zero axial field component, $H_\perp = H$. Vortex density is constant for $H_\perp < H_{\perp t}$ and suppressed for higher bias fields. The suppression starts when the characteristic length scale of the bias field $\xi_{\rm bias} (H_\perp )$ becomes smaller than the relevant Kibble-Zurek length. Solid lines correspond to theoretical model (see text for details). The dashed line shows where the inter-vortex distance becomes comparable with the container size. \textbf{b} The extracted threshold bias length $\xi_{\rm t}$ as a function of $l_{\rm KZ}$ with the same symbols. The dashed line corresponds to $\xi_{\rm t} = l_{\rm KZ}$. The patterned gray diamond is the same measurement as the black diamond, but with $l_{\rm KZ}$ on the horizontal axis replaced with an estimation of the transition front thickness $l_{\rm F}$ \cite{KibbleVolovik}. For other measurements, $l_{\rm F}$ lies beyond the right border of the plot.}
\end{figure}

In the rotating experiments with the magnetic field oriented transverse to the nafen strands during cooldown \cite{Autti2016}, the NMR satellite peak related to HQV-bound solitons was absent -- consistent with the absence of HQVs. However, since the angular velocity is conserved in the superfluid phase transition, this implies that SQVs were created. The effect of the applied magnetic field on HQV creation was further studied in stationary (non-rotating) measurements in Ref.~\cite{KZSupp}, where HQVs were created by temperature quenches via the Kibble-Zurek mechanism. The conclusion was that the HQV density is suppressed at the normal-polar phase transition when the soliton width, controlled by the magnetic field, becomes smaller than the Kibble-Zurek length $l_{\rm KZ}$, c.f. Fig.~\ref{SuppressedKZ} and Eq.~(\ref{eq:lKZ}). One way to understand this observation is by assuming that the applied magnetic field then fixes the $\hat {\bf d}$ vector on a characteristic length scale already during the phase transition. If the size of this length scale is smaller than $l_{\rm KZ}$, the $\hat{\bf d}$ vector orientation becomes correlated at length scales exceeding $l_{\rm KZ}$ -- in contrast to the original idea assuming a random realization of the order parameter, i.e. lack of correlation, at such length scales. Therefore, the very same properties that define the characteristic length scales of topological solitons also play an integral role in formation of topological defects during phase transitions.

Additionally, one might ask what is the interplay of the Kibble-Zurek mechanism (KZM) \cite{Kibble1976,Zurek1985} and rotation. The KZM is expected to create various order-parameter defects, including vortices of all possible types \cite{KZ_nature,KZ_nature2,PhysRevB.90.024508,ProgLowTempPhys_page9}. In the transition, the inter-vortex distance is set by the KZ length
\begin{equation} \label{eq:lKZ}
 l_{\mathrm KZ} = \xi_0 (\tau_{\mathrm{Q}}/\tau_0)^{1/4},
\end{equation}
where $\tau_{\mathrm{Q}}^{-1}=\left. -\frac{\mathrm{d}(T/T_{\mathrm c})}{\mathrm{d} t} \right|_{T=T_{\mathrm c}}$ is the cooldown rate at $T_{\mathrm c}$, $\xi_0 = \xi(T=0)$ and the order-parameter relaxation time $\tau_0 \sim 1\,$ns. For HQVs the inter-vortex distance sets the length of the interconnecting solitons and thus the amplitude of the satellite signal. The resulting dependence $I_{\mathrm{sat}}\propto n_{\mathrm{v}} l_{\mathrm{KZ}} \propto l_{\mathrm{KZ}}^{-1} \propto \tau_{\mathrm{Q}}^{-1/4}$ is indeed observed in the experiment, Fig.~\ref{intensityOmega_KZ}b. The magnitude of the signal corresponds to the averaged soliton length of $1.4~ l_{\mathrm{KZ}}$, as has been estimated also in the B phase of ${}^3$He \cite{ProgLowTempPhys_page9,Bauerle1998}. The shift of experimental data in Fig.~\ref{intensityOmega_KZ}a above the theoretical expectation indicates that the KZM is important also in cooldowns with applied rotation.

\begin{figure}[tb]
\centerline{\includegraphics[width=\linewidth]{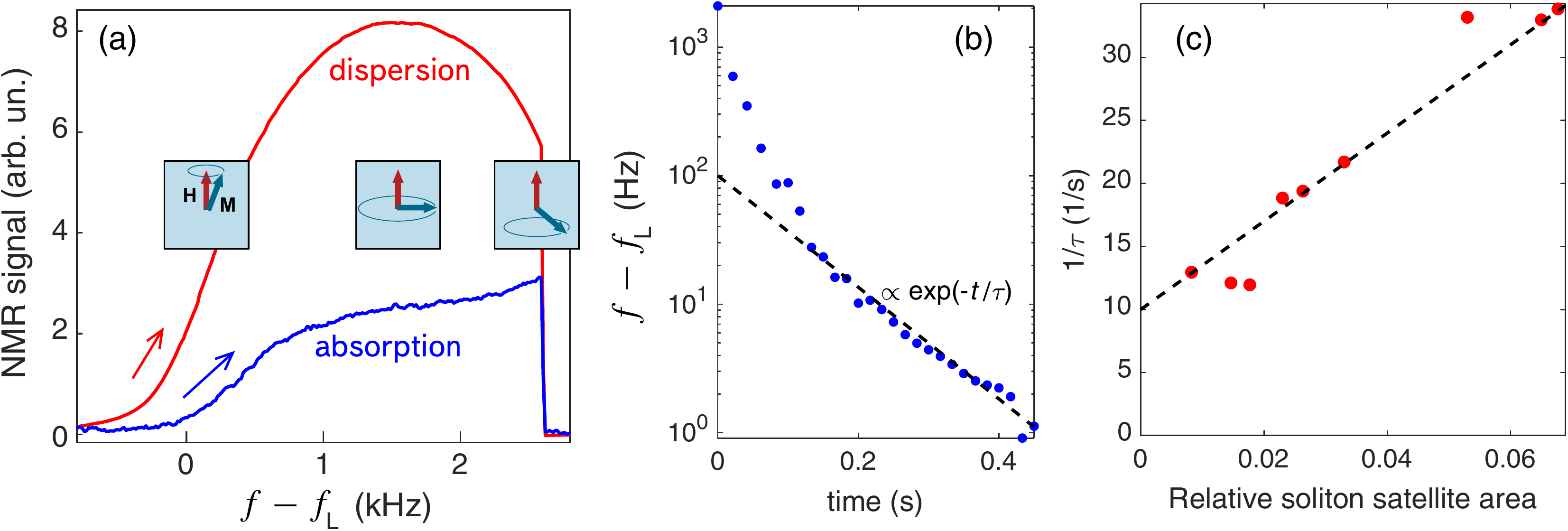}}
\caption{\textbf{Detection of vorex-bound solitons using magnon BEC.} (a) In continuous-wave NMR, coherently precessing magnetization state is formed with the sweep of the frequency of the pumping field $f$ of sufficient magnitude via Larmor frequency $f_{\rm L}$ in the upward direction \cite{BFS}. It is manifested by a characteristic non-linear response. (b) When external pumping is switched off, the pumped magnons decay, but magnetization of the sample continues to precess coherently while the precession frequency $f$ returns to $f_{\rm L}$ in exponential decay with time constant $\tau$. (c) When solitons are present in the sample, the relaxation rate $\tau^{-1}$ increases proportionally to the volume occupied by the solitons, which is in the plot is characterized by the intensity of the soliton satellite in the NMR spectrum measured independently. The measurements were performed with the nafen-243 sample at $P=7\,$bar and $T=0.4\,T_{\rm c}$, where the polar phase is the equilibrium phase.
}

\label{relaxmeas}
\end{figure}

Alternative method of observation of the solitons bounded by half-quantum vortices in the polar phase utilizes coherent magnetization precession state forming Bose-Einstein condensate of magnon quasiparticles, Fig.~\ref{relaxmeas}. The relaxation rate of the magnon condensate increases proportionally to the volume occupied by the solitons. This method is especially useful with small amounts of solitons present in the sample, as accurate measurement of the satellite intensity in cw NMR might require hours of averaging while the relaxation rate can be measured with sufficient precision in seconds. At larger soliton densities, though, the relaxation rate becomes too fast to be reliably measured.

%%%%%%%%%%%%%%%%%%%%%%%%%%%%%%%%%%%%%%%%%
%%%%%%%%%%%%%%%%%%%%%%%%%%%%%%%%%%%%%%%%%
%%%%%%%%%%%%%%%%%%%%%%%%%%%%%%%%%%%%%%%%%
%%%%%%%%%%%%%%%%%%%%%%%%%%%%%%%%%%%%%%%%%

\subsection{HQVs in the PdA phase}

The transverse resonance frequency of the bulk fluid in the PdA phase with magnetic field in the direction parallel to the strand orientation, i.e. for $\mu=0$, is \cite{PhysRevLett.115.165304}
\begin{equation} \label{eq:distAmain}
 \Delta \omega_{\mathrm{PdA}} = \omega_{\mathrm{PdA}}-\omega_{\mathrm{L}} \approx \frac{\Omega_{\mathrm{PdA}}^{2}}{2 \omega_{\mathrm{L}}},
\end{equation}
where $\Omega_{\mathrm{PdA}}$ is the frequency of the longitudinal resonance in the PdA phase. The NMR line retains its shape during the second order phase transition from the polar phase but renormalizes the longitudinal resonance frequency due to appearance of the order parameter component $b\,\hat{\mathbf{n}}$ in Eq.~(\ref{opPdA}).

Since the $\hat{\mathbf{m}}$ vector is fixed by nafen parallel to the anisotropy axis, the $\hat{\mathbf{l}}$ vector then lies on the plane perpendicular to it, prohibiting the formation of continuous vorticity \cite{PhysRevLett.36.594} like the double-quantum vortex in $^3$He-A \cite{doublequantum}. Some planar structures in the $\hat{\mathbf{l}}$-vector field, such as domain walls \cite{Kasai2018} or disclinations, remain possible but the effect of the $\hat{\mathbf{l}}$ texture on the trapping potential for spin waves is negligible due to the large polar distortion \cite{PhysRevLett.115.165304} (i.e. for $|b| \ll 1$).

%%%%%%%%%%%%%%%%%%%%%%%%%%%%%%%%%%%%%%%%%%%%%%%%%%%%%%%%%%%%%
%%%%%%%%%%%%%%%%%%%%%%%%%%%%%%%%%%%%%%%%%%%%%%%%%%%%%%%%%%%%%
\begin{figure}[t]
\centering
\includegraphics[width=\linewidth]{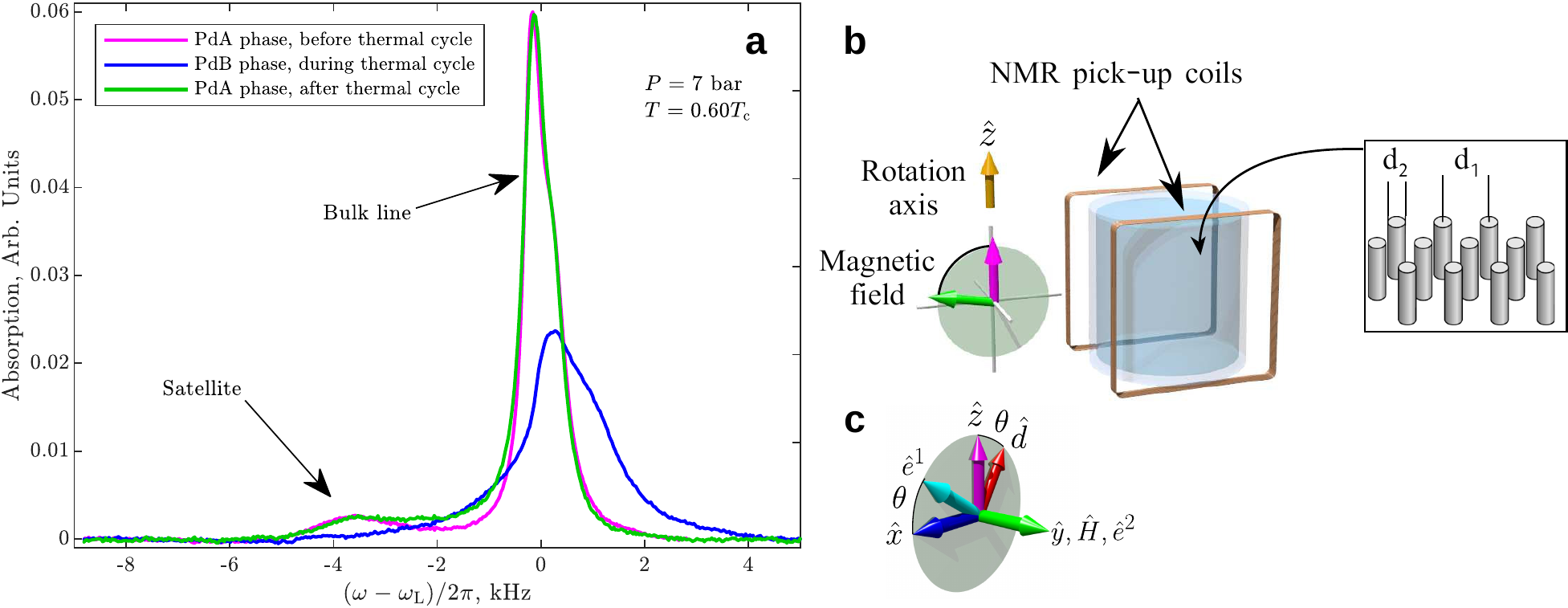}
 \caption{{\textbf{HQVs in thermal cycling.}} {\textbf a} The plot shows the measured NMR spectra in transverse ($\mu=\pi/2$) magnetic field in the presence of HQVs. HQVs were created by rotation with 2.5~rad/s during the transition from normal phase to the polar phase. The NMR spectrum includes the response of the bulk liquid and the $\hat{\mathbf{d}}$-solitons, which appear as a characteristic satellite peak at lower frequency. The satellite intensity in the PdA phase remains unchanged after thermal cycling. The NMR spectrum in the PdB phase at the same temperature, measured between the two measurements in the PdA phase, is shown for reference \cite{Makinen2019}. {\textbf b} The $^3$He sample is confined within a cylindrical container filled with nafen-90, which consists of nearly parallel Al$_2$O$_3$ strands with $d_2\approx 8\,$nm diameter, separated by $d_1 \approx 50\,$nm on average. The strands are oriented predominantly along the axis denoted $\hat {\bf z}$. The sample can be rotated with angular velocities up to 3$\,$rad$\,$s$^{-1}$ around the same axis. The sample is surrounded by rectangular nuclear magnetic resonance (NMR) pick-up coils. The static magnetic field transverse to the NMR coils can be oriented at an arbitrary angle $\mu$ with respect to the $\hat {\bf z}$ axis. {\textbf c} The magnetic field, oriented along the $y$-direction ($\mu = \pi/2$) in this figure, locks the $\hat {\bf e}^2$-vector in the polar-distorted B phase order parameter, Eq.~(\ref{eq:distBop}). Vectors $\hat {\bf d}$ and $\hat {\bf e}^1$ are free to rotate in the $xz$-plane by angle $\theta$.} 
 \label{dist_A_spectra_before_and_after}
\end{figure}
%%%%%%%%%%%%%%%%%%%%%%%%%%%%%%%%%%%%%%%%%%%%%%%%%%%%%%%%%%%%%
%%%%%%%%%%%%%%%%%%%%%%%%%%%%%%%%%%%%%%%%%%%%%%%%%%%%%%%%%%%%%

\begin{figure}%[htt]
\centering
 \includegraphics[width=0.9\linewidth]{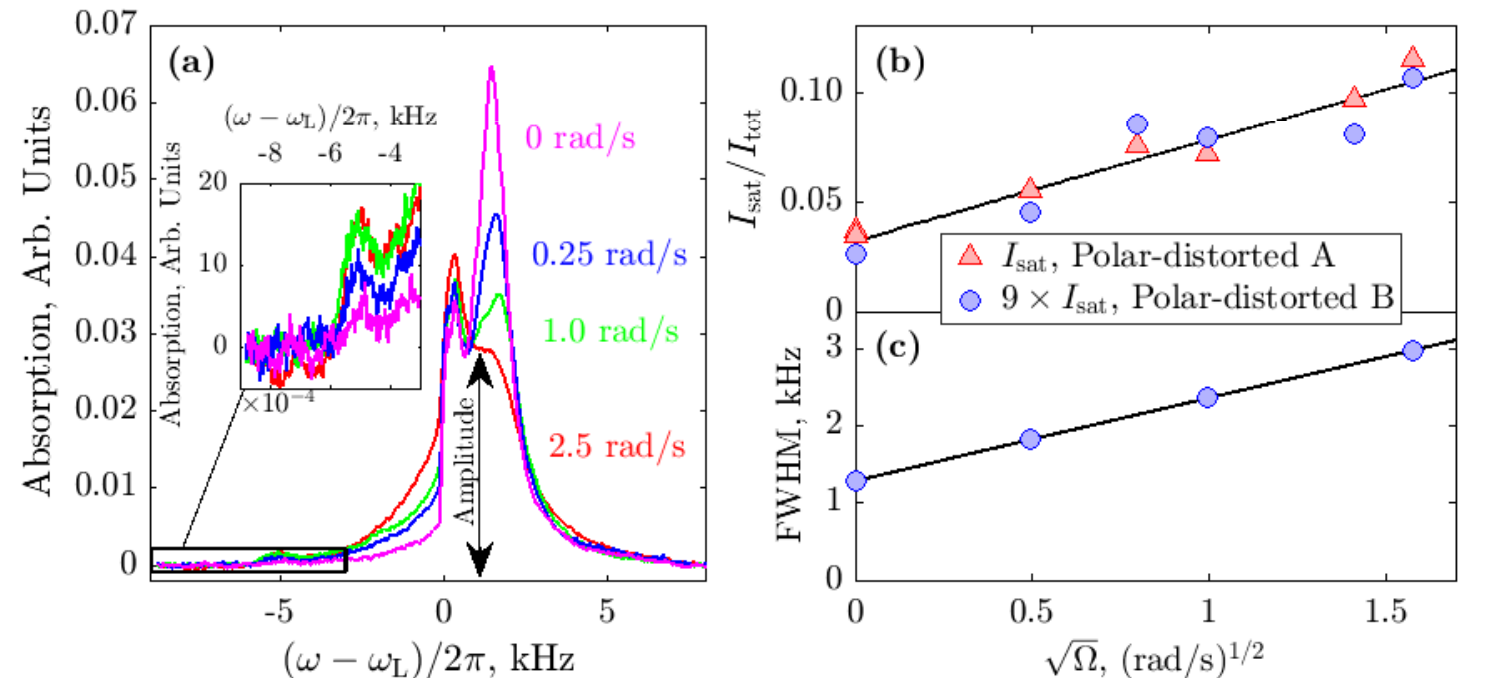}
\caption{ \textbf{NMR spectra in the PdB phase.} (a) The plot shows the measured NMR spectrum in the PdB phase at 0.38~$T_{\mathrm{c}}$ for different HQV densities, controlled by the angular velocity $\Omega$ at the time of crossing the $T_{\mathrm c}$ \cite{Makinen2019}. The presence of  KLS  walls produces characteristic features seen both as widening of the main line (located at small positive frequency shift) and as a satellite peak with a characteristic negative frequency shift. The inset shows magnified view of the satellite peak. (b) The satellite intensity in the PdA phase at $0.60 T_{\mathrm{c}}$ (blue circles) and in the PdB phase multiplied by a factor of 9 (red triangles) at $0.38 T_{\mathrm{c}}$ show the expected $\sqrt{\Omega}$-scaling. The solid black line is a linear fit to the measurements including data from both phases. The non-zero $\Omega=0$ intersection corresponds to vortices created by the Kibble-Zurek mechanism \cite{Autti2016,Kibble1976,Zurek1985}. (c) The FWHM of the main line, determined from the spectrum in panel (a), gives FWHM $\approx 3$~kHz for 2.5~rad/s. FWHM for other angular velocities is recalculated from the amplitude of the main NMR line, shown in panel (a), assuming constant area.} 
 \label{fig:distB_satellites}
\end{figure}

In the presence of HQVs the excitation of standing spin waves localized on the soliton leads to a characteristic NMR satellite peak in transverse ($\mu=\pi/2$) magnetic field, c.f. Fig.~\ref{dist_A_spectra_before_and_after}, with frequency shift
\begin{equation} \label{eq:lambda}
 \Delta \omega_{\mathrm{PdAsat}} = \omega_{\mathrm{PdAsat}} - \omega_{\mathrm{L}} \approx \lambda_{\mathrm{PdA}} \frac{\Omega_{\mathrm{PdA}}^{2}}{2 \omega_{\mathrm{L}}},
\end{equation}
where $\lambda_{\mathrm{PdA}}$ is a dimensionless parameter dependent on the spatial profile (texture) of the order parameter across the soliton. For an infinite $\hat{\mathbf{d}}$-soliton, one has $\lambda_{\mathrm{PdA}} = -1$, corresponding to the zero-mode of the soliton, as in the polar phase \cite{Mineev2016,SlavaPolar,HuMaki1987,RevModPhys.59.533}. The measurements in the supercooled PdA phase, Fig.~\ref{dist_A_spectra_before_and_after}, at temperatures close to the transition to the PdB phase give value $\lambda_{\mathrm{PdA}} \approx -0.9$, which is in good agreement with theoretical predictions and the polar phase measurements with the 243 mg/cm$^3$ nafen sample. This confirms that the structure of the $\hat{\mathbf{d}}$-solitons connecting the HQVs is similar in polar and PdA phases and the effect of the orbital part to the trapping potential can be neglected. Furthermore, the satellite intensity shown in Fig~\ref{fig:distB_satellites}~(b) scales with $\sqrt{\Omega}$ as in the polar phase, indicating that the observed signal is linked to topological solitons bounded by HQVs also in the PdA phase. 

%%%%%%%%%%%%%%%%%%%%%%%%%%%%%%%%%%%%%%%%%%%%%%%%%%%%%%%%%%%%%
%%%%%%%%%%%%%%%%%%%%%%%%%%%%%%%%%%%%%%%%%%%%%%%%%%%%%%%%%%%%%
%%%%%%%%%%%%%%%%%%%%%%%%%%%%%%%%%%%%%%%%%%%%%%%%%%%%%%%%%%%%%
%%%%%%%%%%%%%%%%%%%%%%%%%%%%%%%%%%%%%%%%%%%%%%%%%%%%%%%%%%%%%

\subsection{Walls bounded by strings in the PdB phase}
\label{PdBSolitonNMR}

As established in Sec.~\ref{SolitonNexusPdB}, isolated HQVs cease to be protected by topology in the PdB phase as its order parameter lacks the relevant $\mathbb{Z}_{2 (\phi+\mathbf{S})}$ symmetry. The experimental data, Fig.~\ref{dist_A_spectra_before_and_after}, strongly suggests that HQVs survive the phase transition to the PdB phase as composite defects, walls bounded by strings (or KLS walls), see Sec.~\ref{ExactSequence}. Let us now take a closer look at the experiments that led to this conclusion.

For a magnetic field oriented transverse to the uniaxial nafen anisotropy axis $\hat{\mathbf{z}}$, the order parameter of the PdB phase is given by Eq.~(\ref{eq:distBop}). We denote with $\theta$ the rotation of the spin space with respect to the orbital space, with $\sin \theta_{0} = q_{2}(2-2q_{1})^{-1}$ corresponding to the minimum energy configuration for $\theta$. The transverse frequency shift with uniform $\theta  = \theta_{0}$ (i.e. the response of the bulk) is given by \cite{DistB}
\begin{equation} \label{eq:freq_pdb_tra}
\frac{\omega_{\perp}^2 -\omega_{\mathrm L}^2}{\Omega_{\mathrm{PdB}}^2} = q_1-q_2^2\,.
\end{equation}

In the axial field, with $\mathbf{H}$ oriented along the anisotropy axis, the homogeneous transverse frequency shift with uniform $\theta  = \theta_{0, \parallel} = \textrm{sgn}(q_2) \pi/2$ is given by \cite{DistB}
\begin{equation} \label{eq:freq_pdb_ax}
\frac{\omega_{\parallel}^2 -\omega_{\mathrm L}^2}{\Omega_{\mathrm{PdB}}^2} = 1+\frac{5}{2} \vert q_2 \vert \,.
\end{equation}

%%%%%%%%%%%%%%%%%%%%%%%%%%%%%%%%%%%%%%%%%%%%%%%%%%%%%%%%%%%%%
%%%%%%%%%%%%%%%%%%%%%%%%%%%%%%%%%%%%%%%%%%%%%%%%%%%%%%%%%%%%%
\begin{figure}
\centering
 \includegraphics[width=0.5\linewidth]{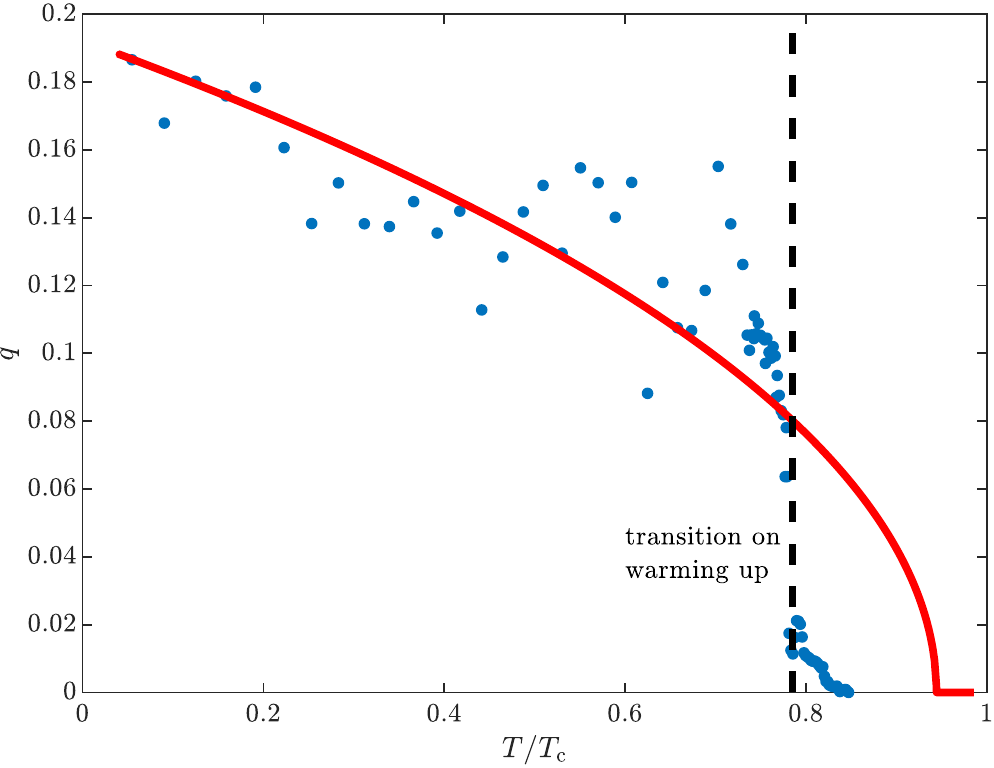}
 \caption{ \textbf{The measured distortion parameter $\bm{q}$ as a function of temperature.} The dots represent the measured values for $q$. The solid red line is an estimation of $q$, calculated based on Ginzburg-Landau theory with strong-coupling corrections using two fitting parameters in the spirit of Ref.~\cite{DistB} and taking $\beta$ parameter values from Ref.~\cite{GLstrong}. The PdB phase critical temperature is shown by dashed line for the transition to the PdA phase on warming. The jump in $q$ at this temperature reflects the fact that the PdB-PdA transition is of the 1st order.} 
 \label{gap_ratio}
\end{figure}
%%%%%%%%%%%%%%%%%%%%%%%%%%%%%%%%%%%%%%%%%%%%%%%%%%%%%%%%%%%%%
%%%%%%%%%%%%%%%%%%%%%%%%%%%%%%%%%%%%%%%%%%%%%%%%%%%%%%%%%%%%%

The $q$-parameter value is determined from the frequency shifts in Eqs.~(\ref{eq:freq_pdb_tra}) and (\ref{eq:freq_pdb_ax}), following a method described in Ref.~\cite{DistB}. In the experimental region of interest, the distortion factor is given by
\begin{equation}
 q = \frac{2-5C}{4} - \frac{1}{4}\sqrt{25C^{2} - 36C + 4}, \label{eq:qeq}
\end{equation}
where
\begin{equation}
 C = \frac{\omega_{\perp} - \omega_{\mathrm{L}}}{\omega_{\parallel} - \omega_{\mathrm{L}}}.
\end{equation}
The expression~(\ref{eq:qeq}) is valid in the range $q \in [0,(\sqrt{14}-2)/5 \approx 0.348]$. To measure $q$, we carefully prepare the state by cooling the sample through the superfluid transition temperature at zero rotation in the transverse magnetic field to avoid creation of half-quantum vortices. Then we cool the sample down to the lowest temperatures and start warming it up slowly, continuously monitoring the NMR spectrum. We perform two temperature sweeps, first in the axial and then in the transverse field. This way we can measure the $q$ parameter in the coexistence region of the PdA and PdB phases. The results of our measurements are shown in Fig.~\ref{gap_ratio}.

In the transverse magnetic field $\mathbf{H}$ exceeding the dipolar field $\sim 3$~mT, the order parameter vector $\hat{\mathbf{e}}^2$ in Eq.~(\ref{eq:distBop}) becomes locked along the field, while vectors $\hat{\mathbf{d}}$ and $\hat{\mathbf{e}}^1$ are free to rotate around the axis $\hat{\mathbf{y}}$, directed along $\mathbf{H}$, with the angle $\theta$ between $\hat{\mathbf{d}}$ and $\hat{\mathbf{z}}$. The order parameter of the PdB phase on a loop around a HQV has the following properties. The phase $\phi$ around the HQV core changes by $\pi$ and the angle $\theta$ (and thus vectors $\hat{\mathbf{d}}$ and $\hat{\mathbf{e}}^1$) winds by $\pi$. Consequently, there is a phase jump $\phi \rightarrow \phi + \pi$ and related sign flips of vectors $\hat{\mathbf{d}}$ and $\hat{\mathbf{e}}^1$ along a perpendicular direction. In the presence of order-parameter components with $q>0$, Eq.~(\ref{eq:distBop}) remains single-valued if, and only if, $q_{2}$ also changes sign. We conclude that the resulting domain wall separates the degenerate states with $q_2 = \pm q$ and together with the bounding HQVs has a structure identical to the domain wall bounded by strings -- the KLS wall -- proposed by Kibble, Lazarides, and Shafi in cosmological context \cite{Kibble1982a,Kibble1982b} and discussed in detail in Sec.~\ref{SolitonNexusPdB}.

\begin{figure*}%[htt]
\includegraphics[width=\linewidth]{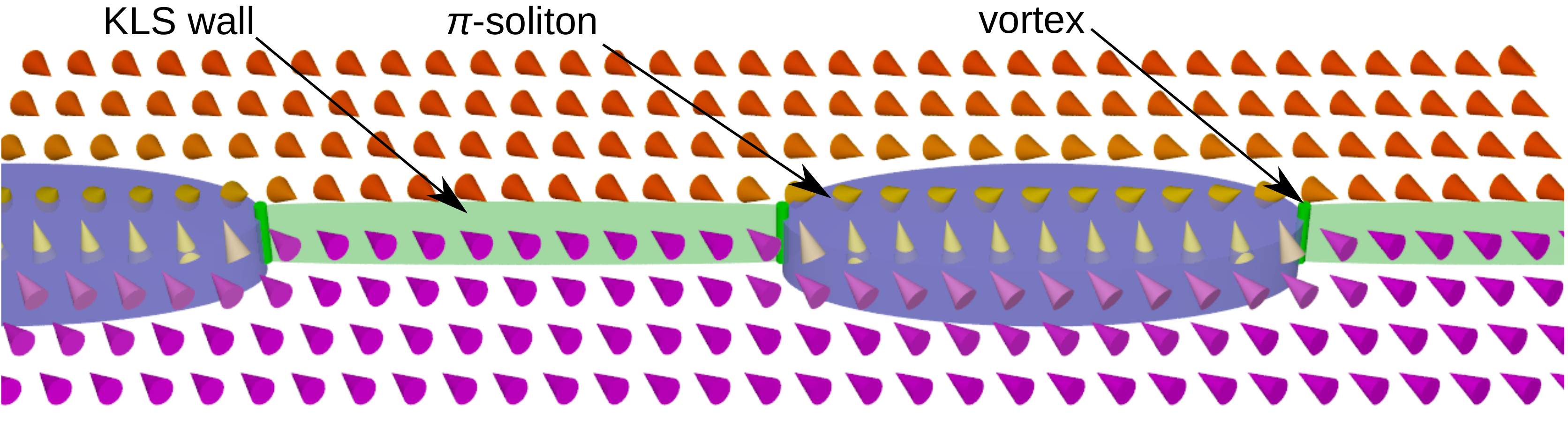}%[width=0.8\linewidth]
\caption[Dummy.]{\textbf{Composite defect of KLS wall and spin ${\pi}$-soliton connected by HQVs.} Each HQV core terminates one soliton - reorientation of the spin part of the order parameter denoted by the angle $\theta$ - and one KLS wall. The orientation of the $\hat{\mathbf{d}}$-vector is shown as cones where their color indicates the angle $\theta$, based on numerical calculations. The $\pi$-soliton (blue), responsible for the observed NMR feature, corresponds to a situation where the  KLS  wall (light green) is bound between a different pair of HQV cores (green) than the soliton. The order parameter is continuous as the KLS walls are accompanied by virtual jumps, where $\phi \rightarrow \phi + \pi$, $\theta \rightarrow \theta + \pi$, and $q_2 \rightarrow -q_2$.}
\label{WallFigCombined}
\end{figure*}

The KLS  wall and the topological soliton  have distinct defining length scales \cite{Volovik1990,Thuneberg2014}, see Fig.~\ref{Degenerate_Spaces}. The KLS wall has a hard core of the order of $q^{-1} \xi_0$ and the soliton has a soft core of the size of the dipole length $\xi_{\mathrm{D}} \gg q^{-1} \xi_0$. The combination of these two objects may emerge in two different configurations, see Fig.~\ref{IllustrationsOfInseparableAndSeparableSolitons}. The observed frequency shift suggests the configuration, depicted in Fig.~\ref{WallFigCombined}, where $\pi$-soliton and KLS wall connect at the HQV acting as nexus.

%%%%%%%%%%%%%%%%%%%%%%%%%%%%%%%%%%%%%%%%%%%%%%%%%%%%%%%%%%%%%
%%%%%%%%%%%%%%%%%%%%%%%%%%%%%%%%%%%%%%%%%%%%%%%%%%%%%%%%%%%%%
%%%%%%%%%%%%%%%%%%%%%%%%%%%%%%%%%%%%%%%%%%%%%%%%%%%%%%%%%%%%%
%%%%%%%%%%%%%%%%%%%%%%%%%%%%%%%%%%%%%%%%%%%%%%%%%%%%%%%%%%%%%

\begin{figure*}%[htt]
\centering
 \includegraphics[width=0.7\linewidth]{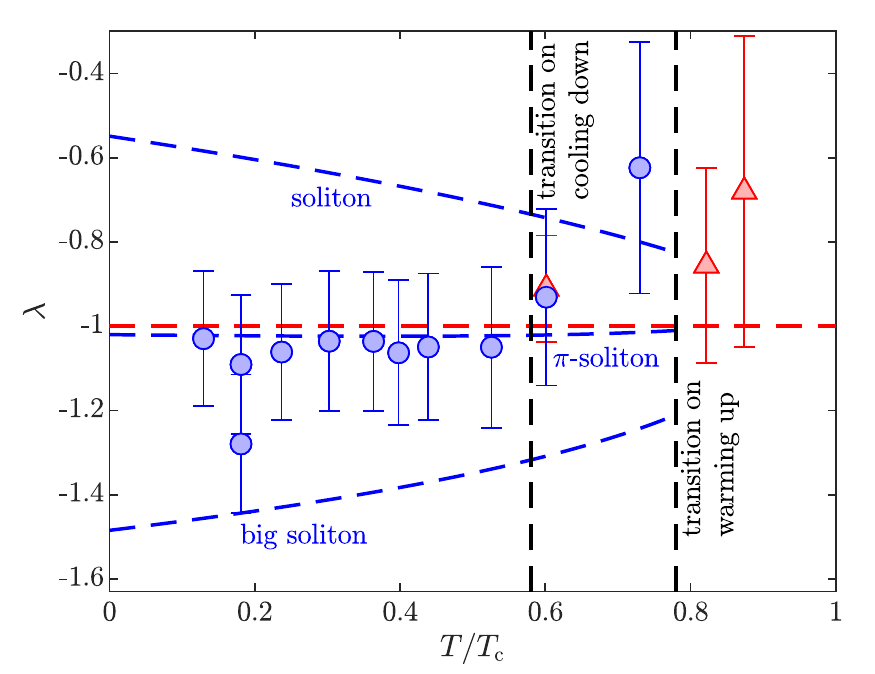}
\caption{\textbf{Frequency shift the HQV satellite as a function of temperature in the polar-distorted phases.} In the PdA phase the measured values are in reasonable agreement with the theoretical prediction for a $\hat{\mathbf d}$-soliton with $\pi$ winding, shown as the red dashed line. The corresponding values in the PdB phase for the lowest-energy $\hat{\mathbf{d}}$-soliton (soliton) and its antisoliton (big soliton), as well as the KLS soliton and the $\pi$-soliton (see text) are shown as dashed blue lines. The experimental data is taken from \cite{Makinen2019} and the dashed lines are based on 2D numerical calculations, Sec.~\ref{PdBshift}, using $D = 20 \xi_{\rm D}$. The error bars denote the uncertainty in the position of the satellite peak by 1.0~kHz and 0.5~kHz in the PdB and PdA phases, respectively. The uncertainty is taken as the full width at half maximum (FWHM) of the satellite peak in the PdB phase and as half of the FWHM due to improved signal-to-noise ratio in the PdA phase.} 
 \label{fig:pdb_lambda}
\end{figure*}

The appearance of KLS walls and the associated $\hat{\mathbf{d}}$-solitons leads to a characteristic frequency shift
\begin{equation}
 \Delta \omega_{\mathrm{PdBsat}} = \omega_{\mathrm{PdBsat}} - \omega_{\mathrm{L}} \approx \lambda_{\mathrm{PdB}} \frac{\Omega_{\mathrm{PdB}}^{2}}{2 \omega_{\mathrm{L}}},
\end{equation}
The dimensionless parameter $\lambda_{\mathrm{PdB}}$ has been calculated in Sec.~\ref{PdBshift}. Comparison of calculations with the experiment for all possible soliton structures from Fig.~\ref{ClassesOfSolitons} is shown in Fig.~\ref{fig:pdb_lambda}. Numerical calculations give the low-temperature values $\lambda_{\mathrm{soliton}} \sim -0.6$ for the soliton ($\Delta \theta = \pi-2\theta_{0}$) and $\lambda_{\mathrm{big}} \sim -1.5$ for its antisoliton, the big soliton ($\Delta \theta = \pi + 2\theta_{0}$). The KLS-soliton ($\Delta \theta = 2 \theta_{0}$) provides too shallow potential to have a bound spin-wave state distinguishable from the main line. The last possibility, the inseparable $\pi$-soliton, gives $\lambda_\pi \approx -1$ for the accessible temperature range, in good agreement with the measured value, $\lambda_{\mathrm{PdB}} \sim -1$. The measured values for $\lambda_{\mathrm{PdB}}$, together with the fact that the total winding of the $\hat{\mathbf{d}}$-vector is also equal to $\pi$ in the PdA and polar phases, suggest that the observed soliton structure in the PdB phase corresponds to the $\pi$-soliton in the presence of a KLS wall.

%%%%%%%%%%%%%%%%%%%%%%%%%%%%%%%%%%%%%%%%%%%%%%%%%%%%%%%%%%%%%
%%%%%%%%%%%%%%%%%%%%%%%%%%%%%%%%%%%%%%%%%%%%%%%%%%%%%%%%%%%%%
%%%%%%%%%%%%%%%%%%%%%%%%%%%%%%%%%%%%%%%%%%%%%%%%%%%%%%%%%%%%%
%%%%%%%%%%%%%%%%%%%%%%%%%%%%%%%%%%%%%%%%%%%%%%%%%%%%%%%%%%%%%

In addition, the  KLS  wall possesses a tension $\sim \xi q^{2} \Delta_{\mathrm{PdB}} ^{2} N_{0}$ \cite{Thuneberg2014,SalomaaVolovik1988}, where $N_{0}$ is the density of states. Thus the presence of KLS walls applies a force pulling the two HQVs at its ends towards each other. The fact that the number of HQVs remains unchanged through the phase transitions signifies that the  KLS  wall tension does not exceed the maximum pinning force in the studied nafen sample. Strong pinning of single-quantum vortices in B-like phase in silica aerogel has also been observed previously \cite{PhysRevLett.94.075301}. An alternative way to remove a  KLS  wall is to create a hole within it, bounded by a HQV \cite{Kibble1982a}. However, growth of such a HQV ring is also prohibited by the strong pinning by the nafen strands. We also note that for significantly larger values of $q$ creation of a HQV vortex-antivortex pair within the KLS wall may become energetically favorable and as a result the HQV pairs bounded by KLS walls would eventually shrink to singly-quantized vortices.

The satellite intensity in the PdB phase, c.f. Fig~\ref{fig:distB_satellites}~(b), scales as $\sqrt{\Omega}$ -- as in the polar and PdA phases. Although the scaling is identical, one striking difference appears in the PdB phase -- the satellite intensity normalized to the total absorption integral in the PdB phase is smaller by a factor of $\sim 9$ compared to the PdA phase. Simultaneously, the original satellite intensity in the PdA phase is restored after a thermal cycle. There is currently no explanation for this observation.

Another effect of rotation in the PdB phase transverse ($\mu = \pi/2)$ NMR spectrum is observed at the main peak, c.f. Fig.~\ref{fig:distB_satellites}~(a). The full-width-at-half-maximum (FWHM), extracted from the amplitude of the main peak assuming $w\cdot h=\mathrm{const}$, where $w$ is its width and $h$ is height, scales as $\propto \sqrt{\Omega}$; Fig.~\ref{fig:distB_satellites}~(c). Increase in the FWHM may indicate that the presence of  KLS  walls enhances scattering of spin waves and thus results in increased dissipation. The small frequency shift of this feature may originate from spin waves weakly bound between KLS walls.

\section{Conclusions and outlook} \label{sec:conclusions}

In superfluid $^3$He topological solitons are manifested as localized winding of the order parameter anisotropy vectors, connecting two energetically degenerate regions with one another. The soliton width is set by the magnetic or the spin-orbit coupling energy arising from deflection of the order parameter from the lowest-energy configuration within the soliton. The topological protection of the soliton structure is given by the $\pi_1$ homotopy group, which is the same as for linear topological defects, quantized vortices. Thus solitons may terminate in bulk liquid on linear topological defects, in addition to the possibility to terminate at the sample boundary. Vortex cores typically are of coherence length size, and are too small to be directly observed with the available experimental methods. Vortex-bound solitons provide an experimental tool for accessing properties of a range of exotic objects via NMR methods. 

In this review article, we have shown the connection between solitons and the underlying order parameter symmetries, and how solitons are manifested in the experiments. So far experiments have demonstrated three well-identified cases of solitons terminated at the quantized vortices: Solitons bounded by HQVs in the polar and PdA phases, solitons accompanying the KLS-walls in the PdB phase, and solitons connecting spin-mass vortices in the B phase of $^3$He. The frequency shift of the associated feature in the NMR spectrum in all cases agrees with theoretical predictions. This observation, together with the scaling of the soliton peak intensity with the angular velocity of the sample container confirmed the existence of HQVs \cite{Autti2016} and of the spin-mass vortices \cite{Kondo1992}. The soliton peak has also been used to study the Kibble-Zurek mechanism of topological defect formation, in particular to demonstrate variety of defect types created by the KZM \cite{Eltsov2000} and to find modification of the KZM in the presence of a symmetry-breaking bias field \cite{KZSupp}, where the number of defects created in the phase transition was suppressed by applying magnetic field tilted with respect to the system symmetry axis during the superfluid transition. In the PdB phase, the NMR properties of the spin soliton were used to identify the type of soliton accompanying the KLS walls, where multiple possibilities existed \cite{Zhang2020, Makinen2019} -- the experiments and numerics are in beautiful agreement. More recent numerical work on a spinor BEC \cite{Hiromitsu2022,Rudd2021} studies the effect of the strong coupling correction on the domain wall structure and the instability of the domain wall induced by spin current.

In conclusion, vortex-bound solitons are ubiquitous in different superfluid phases of $^3$He. They have proven important for experimental identification of composite topological defects. Some composite topological defects, such as the KLS walls bounded by strings, require a specific hierarchy of symmetry-breaking phase transitions in order to be realized. Simply proving their existence in the cosmological vacuum would be a major step forward in the study of the early times of our universe, immediately ruling out some grand unified theories \cite{Kibble1982a}. Additionally, it has been proposed that solitons could provide observable signatures of axion dark matter \cite{AxionSolitons}. Recently, topological classifications of Yang-Mills solitons, non-Abelian sine-Gordon solitons, and skyrmions in quantum field theories and quantum chromodynamics have also been discussed in terms of composite objects \cite{Nitta2022a}. HQVs, identified by the NMR signature of the vortex-bound solitons, hold promise for topological, or error-tolerant, quantum computation with non-Abelian Majorana core-bound states \cite{PhysRevLett.86.268}. Therefore, understanding and identifying vortex-bound solitons is important for a number of research directions.

\ack

We thank Grigori Volovik, Jaakko Nissinen, George Lazarides and Qaisar Shafi for useful discussions. This work was supported in part by Academy of Finland (grant 332964), by the European Union’s Horizon 2020 research and innovation programme under Grant Agreement No. 824109, and by UK STFC grant number ST/T00682X/1.

\section*{Data availability statement}

The data that support the findings of this study are available upon request from the authors.

\appendix

\section{$A_{1}$ phase -- Order Parameter and Remaining Symmetry}
\label{A1Phase}
In bulk fluid, the A$_1$ phase exists only at a very narrow temperature region close to $T_{\mathrm{c}}$ in the presence of magnetic field. The A$_1$ phase consists of Cooper pairs with both spins oriented along the direction of the magnetic field. The order parameter of the A$_1$ phase can be written as
\begin{equation}
 A_{\mu j}^{\mathrm{A_{1}}} = \Delta_{\mathrm{A_1}} e^{i \phi} (\hat{\mathbf{d}}_\mu + i \hat{\mathbf{e}}_\mu^1 ) (\hat{\mathbf{m}}_{j} + i\hat{\mathbf{n}}_{j}),
\end{equation}
where $\hat{\mathbf{d}}$ and $\hat{\mathbf{e}}^1$ form an orthogonal triad in spin space with vector $\hat{\mathbf{e}}^2 = \hat{\mathbf{d}} \times \hat{\mathbf{e}}^1$ and $\Delta_{\mathrm{A_{1}}}$ is the maximum gap in the A$_1$ phase. Similar to the A phase, a gauge transformation $A_{\mu j}^{\mathrm{A_{1}}} \rightarrow e^{i \Delta \phi} A_{\mu j}^{\mathrm{A_{1}}}$ can be compensated by simultaneous rotation of the orbital space by $- \Delta \phi$ about $\hat{\mathbf{l}}$, corresponding to remaining $U(1)_{\phi+\mathbf{L}}$ symmetry. On the other hand, the spin part of the order parameter has identical structure to the orbital part, and the gauge transformation can be compensated also by rotation of the spin space around $\hat{\mathbf{e}}^2$ by $- \Delta \phi$, resulting in a conserved $U(1)_{\phi+\mathbf{S}}$ corresponding to simultaneous rotation of the spin space and change of phase. Alternatively, one of the $U(1)$ symmetries may be interpreted as a simultaneous rotation of the orbital and spin spaces. Additionally, the time-reversal symmetry is partially broken due to the non-zero imaginary part both in orbital and spin spaces, reducing it to the combined discrete symmetry $\mathbb{Z}_{2 (T+\mathbf{J})}$ corresponding to simultaneous time reversal, $\pi$-rotation of the orbital space about $\hat{\mathbf{m}}$, and $\pi$ rotation of the spin space about $\hat{\mathbf{d}}$. Thus, the order parameter of the A$_1$ phase is invariant under \cite{vollhardt2013superfluid}
\begin{equation}
 H_{\mathrm{A_1}} = U(1)_{\phi+\mathbf{L}} \times U(1)_{\phi+\mathbf{S}} \times \mathbb{Z}_{2 (T+\mathbf{J})} \times C \times PU_\pi\,.
\end{equation}

\section{Length Scales in The Presence of Orientation Energies}
\label{LengthScales}
The magnetic healing length $\xi_{\rm H}$ is determined by the competition of the gradient energy density 
\begin{equation}
  \hspace{-2.2 cm} f_{\nabla} = \frac{1}{2} K_1 \partial_{i} A_{\alpha j}^{\rm PdB} \partial_{i} \left( A^{{\rm PdB}}_{\alpha j} \right)^* +\frac{1}{2} K_2 \partial_{j} A_{\alpha i}^{\rm PdB} \partial_{i} \left( A^{{\rm PdB}}_{\alpha j} \right)^* + \frac{1}{2} K_3 \partial_{i} A_{\alpha i}^{\rm PdB} \partial_{j} \left( A^{{\rm PdB}}_{\alpha j} \right)^*,
\label{GradientEnergy}
\end{equation}    
where $K_{1}=K_{2}=K_{3}$ \cite{vollhardt2013superfluid}, with the magnetic energy density, which is
\begin{equation}
f_{\rm H} = -\frac{1}{2} \chi_{\alpha \beta} H_{\alpha} H_{\beta}=\frac{1}{2}\gamma^{2} S_{a}S_{b}(\chi^{-1})_{ab}- \gamma H_{a}S_{a},
\label{MagneticEnergy}
\end{equation}
where the $\chi_{\alpha \beta}$ is the uniaxial magnetic susceptibility tensor of the PdB phase, $H_{\alpha}$ are magnetic field strengths with $\alpha=1,2,3$, $S_{a}$ are spin densities with $a=1,2,3$, and $\gamma$ is the gyromagnetic ratio of $^3$He \cite{vollhardt2013superfluid}. Using Eqs.~(\ref{GradientEnergy}) and (\ref{MagneticEnergy}) allows writing the magnetic length as 
\begin{equation}
\xi_{\rm H}= \sqrt \frac{K_{1}\Delta_{\rm PdB}^{2}}{(\chi_{\bot}-\chi_{\|})H^{2}},
\label{MagneticLength}
\end{equation}
where $\chi_{\bot}$ and $\chi_{\|}$ are transverse and longitude spin magnetic susceptibilities of PdB phase.
Following the same idea, the dipole length $\xi_{\rm D}$ is determined by the gradient energy density $f_{\nabla}$ and the spin-orbit coupling (SOC) energy density 
\begin{equation}
f_{\rm SOC} = \frac{3}{5}g_{\rm D} \left( \left( A^{\rm PdB}_{ii} \right) ^* A_{jj}+  \left( A^{\rm PdB}_{ij} \right)^* A_{ji} - \frac{2}{3} \left( A^{\rm PdB}_{ij} \right) ^* A_{ij} \right), 
\label{EnergyDensityOfSOC}
\end{equation}
where $g_{\rm D}$ is the strength of the spin-orbit coupling. Then we have
\begin{equation}
\xi_{\rm D}= \sqrt{ \frac{5K_{1}}{6g_{\rm D}} }.
\label{DipoleLength}
\end{equation}

\section{Relative Homotopy Groups and Exact Sequences} 
\label{Appendix} 
The homotopy groups and relative homotopy groups of vacuum manifolds $R_{1}$ and $R_{2}$ form a long exact sequence (LES) \cite{NashBook1988}
\begin{equation}
\xymatrix@R=5pt{
... \,\, \pi_{k}(R_2) \ar[r]^(0.6){x^{n-1}_{*}} & \pi_{k}(R_1) \ar[r]^(0.4){x^{n}_{*}} & \pi_{k}( R_1, R_2) \ar[r]^(0.5){{\partial}^{*}} & \pi_{k-1}(R_{2}) \,\,... 
}
\label{LongExactSequenceDefination}
\end{equation}
The exact sequence of (relative) homotopy groups means that the image of any homomorphism $x^{n-1}_{*}:M \rightarrow N$ in Eq.~(\ref{LongExactSequenceDefination}) (the sets of the elements of the group $N$ into which the elements of $A$ are mapped) is the kernel of the next homomorphism $x_{*}^{n+1}:N \rightarrow W$ (the sets of the elements of $N$ which are mapped to the zero or unit element of $W$), i.e. ${\rm Im}\,x^{n}_{*} \cong {\ker}\,x^{n+1}_{*}$ with $n\in \mathbb{Z}$ \cite{NashBook1988}. 

The relative homotopy classes of $\pi_{k+1}(R_{1},R_{2})$ are mapped to the homotopy classes of $\pi_{k}(R_{2})$ by mapping the $k$-dimensional subspace of $k+1$ sphere, which surrounds the defects, into $R_{2}$. This mapping between two homotopy classes with different dimensions is called boundary homomorphism $\partial^{*}$ \cite{NashBook1988}. When this mapping is not trivial, i.e. ${\rm Im}\,\partial^{*} \neq 0$, topological defects given by $\pi_{k}(R_1,R_2)$ can be mapped to those given by $\pi_{k-1}(R_{2})$, living on the $k+1$ sphere enclosing the original defects. In other words, boundary homomorphism describes how topological objects with different dimensionalities connect to one another -- it is therefore a convenient tool for describing composite topological objects. In contrast, the commonly used homotopy group $\pi_{k}(R_{1})$ lacks boundary homomorphism and therefore does not provide information about such connections.

The LES, Eq.~(\ref{LongExactSequenceDefination}), has infinite number of terms. It is therefore useful to split it up to short exact sequences (SES) \cite{suzuki1982}. For every relative homotopy group $\pi_{1}(R_{1},R_{2})$, the LES can be split as
\begin{equation}
\xymatrix@R=8pt{
0 \ar[r] & {\rm Im}\,x^{n}_{*} \ar[r]^-{x^{n}_{*}} & \pi_{k}(R_{1},R_{2}) \ar[r]^-{{\partial}^{*}} & {\rm Im}\,{\partial}^{*} \ar[r] & 0
} 
\label{SESPi2R1R2}
\end{equation}
by the image of $\partial^{*}$ and $x^{n}_{*}$. In this case, the relative homotopy group $\pi_{k}(R_{1},R_{2})$ is an extension of ${\rm Im}\,x^{n}_{*}$ by ${\rm Im}\,{\partial}^{*}$.

\subsection{SESs of $\pi_{n}(R^{\rm PdA}_{1},R^{\rm PdA}_{2})$}
\label{SESsPdA}

Following the definition of LES in Eq.~(\ref{LongExactSequenceDefination}) and the vacuum manifolds mentioned in Sec.~\ref{NafenConfinement}, we have the LESs
\begin{equation}
\xymatrix@R=5pt{
... \,\, \pi_{1}(R^{\rm PdA}_2) \ar[r]^(0.6){x^{n-1}_{*}} & \pi_{1}(R^{\rm PdA}_1) \ar[r]^(0.4){x^{n}_{*}} & \pi_{1}( R^{\rm PdA}_1, R^{\rm PdA}_2) \ar[r]^{{\partial}_1^{*}} & \pi_{0}(R^{\rm PdA}_{2}) \,\,... 
}
\end{equation}
and
\begin{equation}
\xymatrix@R=5pt{
... \,\, \pi_{2}(R^{\rm PdA}_2) \ar[r]^(0.6){x^{n-1}_{*}} & \pi_{2}(R^{\rm PdA}_1) \ar[r]^(0.4){x^{n}_{*}} & \pi_{2}( R^{\rm PdA}_1, R^{\rm PdA}_2) \ar[r]^{{\partial}_2^{*}} & \pi_{1}(R^{\rm PdA}_{2}) \,\,..., 
}
\end{equation}
which in fact are
\begin{equation}
\xymatrix@R=5pt{
... \,\, \mathbb{Z}_{\mathbf{L}} \ar[r]^{x^{n-1}_{*}} & \mathbb{Z}_{\mathbf{L}} \times \mathbb{Z}_{\mathbf{\phi}} \ar[r]^(0.4){x^{n}_{*}} & \pi_{1}( R^{\rm PdA}_1, R^{\rm PdA}_2) \ar[r]^(0.7){{\partial}_1^{*}} & 0 \,\,... 
}
\label{p1LESPdA}
\end{equation}
and
\begin{equation}
\xymatrix@R=5pt{
... \,\, 0 \ar[r]^(0.6){x^{n-1}_{*}} & \mathbb{Z}_{\mathbf{S}} \ar[r]^(0.3){x^{n}_{*}} & \pi_{2}( R^{\rm PdA}_1, R^{\rm PdA}_2) \ar[r]^(0.65){{\partial}_2^{*}} & \mathbb{Z}_{\mathbf{L}} \,\,...
}
\label{p2LESPdA}
\end{equation}
It is easy to see that in Eq.~(\ref{p1LESPdA}) ${\rm Im}\,\partial^{*}_{1} = 0$, while ${\rm Im}\,x^{n-1}_{*} = \mathbb{Z}_{\mathbf{L}}$ as the orbital vertices formed by $\hat{\mathbf{n}}$ (or $\hat{\mathbf{l}}=\hat{\mathbf{m}}\times\hat{\mathbf{n}}$) form same group  $\mathbb{Z}_{\mathbf{L}}$. It follows that $\ker\,x^{n}_{*} \cong \mathbb{Z}_{\mathbf{L}}$ and
\begin{equation}
{\rm Im}\,x^{n}_{*} \cong (\mathbb{Z}_{\mathbf{L}} \times \mathbb{Z}_{\mathbf{\phi}})/\mathbb{Z}_{\mathbf{L}} = \mathbb{Z}_{\mathbf{\phi}},
\end{equation}
leading to the SES of $\pi_{1}( R^{\rm PdA}_1, R^{\rm PdA}_2)$
\begin{equation}
\xymatrix@R=8pt{
0 \ar[r] & \mathbb{Z}_{\phi} \ar[r] & \pi_{1}( R^{\rm PdA}_{1},R^{\rm PdA}_{2}) \ar[r]^-{\partial^{*}} & 0 \ar[r] & 0.
}
\end{equation}
Following similar reasoning, we have ${\rm Im}\,\partial^{*}_{2} = 0$, where $0$ is the kernel of the mapping $x^{n-1}_{*}:\mathbb{Z}_{\mathbf{L}} \mapsto \mathbb{Z}_{\mathbf{L}} \times \mathbb{Z}_{\mathbf{\phi}}$, resulting in
\begin{equation}
\xymatrix@R=8pt{
0 \ar[r] & \mathbb{Z}_{\mathbf{S}} \ar[r] & \pi_{2}( R^{\rm PdA}_{1},R^{\rm PdA}_{2}) \ar[r]^-{\partial^{*}} & 0 \ar[r] & 0.
}
\end{equation}

\subsection{SESs of $\pi_{n}(R^{\rm PdB}_{1},R^{\rm PdB}_{2})$}
\label{SESsPdB}

We now conduct the calculations of SESs for the PdB phase. Similarly to the previous section, we start from
\begin{equation} 
\xymatrix@R=5pt{
... \,\, \mathbb{Z}_{\mathbf{J}} \ar[r]^{x^{n-1}_{*}} & \mathbb{Z}_{2\mathbf{J}} \times \mathbb{Z}_{\mathbf{\phi}} \ar[r]^(0.4){x^{n}_{*}} & \pi_{1}( R^{\rm PdB}_1, R^{\rm PdB}_2) \ar[r]^(0.6){{\partial}_1^{*}} & \mathbb{Z}_{2(\phi + \mathbf{S})} \ar[r] & 0\,\,... 
}
\label{p1LESPdB}
\end{equation}
and
\begin{equation}
\xymatrix@R=5pt{
... \,\, 0 \ar[r]^(0.6){x^{n-1}_{*}} & 0 \ar[r]^(0.25){x^{n}_{*}} & \pi_{2}( R^{\rm PdA}_1, R^{\rm PdA}_2) \ar[r]^(0.7){{\partial}_2^{*}} & \mathbb{Z}_{\mathbf{J}} \ar[r]^(0.4){x^{n+1}_{*}} & \mathbb{Z}_{\phi} \times \mathbb{Z}_{2\mathbf{J}} \,\,...\,,
}
\label{p2LESPdB}
\end{equation}
where
\begin{equation}
\mathbb{Z}_{\mathbf{J}} = \pi_{1}(R^{\rm PdB}_{2}),\,  \mathbb{Z}_{\mathbf{\phi}} \times \mathbb{Z}_{2\mathbf{J}} = \pi_{1}(R^{\rm PdB}_{1}),\, 0 = \pi_{0}(R^{\rm PdB}_{2}) = \pi_{0}(R^{\rm PdB}_{1}).  
\end{equation}
Mapping $ x^{n-1}_{*}:\pi_{1}(R^{\rm PdB}_{2}) \mapsto \pi_{1}(R^{\rm PdB}_{2})$ ($x^{n+1}_{*}$ in Eq.~(\ref{p2LESPdB})) plays a significant role in splitting Eqs.~(\ref{p1LESPdB}) and (\ref{p2LESPdB}). The $\mathbb{Z}_{\mathbf{J}}$ vertices are mapped into $\mathbb{Z}_{2\mathbf{J}}$ through a surjection, i.e. ${\rm Im}\,x^{n-1}_{*} = \mathbb{Z}_{\mathbf{J}}$. A natural choice of $\ker\,x^{n-1}_{*}$ is $2\mathbb{Z}_{\mathbf{J}}$, which is a group of vortices with even winding number \cite{Zhang2020b}. As a result, ${\rm Im}\,\partial^{*}$ in Eqs.~(\ref{p1LESPdB}) and (\ref{p2LESPdB}) are $\mathbb{Z}_{2(\phi + \mathbf{S})}$ and $2\mathbb{Z}_{\mathbf{J}}$, respectively. Since ${\rm Im}\,x^{n-1}_{*} = \mathbb{Z}_{2\mathbf{J}} = \ker\,x^{n}_{*}$ in Eq.~(\ref{p1LESPdB}), we also have 
\begin{equation}
{\rm Im}\,x^{n}_{*} = (\mathbb{Z}_{2\mathbf{J}} \times \mathbb{Z}_{\mathbf{\phi}})/\ker\,x^{n}_{*} = \mathbb{Z}_{\mathbf{\phi}}.
\end{equation}  
Then we get the SES
\begin{equation}
\xymatrix@R=8pt{
0 \ar[r] & \mathbb{Z}_{\phi} \ar[r] & \pi_{1}( R^{\rm PdB}_{1},R^{\rm PdB}_{2}) \ar[r]^-{\partial^{*}} & \mathbb{Z}_{2 (\phi + \mathbf{S})} \ar[r] & 0
}
\label{SESPi1PdBa}
\end{equation}
for $\pi_{1}( R^{\rm PdB}_1, R^{\rm PdB}_2)$ and
\begin{equation}
\xymatrix@R=8pt{
0 \ar[r] & 0 \ar[r] & \pi_{2}( R^{\rm PdB}_{1},R^{\rm PdB}_{2}) \ar[r]^-{\partial^{*}} & 2\mathbb{Z}_{\mathbf{J}} \ar[r] & 0
}
\label{SESPi2PdBa}
\end{equation}
for $\pi_{2}( R^{\rm PdB}_1, R^{\rm PdB}_2)$.

\subsection{SES of $\pi_{1}( R_{1}^{\rm H}, \tilde{R}_{1}^{\rm SOC})$}
\label{SESpi1RHRSOC}

The long exact sequence for $\pi_{1}( R_{1}^{\rm H}, \tilde{R}_{1}^{\rm SOC})$ with reduced vacuum manifolds is written as
\begin{equation} \hspace{-1cm}
\xymatrix@R=3pt{
\pi_{1}(\tilde{R}_{1}^{\rm SOC}) \ar[r] & \pi_{1}(R_{1}^{\rm H}) \ar[r]^-{x^{n}_{*}} & \pi_{1}(R_{1}^{\rm H},\tilde{R}_{1}^{\rm SOC}) \ar[r]^-{\partial^{*}} & \pi_{0}(\tilde{R}_{1}^{\rm SOC}) \ar[r] & \pi_{0}(R_{1}^{\rm H})
},
\label{LES1}
\end{equation}
where $\partial^{*}$ is the boundary homomorphism \cite{NashBook1988,suzuki1982}. Plugging in the homotopy groups of the reduced manifold, Eq.~(\ref{LES1}), gives
\begin{equation}
  \xymatrix@R=5pt{
  \mathbb{Z}_{\phi} \ar[r] & \mathbb{Z}_{\bf S} \times \mathbb{Z}_{\phi} \ar[r]^-{x^{n}_{*}} & \pi_{1}( R_{1}^{\rm H}, \tilde{R}_{1}^{\rm SOC}) \ar[r]^-{\partial^{*}} & \mathbb{Z}_{4} \ar[r] & 0
  }. 
\end{equation}  
It is worth noting that ${\rm Im}\,\partial^{*} = \mathbb{Z}_{4}$ and ${\rm Im}\,x^{n}_{*} = \mathbb{Z}_{\mathbf{S}}$, resulting in
\begin{equation}
  \xymatrix@R=5pt{
  0 \ar[r] & \mathbb{Z}_{\bf S} \ar[r]^-{x^{n}_{*}} & \pi_{1}( R_{1}^{\rm H}, \tilde{R}_{1}^{\rm SOC}) \ar[r]^-{\partial^{*}} & \mathbb{Z}_{4} \ar[r] & 0
  }. 
\end{equation}

\end{document}